\documentclass[12pt,preprint]{aastex}
\usepackage{graphicx}

\usepackage{natbib}

\newcommand{\be}{\begin{equation}}
\newcommand{\ee}{\end{equation}}
\newcommand{\nn}{\mbox{} \nonumber \\ \mbox{} }
\newcommand{\ba}{\begin{eqnarray}}
\newcommand{\ea}{\end{eqnarray}}
\newcommand{\om}{\omega}
\newcommand{\Alfven}{ Alfv\'{e}n }
\newcommand{\curl}{{\rm curl\, }}
\newcommand{\E}{{\bf E}}
\newcommand{\B}{{\bf B}}
\newcommand{\J}{{\bf J}}

\newcommand{\sech}{{\rm \,sech\,}}
\newcommand{\Bf}{{magnetic field}}
\newcommand{\Bfs}{{magnetic fields}}
\newcommand{\Ef}{{electric  field\,}}
\newcommand{\Efs}{{electric fields\,}}

\newcommand{\ms}{magnetosphere}

\newcommand{\EM}{electromagnetic}

\newcommand{\vdiv}[1]{\spr{\nabla}{#1}}
\newcommand{\spr}[2]{\bmath{#1} \!\cdot\! \bmath{#2}}
\newcommand{\vpr}[2]{\bmath{#1} \!\times\! \bmath{#2}}

\newcommand\eg{it{e.g.}}

\newcommand\lo{\mathrel{\raise.3ex\hbox{$<$}\mkern-14mu\lower0.6ex\hbox{$\sim$}}}
\newcommand\go{\mathrel{\raise.3ex\hbox{$>$}\mkern-14mu\lower0.6ex\hbox{$\sim$}}}

\begin{document}

\title{Tearing  of    charged  current layers}
\author{Maxim Lyutikov\\
Department of Physics and Astronomy,  Purdue University, \\
 525 Northwestern Avenue,
West Lafayette, IN
47907-2036 }

\begin{abstract}
Astrophysical current layers, e.g.,  in pulsar winds,  can be   electrically charged, while  the plasma  is charge-symmetric, $e^\pm$.
  Using PIC simulations,  we investigate  dynamics and  plasmoid formation  (tearing instability) in charged  Harris-type  
and    rotational current layers. Electrically charged current layers, initially in global  force-balance,  are  {\it electrostatically} unstable:   the resulting dynamics is an intricate interplay between electrostatic  Bernstein waves (BWs) and the current  tearing mode.  Besides overall density and \Bf,  plasma temperature is an important factor. 
In  the  charged Harris sheet set-up, the quickly generated BW are  trapped  within the layers (internally reflected at the upper hybrid resonance).
BWs quickly redistribute the charge    modifying  the initial stage of tearing, but without strongly affecting overall plasmoid growth; resulting plasmoids are mildly charged.
 In     rotational current layers: (i) even initially overall uncharged configurations develop large fluctuations of charge density; (ii) overall dynamics depends on the initial overall temperature; (iii) for certain combination of parameters tearing rate  is greatly increased in the charged case.
\end {abstract}

\maketitle

\section{Introduction}

Tearing of current layers is one the most  basic plasma processes \citep{furth,Galeev78,1977PhFl...20.1341D}. 
In relativistic plasmas it was considered in  \cite{1979SvA....23..460Z,zenitani_01,zenitani_07,2025ApJ...979..104D,2025ApJ...992..193G,2025JPlPh..91E..42S}. (Relativistic may mean  plasma temperature and/or highly magnetized plasma, with \Alfven velocity close to the speed  of light.) 

Resistive force-free tearing mode was first considered by \cite{2003MNRAS.346..540L} and simulated by \cite{2007MNRAS.374..415K}.    The results, when expressed in term of \Alfven and  resistive times,  surprising matched the non-relativistic expression. It was observed in \cite{2007MNRAS.374..415K} that force-free equations can be put into an MHD-like form.

In the field of plasma-astrophysics, the tearing mode is bound to be important in a number of set-ups, especially  the pulsar current sheets \cite{1990ApJ...349..538C,2014ApJ...780....3U}. Tearing mode, often called the plasmoid formation, is the dominant model of collisionless plasma reconnection \cite{2007PhPl...14j0703L,2010PhPl...17f2104H}. 

Relativistic tearing mode, especially in astrophysical applications, has another hitherto unexplored property: the current layer may be electrically charged. The prime example is \cite{1973ApJ...180L.133M} solution for pulsar winds. 
In this paper we explore the plasma physics properties of relativistic charged current layers.

A comment on relativistic transformation is due. Locally, a current in one system of reference, by Lorentz transformation, corresponds to a charge density in  some other frame -  we are not interested in those trivial cases. We consider  set-ups, mildly nonlocal, so that even if by Lorentz transformation a local charge density can be made zero, there is a  spatially   distributed currents/charges that cannot  be Lorentz reduced  by a  {\it local} transformation, see more detailed discussion in \S \ref{Michel}.

First, in Sections \ref{setup} and \ref{rotational} we describe the results of numerical simulations, and later in \S \ref{exaplain} we give explanations of  the  results.

\section{Charged current layers in two-fluid set-up}
\label{setup}

Historically, two set-up were used to study tearing mode: rotational force-free current sheet, and pressure supported Harris-type configuration \citep{1962NCim...23..115H}. (In pulsars, the equatorial current sheet is of the Harris type). 
 Here we adapt both set-up to include non-zero charge density (\EM\ shear).  We first start with charged Harris configuration, and in \S \ref{rotational} consider charged rotational current sheet. 

Two steps are required to set-up a charged current sheet: (i) global equilibrium configuration in force balance (this involves \EM\ fields, total charges and current, and for Harris sheet a pressure  prescription); 
(ii) local prescription for velocity,   density  and pressure of   each species to match  the global constraint.  \citep[See also related discussion][]{2026PhRvD.113b3045M}

In case of pair plasma,  the point  (ii) above  has a simplification: both charges contribute equally - both to the current, and to the charge density.

We first discuss the corresponding set-up for a single layer (Sections \ref{rotational} and \ref{ChargedHarris}). 
In simulations, we use double current layer.  The requirement  to use double layers are different, and somewhat subtle in each case, see \S \ref{needfordouble}.

For notations, we use the following abbreviations: Charged Harris Current  sheet (CHCS), Charged Rotational Current Sheet  (CRCS).

\subsection{Charged current layer in pulsar winds  (Michel's equatorial current sheet).}
\label{Michel}

We use an abbreviated notations with elementary charge, electron mass and speed of light set to unity.  In these notation, \Bf\ $B _0 \equiv \omega_B = e B_0/(m_e c)$. We use  natural cgs units  (a factor $1/(4\pi) $  is incorporated into definitions of charge). The corresponding relation look SI-like, with $\epsilon_0$ and $\mu_0$ set to unity. 

\cite{1973ApJ...180L.133M} solution for split-monopole \ms can be written as (in $r-\theta-\phi$  coordinates)
    \ba && 
    \B =\left\{\frac{1}{\tilde{r}^2},0,\frac{\sin (\theta )}{\tilde{r}}\right\} B_0 \times {\rm sign} (\pi/2- \theta)
    \nn &&
    \E =\left\{0,\frac{\sin (\theta )}{\tilde{r}},0\right\} B_0 \times {\rm sign} (\pi/2- \theta)
    \nn &&
    \beta_{EM} = \left\{\frac{\tilde{r}^2 \sin ^2(\theta )}{\tilde{r}^2 \sin ^2(\theta
   )+1},0,-\frac{\tilde{r} \sin (\theta )}{\tilde{r}^2 \sin ^2(\theta )+1}\right\}
   \nn && 
   \rho_e = 2  \frac{B_0 \Omega  |\cos (\theta )| }{  \tilde{r}^2}
   \nn && 
   {\bf j} = 2 \left\{\frac{B_0 \Omega | \cos (\theta )|}{ c \tilde{r}^2},0,0\right\} 
   \nn &&
    \tilde{r} = \frac{ r \Omega}{c} 
     \label{fieldsHarris}
   \ea
   (in regular  cgs units, $\rho_e $ and $j$ need to be divided by $4\pi$.

\begin{figure}[h!]
\includegraphics[width=.99\linewidth]{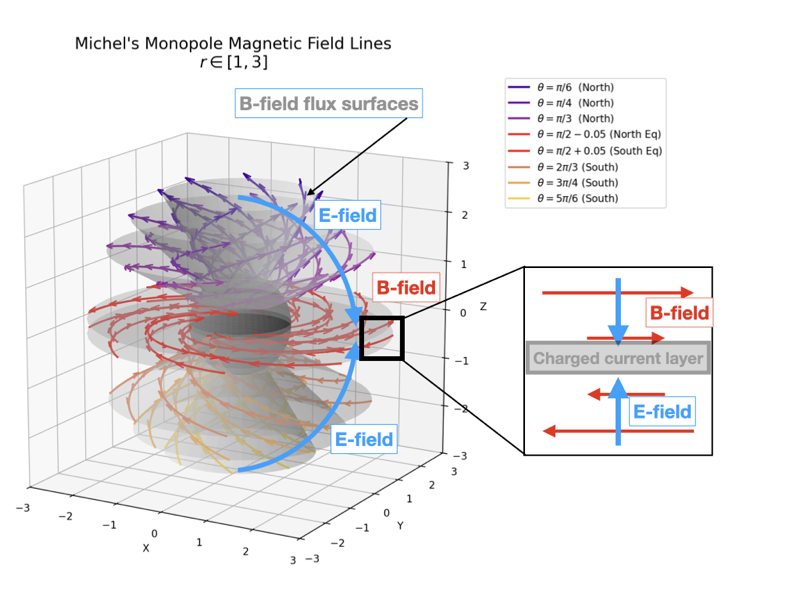}
\caption{Michel's solution for pulsar wind. Zoomed-in panel shows charged Harris-like  current sheet (total \Bf\  is zero at the equator) .}
\label{Tearing-Charged-PI}
\end{figure}

The  electric potential for Michel solution can be written as (assuming $\Phi=0$ on the axis),
  \be
  \Phi =  \frac{B_0 c (1-|\cos (\theta )|)}{\Omega } 
  \ee
  Thus, $ \Phi$ has an extremum in the equatorial plane. In the two-fluid approach,  for the Michels' force-free solution one sign of charges is at the maximum of the electric potential, while  another is at the minimum. As we will demonstrate below, this leads to fast electrostatic instability on Bernstein waves.

   The flow    has equatorial charge density and surface current
   \ba && 
   \sigma_E = 2\frac{B_0}{  \tilde{r}}
   \nn &&
   g=  2 
   \frac{ \sqrt{1+  \tilde{r}^2}}{   \tilde{r}^2}B_0 
   \ea
   
  The \EM\ structure is space-like, $g^2  > \sigma_E^2$.  By boosting  with  $  \beta_{EM} $,  {\it locally} \Ef\ can be  eliminated. But  this cannot be done globally, or within   a finite volume.  For example, consider a local box near the equator $\theta  \approx \pi/2$. Let the region of interest occupy $| \theta - \pi/2 | \leq  (\delta \theta)$, $ (\delta \theta)  \ll 1 $. Making a boost corresponding to  $ \theta_0= \pi/2 \pm   (\delta \theta) $, the resulting \Ef\ is 
  \be
  E_\theta \propto ( \sin \theta   - \cos (\delta \theta) )  \times {\rm sign} (\pi/2- \theta)
  \label{EMichel}
  \ee
  It is zero at  $ \theta = \pi/2 \pm  \theta $ {\it and} zero at $\theta = \pi/2$.

\subsection{Resolved  charged Harris current sheet} 
\label{ChargedHarris}

To approximate the behavior (\ref{EMichel})  (zero \Ef\  at the boundaries and in the middle), in a local simulation box we then set 
\ba &&
\B =  \left\{\tanh \tilde{z},0,0\right\} B_0
\nn &&
\E=\left\{0,0,\tanh \tilde{z} {\rm sech}^2\tilde{z}\right\} \beta_0 B_0
  \ea
  This \Ef\ models the local zero values at  $ \theta  =   \pi/2 \pm (\delta \theta) $ ($ z\to \pm \infty$),   and the requirement  of zero \Ef\ at  $\theta = \pi/2$ ($z=0$).

Assuming globally constant temperature  $\Theta \equiv k_B T /(m_e c^2)$, the corresponding relations for the charged Harris current sheet are
\ba &&
n= n_0 (1\pm f_p)
\nn &&
f_p=\frac{\sigma  }{4 \Theta }  \left(1+ \beta _0^2 \tanh
   ^2\tilde{z} {\rm sech}^2\tilde{z}\right) {\rm sech}^2\tilde{z}
   \nn &&
p= 2 n \Theta
\nn &&
n_\pm = n   \pm ( \delta n)
\nn &&
 ( \delta n)= (2 - \cosh 2  \tilde{z}) \sech^4  \tilde{z} \times \frac{ \beta_0 B_0}{2 L}
\nn &&
\rho_e  =  2  ( \delta n)
\nn &&
\beta_{EM} = \frac{ \E \times \B}{B^2}=  \left\{0,\beta _0 {\rm sech}^2\tilde{z},0\right\}
\nn &&
v_d =\frac{B_0 }{2 L n} {\rm sech}^2\tilde{z}
   \nn &&
\left.   \nabla \times \B \right|_y= (n_p+n_e) v_d
\nn &&
\rho_e \E + ( \nabla \times \B) \times \B + \nabla p =0
\nn &&
\Phi = \frac{L}{2} \sech^2 \tilde{z} \beta_0 B_0
\label{setupHarris}
   \ea
   Drift velocity $v_d$ is positive for positrons and negative for electrons. We stress that for charged  Harris layer, both the overall density $n$ and density of each  component $n_\pm$ are $z$-dependent.  
   
   The 4-current is space-like, $\rho_e^2 - j^2 \leq 0$. Total electric charge is zero in each semi-space;  it is composed of layer of negative charge $|\tilde{z} | \leq 0.65847$ of value $Q=2 \times 0.384 \beta_0 B_0$, and opposite at larger $z$.
   
   Importantly, electric potential $\Phi$ has a maximum in the center. As a result, in two-fluid formulation,  at the center (where \Bf\ is zero)  one sign of charge is at the maximum, another at the minimum.

\begin{figure}[h!]
\includegraphics[width=.53\linewidth]{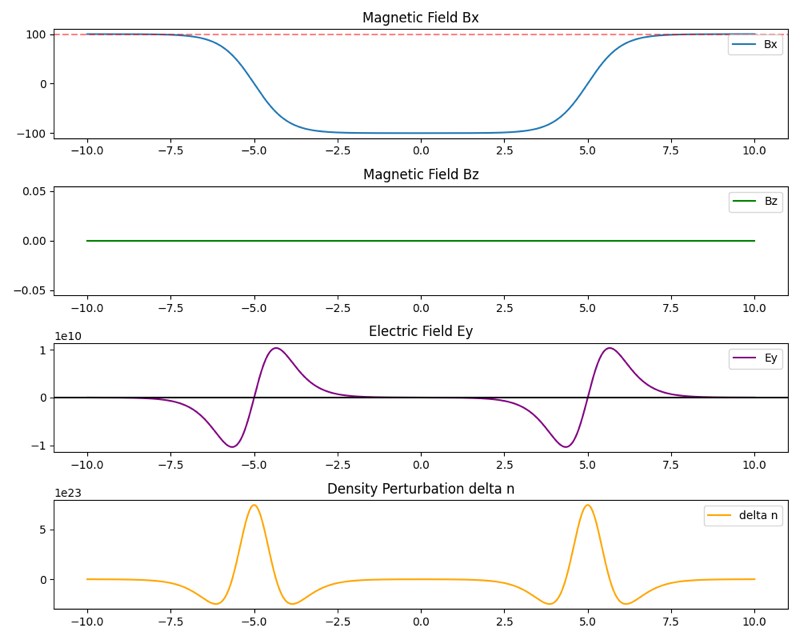}
\includegraphics[width=.42\linewidth]{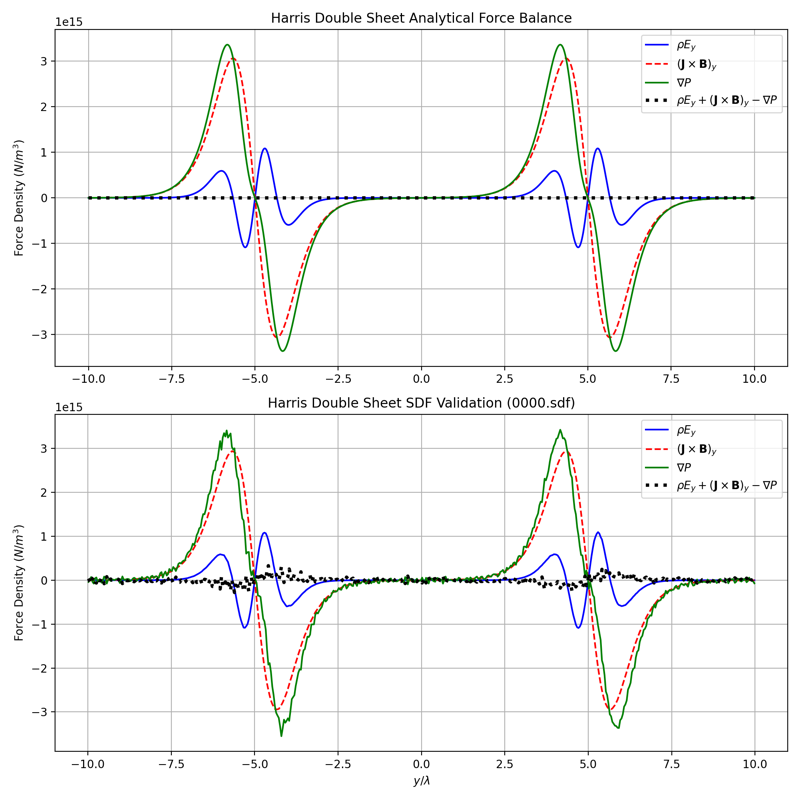}
\caption{Set-ups   for  double  Harris sheet, and a check of initialization.  There is some freedom in choosing the relative orientation of  \Efs\ between the layers that does not significantly affect the results (since two layers are sufficiently  disconnected at the set-up).  }
\label{configurations} 
\end{figure}

Using analytical set-ups for rotational and charged current sheets \S \ref{setup}, we set {\it  double}  current sheet configurations,  Fig. \ref{configurations} and Appendix  \ref{needfordouble}.

\subsection{Results of simulation: charged Harris}

\begin{figure}[h!]
\includegraphics[width=.34\linewidth]{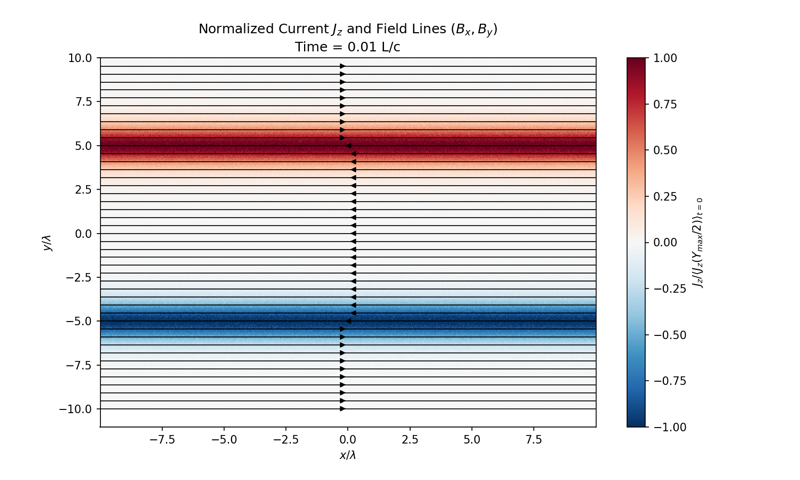} \vline
\includegraphics[width=.34\linewidth]{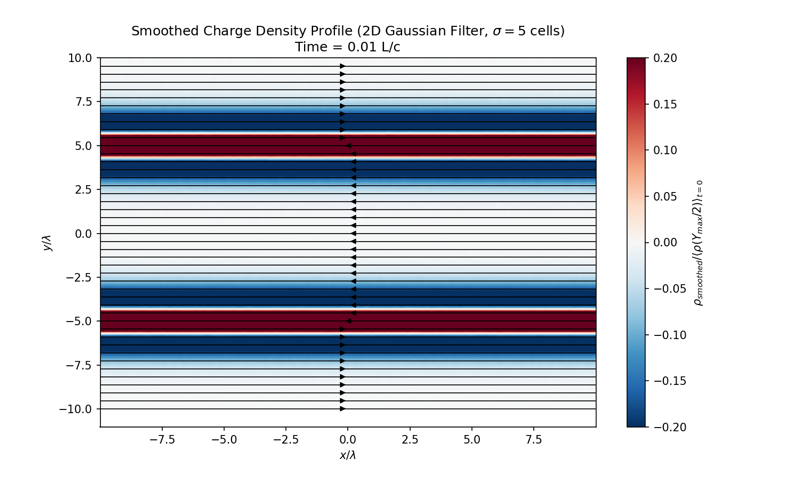}\vline
\includegraphics[width=.3\linewidth]{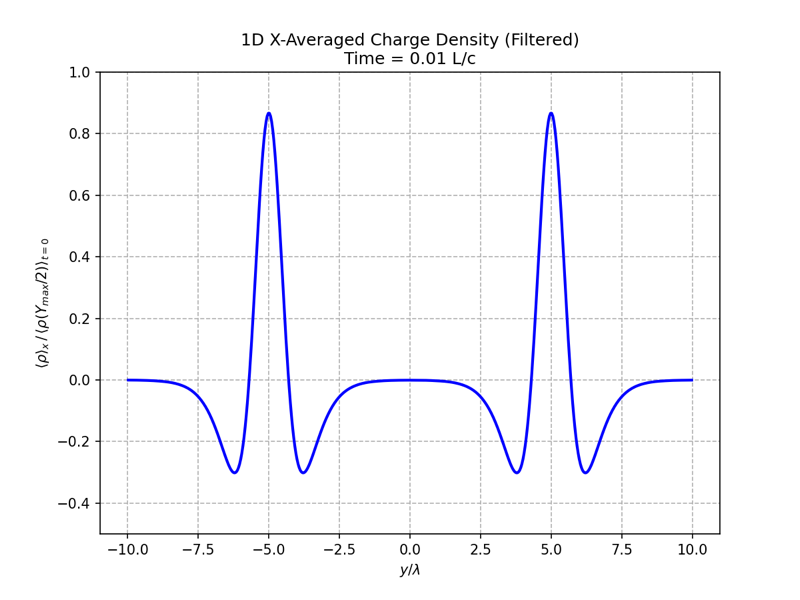}\\
\includegraphics[width=.34\linewidth]{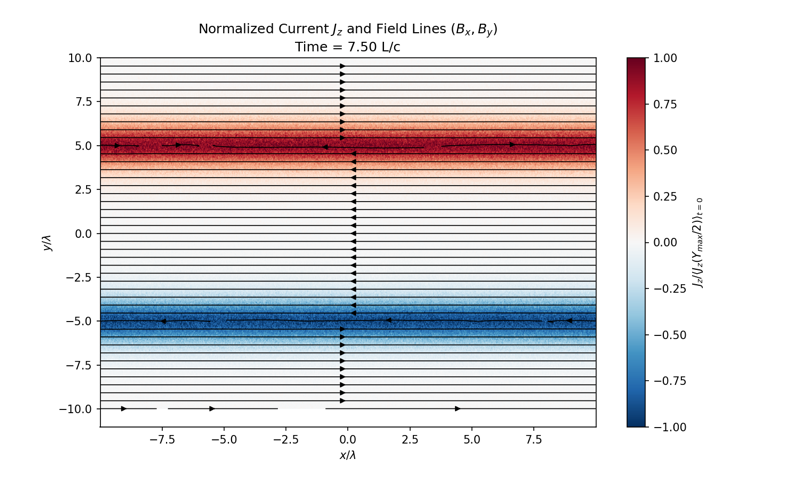}\vline
\includegraphics[width=.34\linewidth]{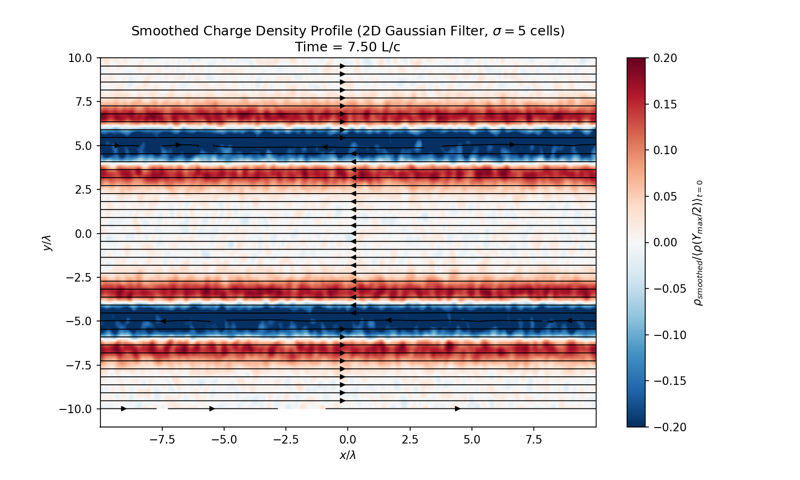}\vline
\includegraphics[width=.3\linewidth]{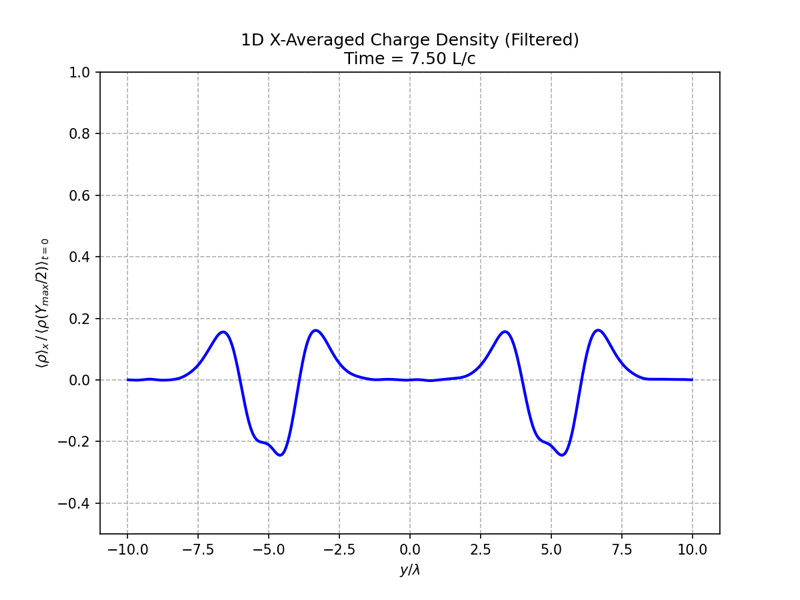}\\
\includegraphics[width=.34\linewidth]{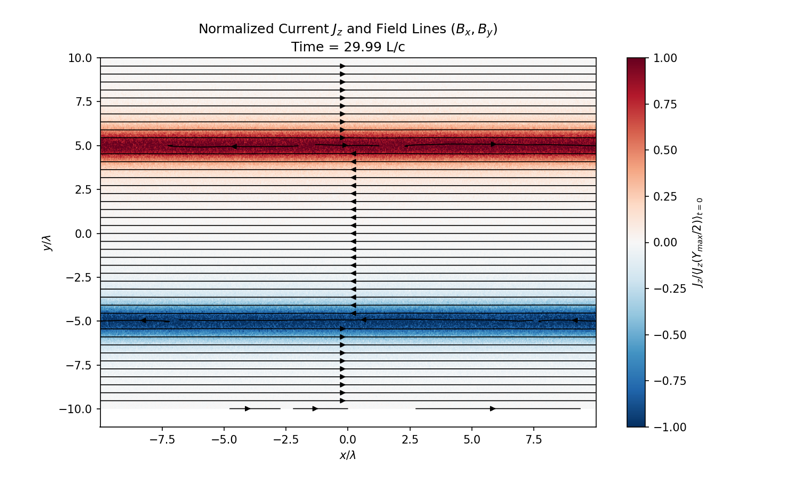}\vline
\includegraphics[width=.34\linewidth]{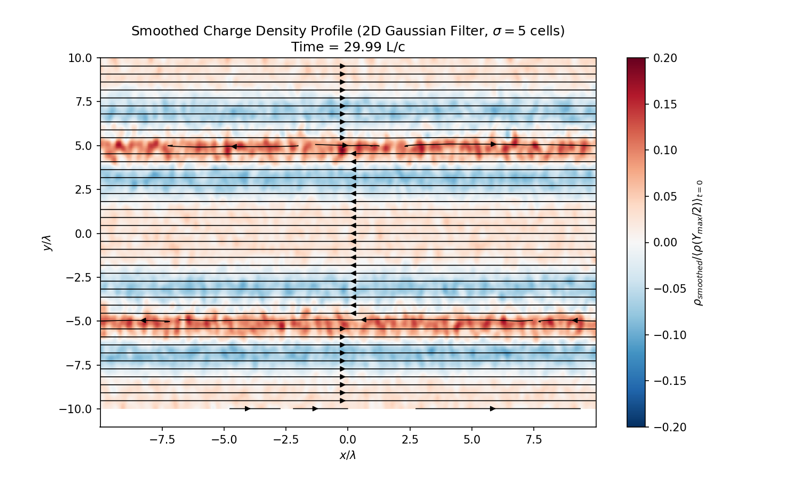}\vline
\includegraphics[width=.3\linewidth]{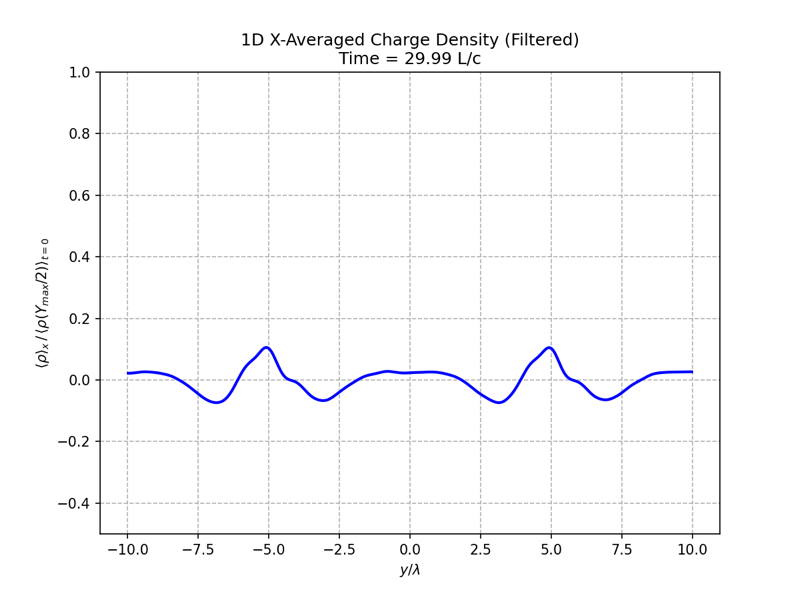}\\
\includegraphics[width=.34\linewidth]{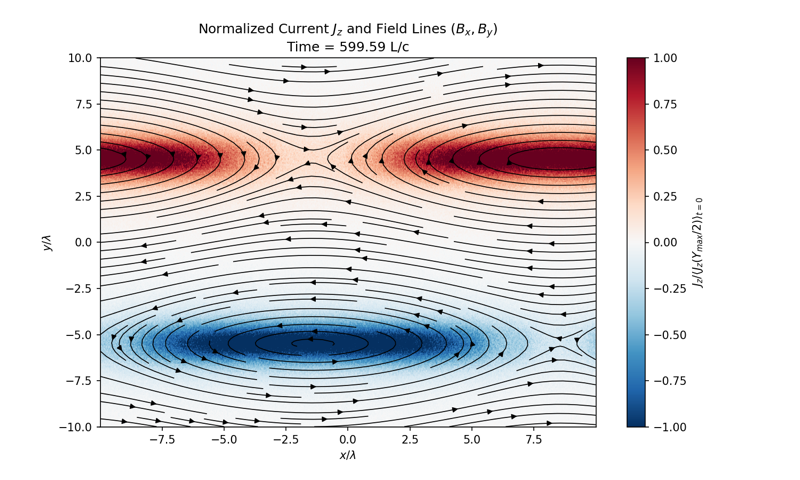}\vline
\includegraphics[width=.34\linewidth]{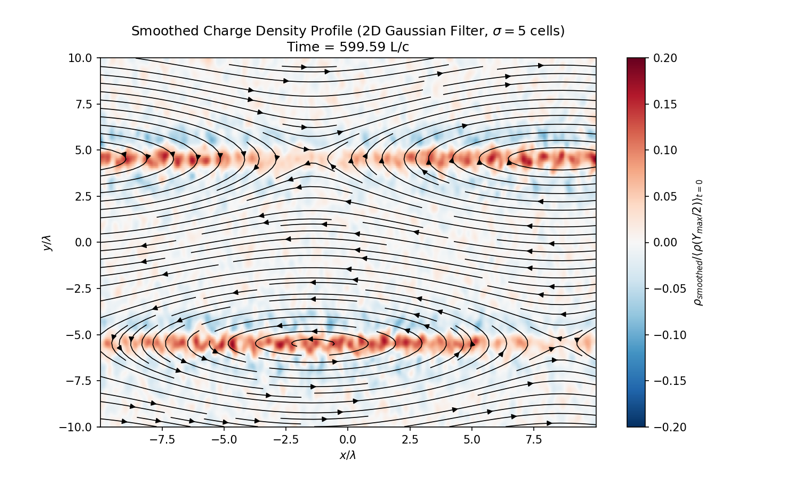}\vline
\includegraphics[width=.3\linewidth]{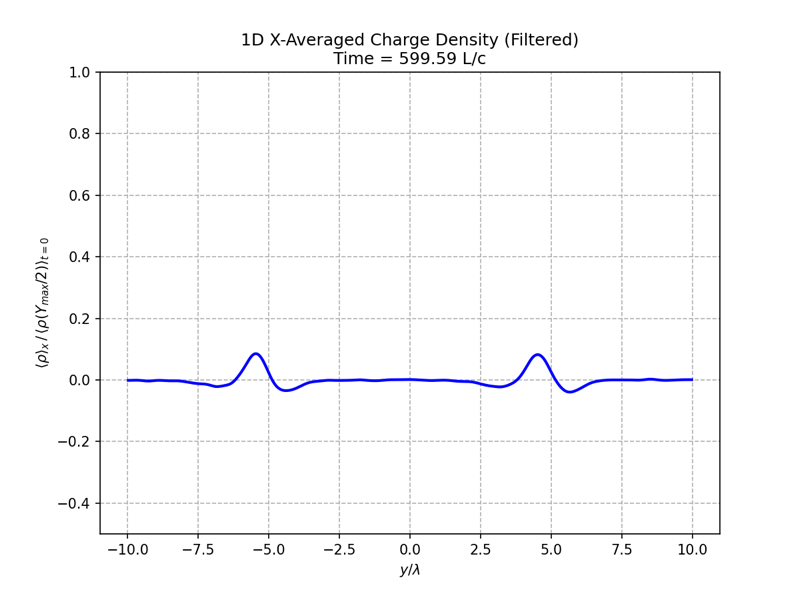}
\caption{Harris CHCS (Cold-1 parameters). Plotted are  out of the plane current (left column), charge density (middle  column), and charge density averaged over $x$  (right  column).  Top row is initial configuration. Times (measured in $L/c$)  are =0, 7.5, 30, 600. }
\label{Data00}
\end{figure}

For 2D simulations, our key result for  Charged Harris Current  sheet (CHCS) is presented in  Figs. \ref{Data00}.
 We observe  that
  the force-balanced  two-fluid charged Harris current sheet is unstable to generation of fast electrostatic oscillations, Bernstein waves, BWs. 
Oscillations originate in the middle of the current sheet and propagate away from the middle. This effect can be understood in terms of electric potential: in the middle it is maximal for one charge, while minimal for another. At the same time the \Bf\ is zero in the middle.  This leads to the generation  fast electrostatic waves, that   appear as trapped Bernstein waves in inhomogeneous warm  pair plasma. 

During  quick charge relaxation, variations of the total current/global structure are minimal. As a result, the current layer first becomes effectively uncharged, while later experiencing the usual tearing instability.
  Overall, for these parameters, 
After quick electrostatic relaxation, the system reaches a state $\rho_e \sim 0$. The ensuring tearing resembles the non-charged case

Fig. \ref{Data00}, charged Harris: at intermediate steps, there is a clear electrostatic dominated evolution: large, coherent charge waves are generated by the current layers.  Charged waves quickly phase-mix, and dissipate, so that at later times overall charge density is minimal, while currents mostly mimic the uncharged  tearing  mode.  Eventually, with periodic boundary conditions, only one plasmoid per half plane remains. It is finally uncharged.

Initially all is in force balance, but an unstable one: in the center, one charge is at potential maximum, another at  potential minimum and B-field is approximately  zero. That sends electrostatic waves (probably something like Bernstein modes, but in inhomogeneous plasma). Modes are trapped within  the layer. As a result there is quick relaxation of the charge.

\begin{figure}[h!]
\includegraphics[width=.5\linewidth]{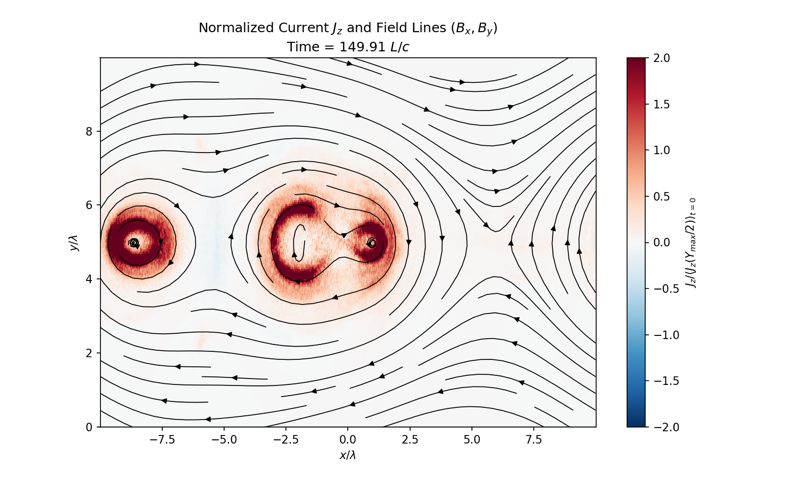}
\includegraphics[width=.5\linewidth]{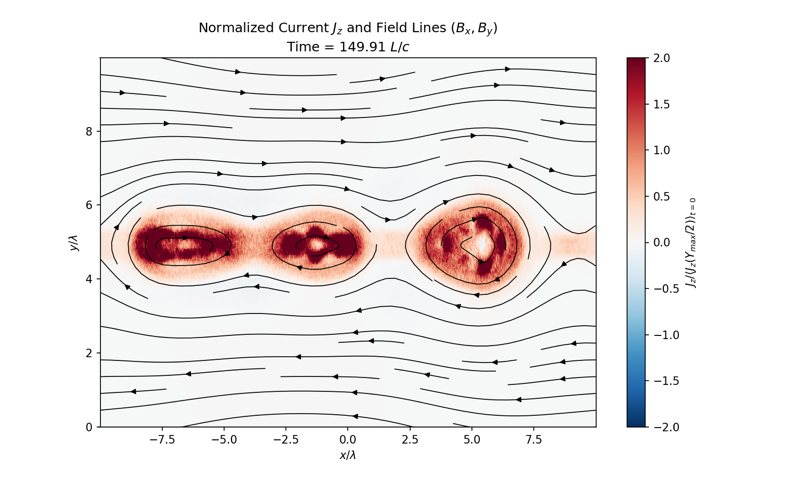}
\includegraphics[width=.5\linewidth]{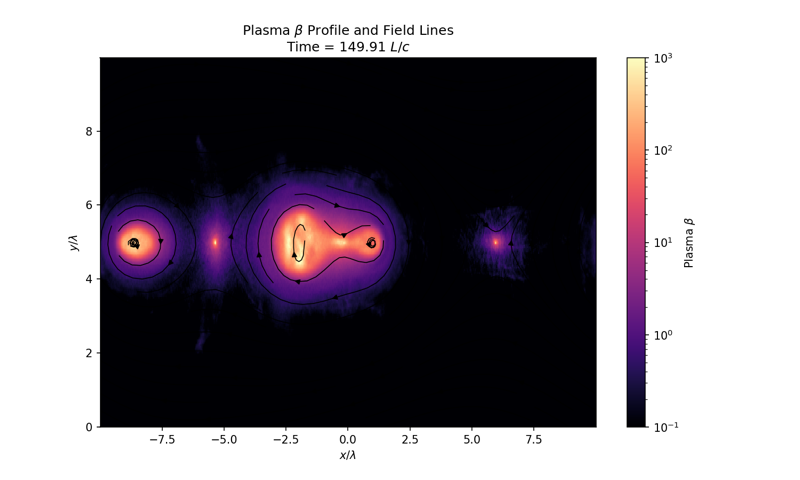}
\includegraphics[width=.5\linewidth]{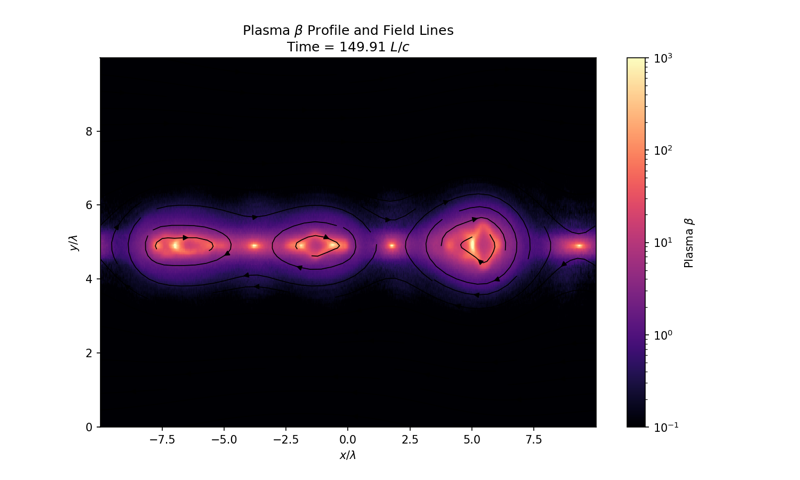}
\caption{Harris sheet. Comparing long-term evolution:  uncharged (left) and  charged  (right). Plotted are half-plane images of out-of the board current  (top row ) and plasma beta (bottom row). Quick electrostatic relaxation does not affect much the longer timescale tearing.  (Run High-B, Table \ref{table}). }
\label{compare}
\end{figure}

The long term evolution of tearing mode is not affected strongly by the initial charge, Fig. \ref{compare}.


\subsection{Properties of electrostatic oscillations  }
Since Bernstein waves' phase speed increases with temperature, one expects that hotter current sheets relax electrostatically faster.
 For Harris sheet, there is no simple way to establish parametric scaling, eg for the period and frequency of charged oscillation, as the plasma is inhomogeneous (both density, \Bf\ and charge density). In addition the amplitude of the initial density perturbation depends on the values of \Bf\ and temperature. 

\begin{figure}[h!]
\includegraphics[width=.5\linewidth]{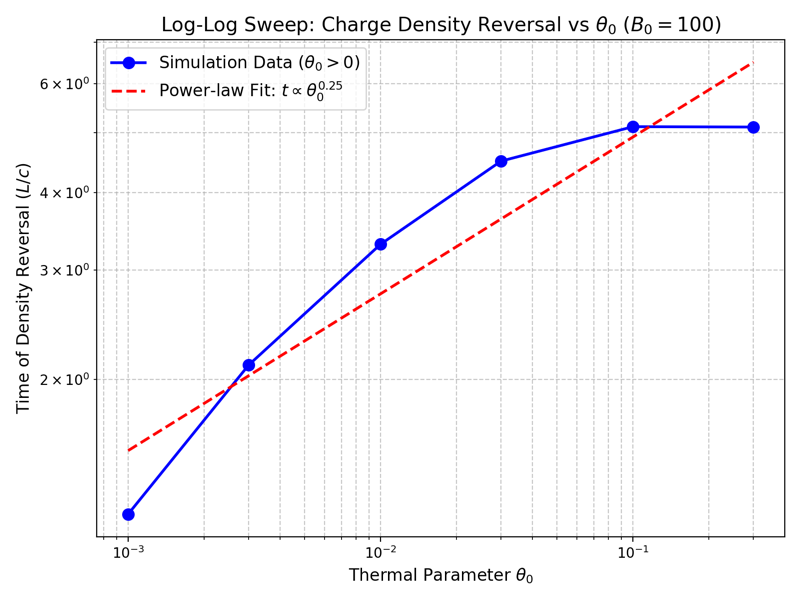}
\includegraphics[width=.5\linewidth]{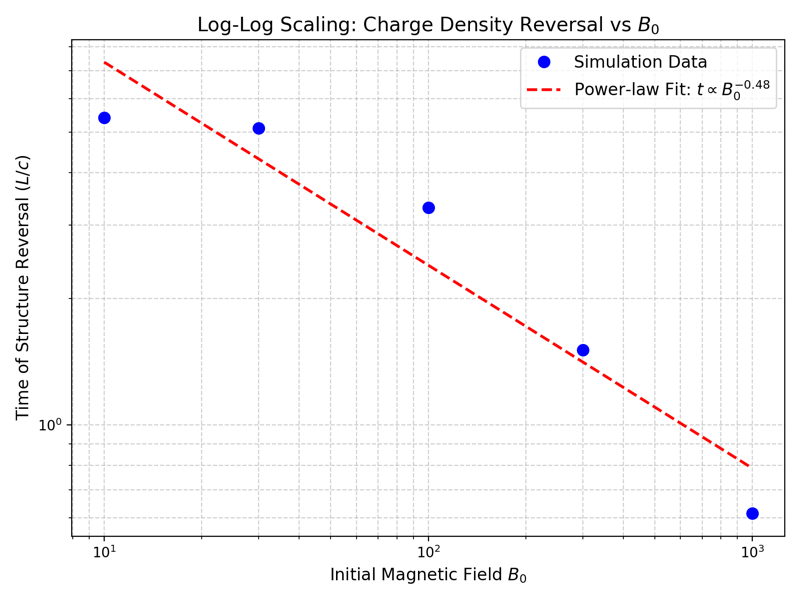}
\caption{Frequency of charge oscillations (proportional to BW's frequency) as function of parameters - temperature (left; \Bf\ parameter $b_0=100$)  and \Bf\ (right;  temperature parameter $\Theta =0.01$). Plotted is time for the charge density to reverse sign at the center of the layer -  this is approximately $1/4$ of the period. As expected, BW oscillation occur faster in warmer plasma, $\propto \sqrt{\Theta}$.  }
\label{plot_b0_scan}
\end{figure}

To test whether the observed electrostatic oscillations are indeed BWs, we performed  the following tests. First, 
In Fig.  \ref{plot_b0_scan} we show results of the parameter scan of the properties of electrostatic oscillations, as a function of the temperature parameter $\Theta$ and overall \Bf\ $b_0$ (the basic run in Fig. \ref{Data00} corresponds to $b_0=100$). Results are clearly consistent with charged BWs.

\begin{figure}[h!]
\includegraphics[width=.24\linewidth]{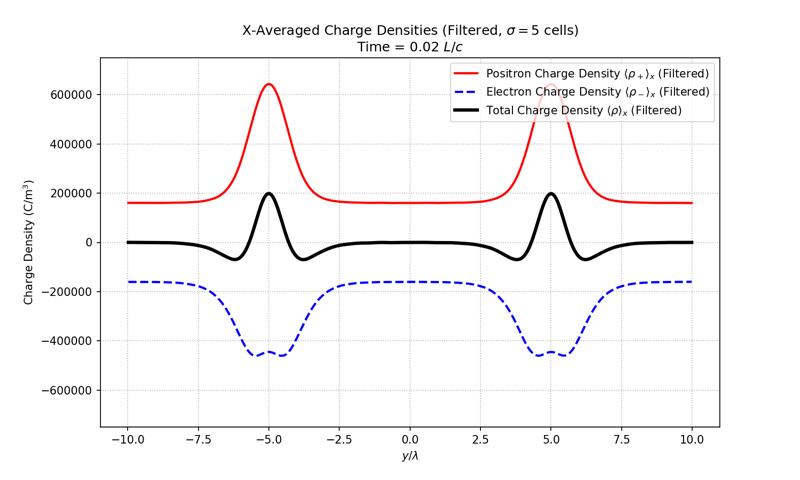}
\includegraphics[width=.24\linewidth]{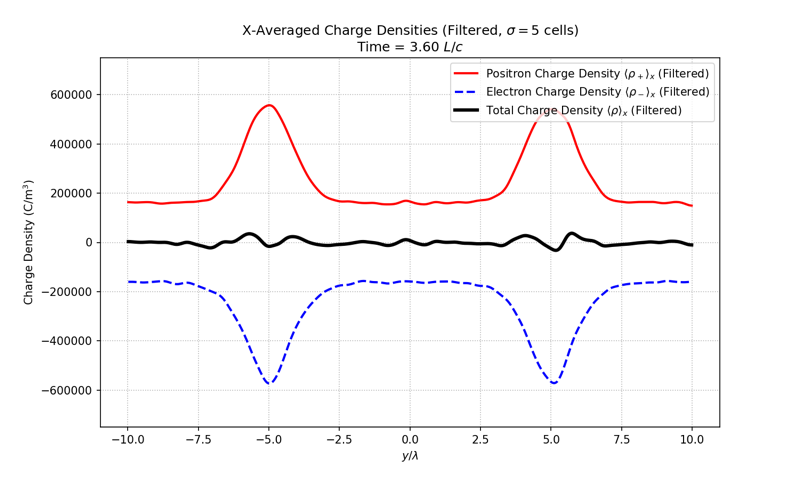}
\includegraphics[width=.24\linewidth]{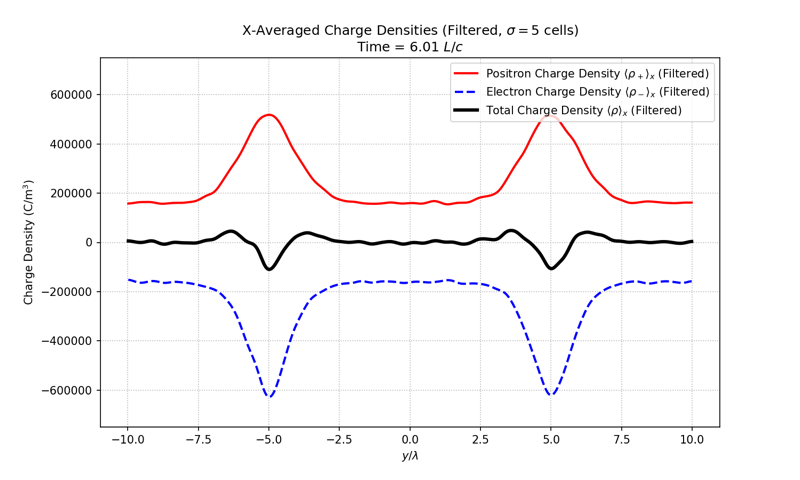}
\includegraphics[width=.24\linewidth]{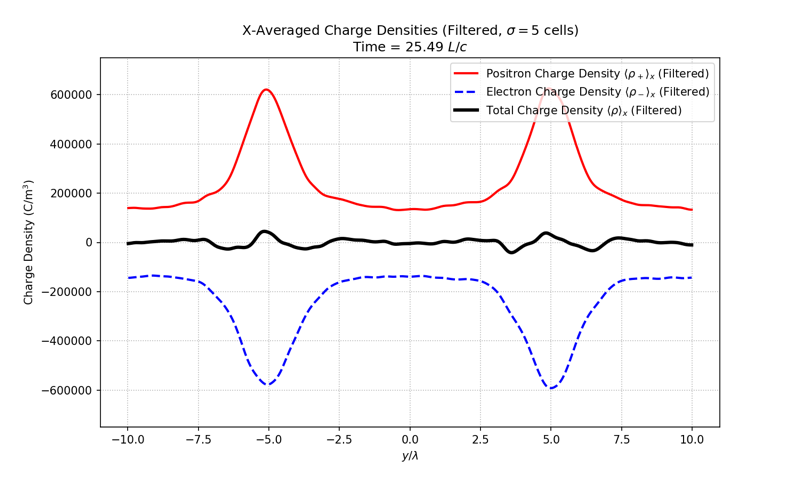}
\caption{Harris. Evolution of charge densities.
Quasi-1D simulations (very small $x$-range)..
}
\label{rho_avg}
\end{figure}

Second,  we performed a set of quasi-1D simulations, where the plasma structure is only 3 cells in $x$ (the minimum allowed by the code), 
Fig \ref{rho_avg} (see also    Fig. \ref{rho_avg-rot}  for a similar test of rotational sheet).  Quasi-1D does not capture plasmoid instability. As we observe, for Harris equilibrium the electrostatic relaxation is very fast, faster than plasmoid formation. Thus, for Harris quasi-1D captures the essence of the electrostatic relaxation.

Fig.  \ref{rho_avg} demonstrates that relatively small charge misbalance is  waves: these are small perturbations of the total charge densities.

\begin{figure}[h!]
\includegraphics[width=.3\linewidth]{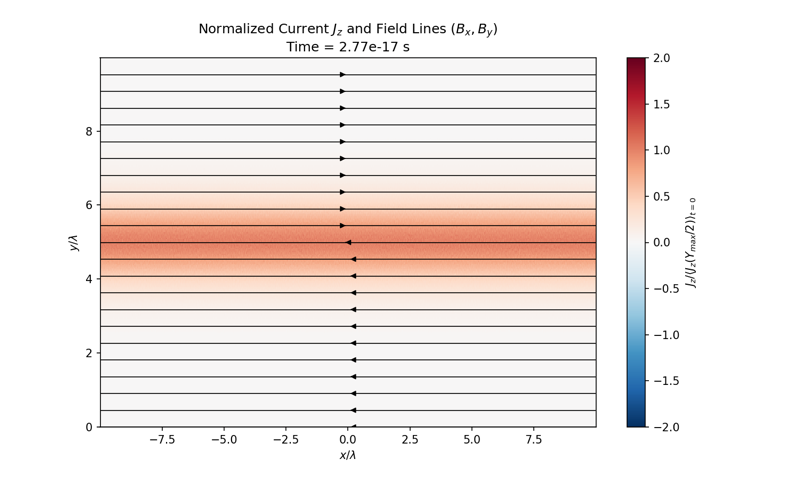}\vline
\includegraphics[width=.3\linewidth]{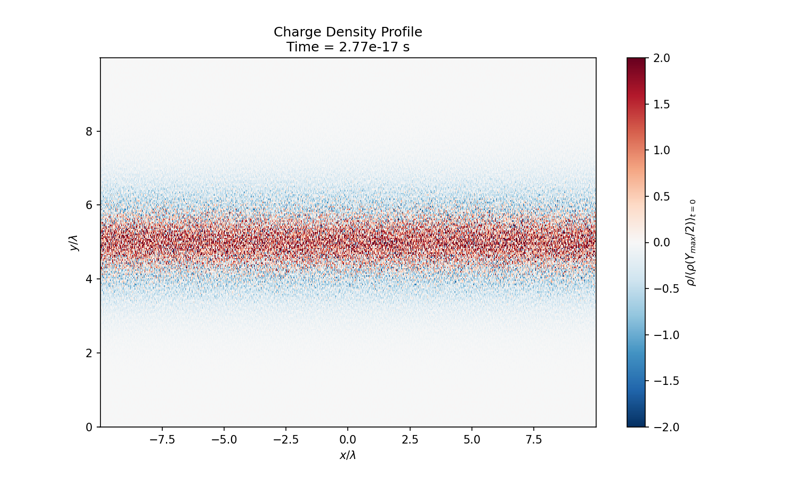}\vline
\includegraphics[width=.25\linewidth]{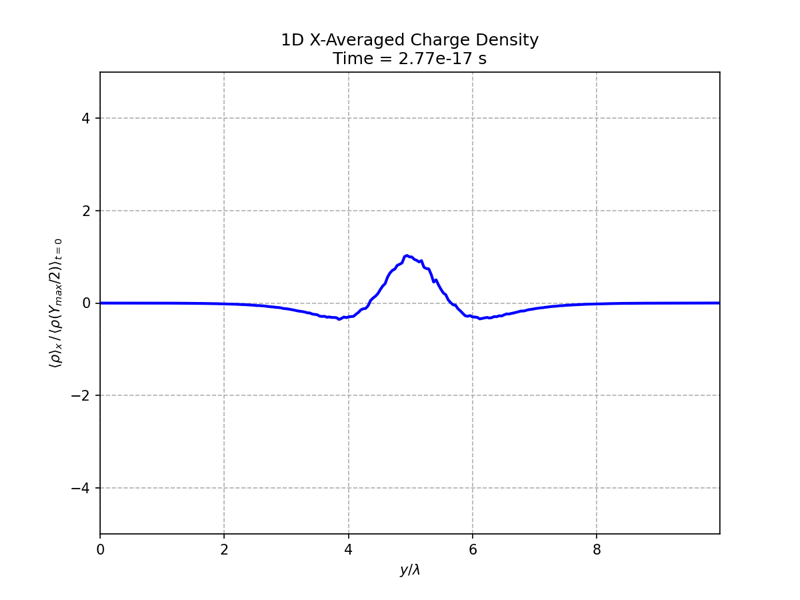}\\
\includegraphics[width=.3\linewidth]{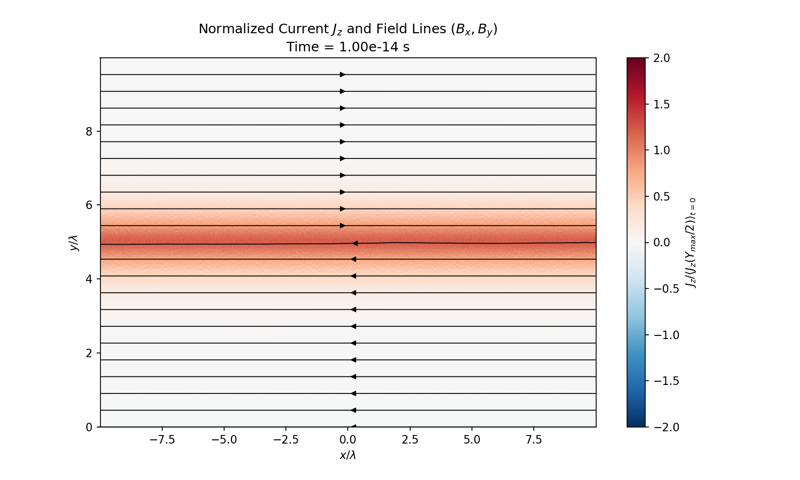}\vline
\includegraphics[width=.3\linewidth]{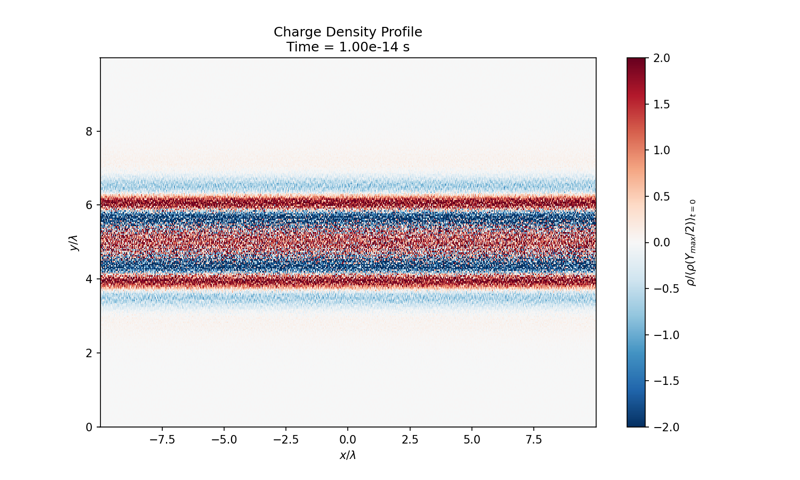}\vline
\includegraphics[width=.25\linewidth]{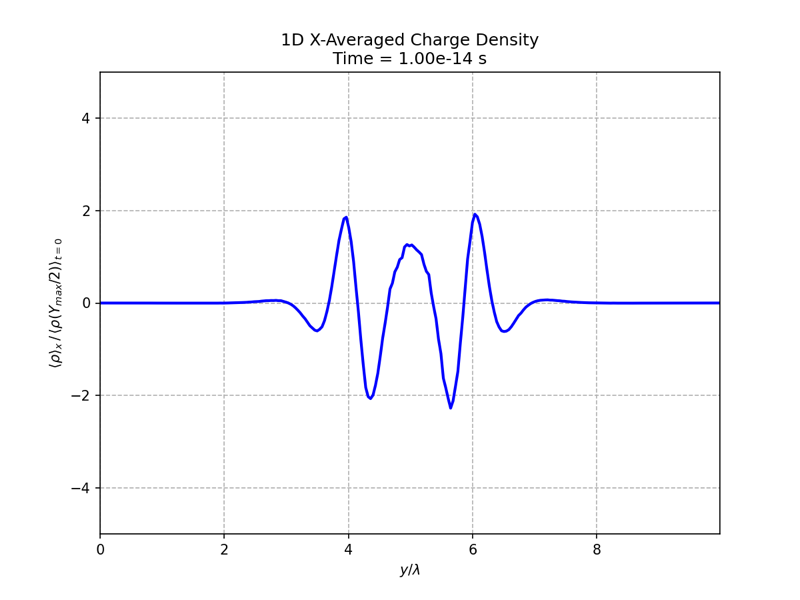}\\
\includegraphics[width=.3\linewidth]{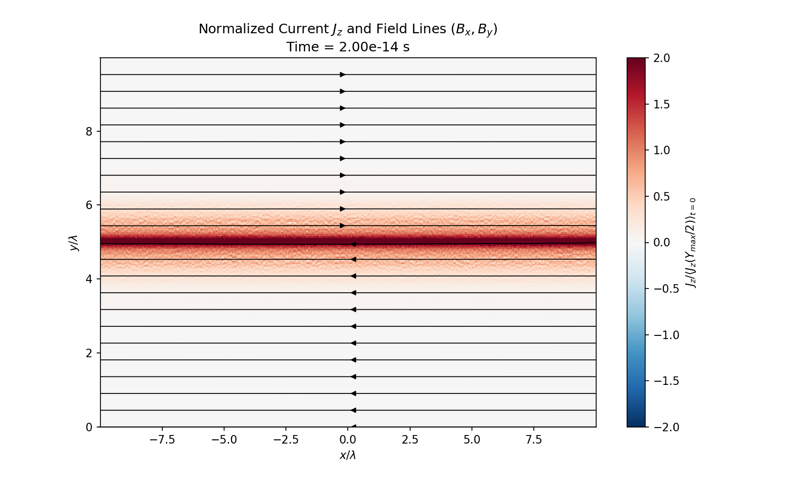}\vline
\includegraphics[width=.3\linewidth]{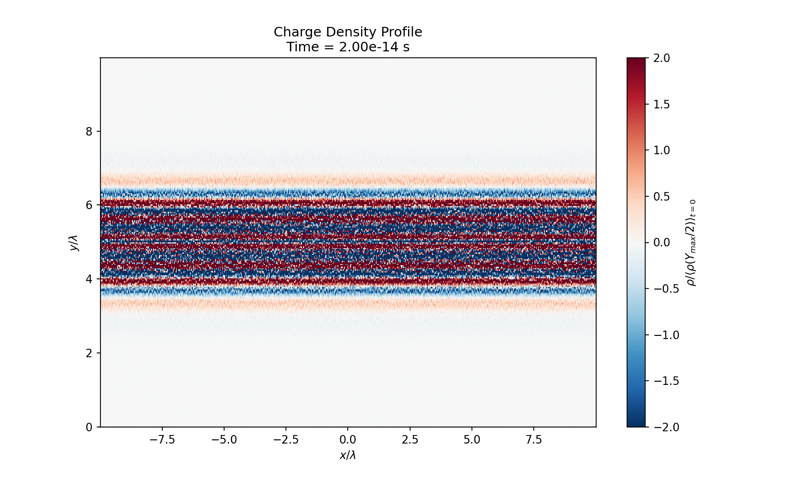}\vline
\includegraphics[width=.25\linewidth]{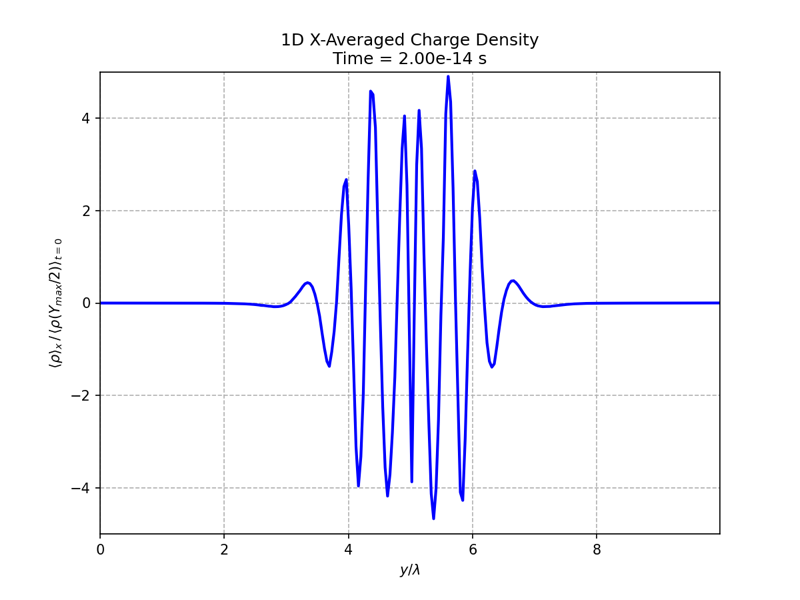}\\
\includegraphics[width=.3\linewidth]{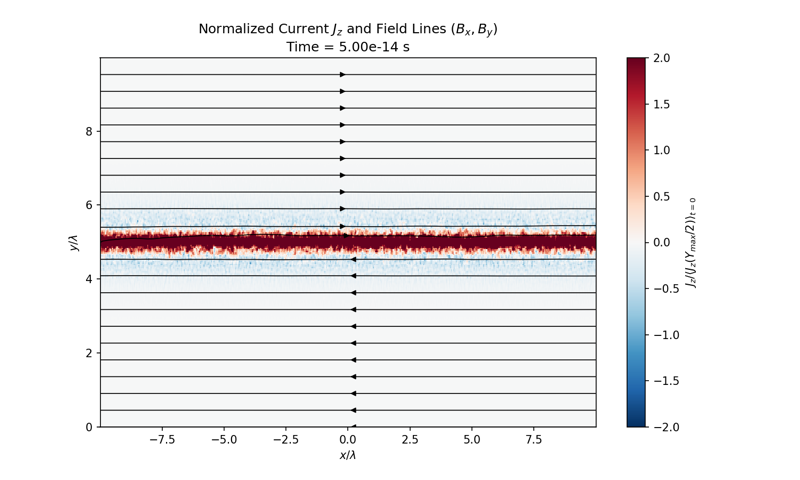}\vline
\includegraphics[width=.3\linewidth]{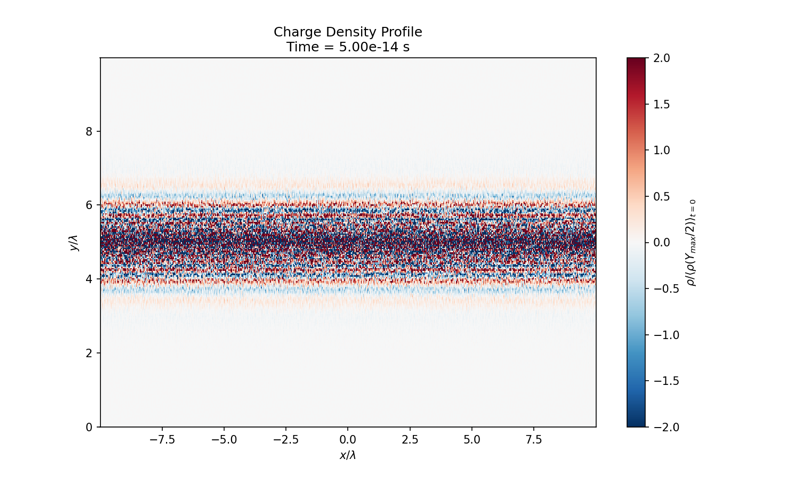}\vline
\includegraphics[width=.25\linewidth]{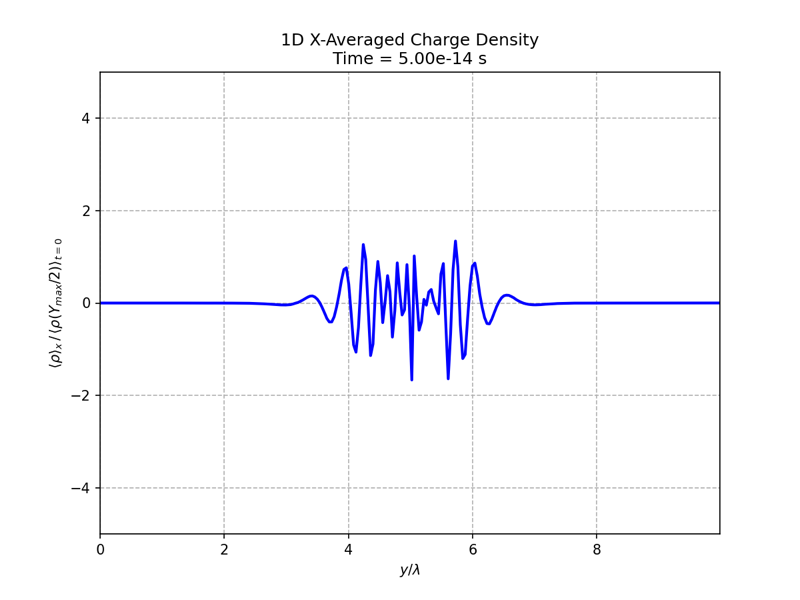}\\
\includegraphics[width=.3\linewidth]{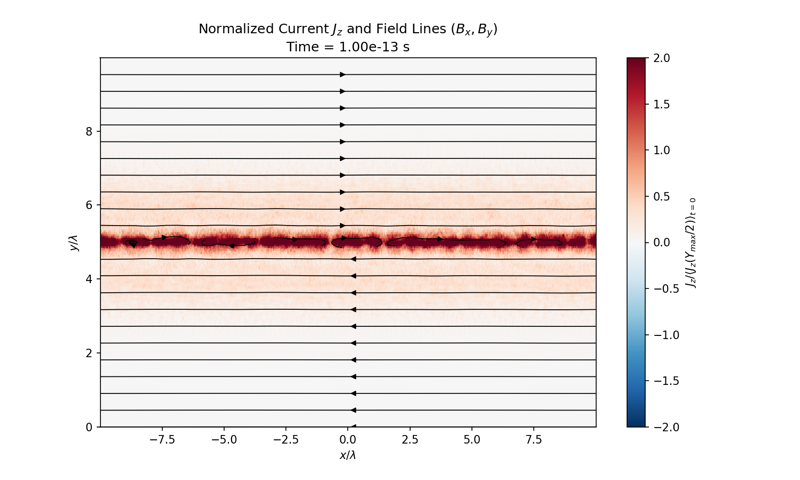}\vline
\includegraphics[width=.3\linewidth]{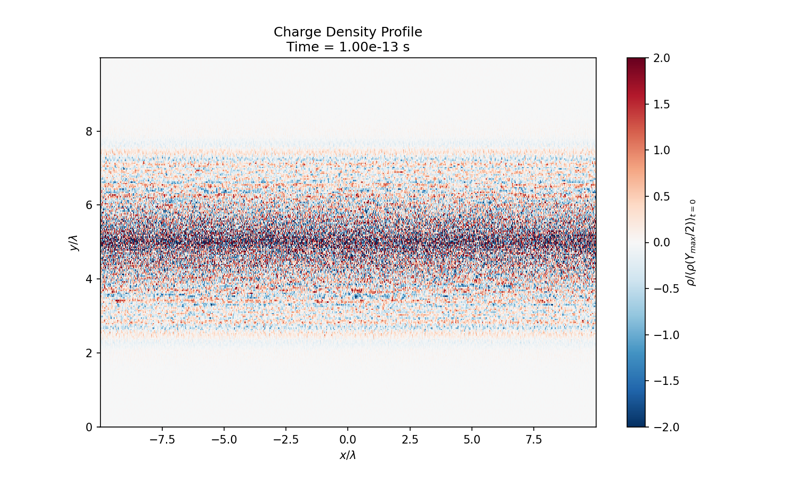}\vline
\includegraphics[width=.25\linewidth]{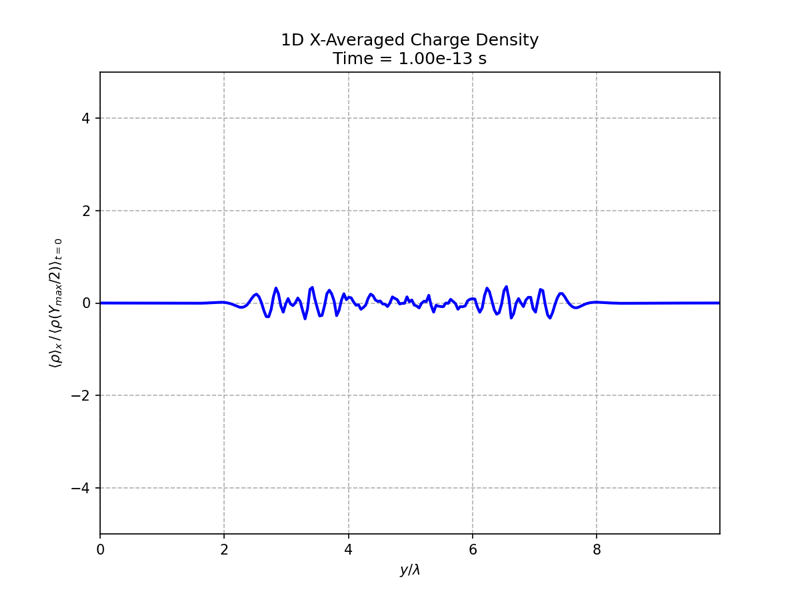}
\caption{Harris CHCS (Run High-B, Table \ref{table}).   Top row is initial configuration. Times (measured in $L/c$)  are =0, 3, 6,15, 50. }
\label{b0-1000}
\end{figure}

Finally in Fig. \ref{b0-1000} we plot a zoomed-in evolution for higher initial \Bf. After approximately 30 $L/c$ the initial charge distribution is erases, while the tearing  did not progress considerably

\newpage
.
\newpage

.
\newpage

\section{Force-free charged (sheared) rotational  current   layer} 
\label{rotational}

\subsection{Global structure} 

Next we turn to charged (sheared) rotational  force-free current   layers  \cite[first  discussed by][]{2003MNRAS.346..540L}. From the force-free  balance condition
\be
B^2 - E^2 ={\rm constant}=B_0^2
\ee
and choosing $B_x \propto \tanh \left(\tilde{z}\right)$,  we find
\ba &&
\B= \left\{\tanh \left(\tilde{z}\right),\gamma _0
   {\rm sech}\left(\tilde{z}\right),0\right\}  B_0 
   \nn && 
   \E= \left\{0,0, - \sqrt{\gamma _0^2-1}
 \times   {\rm sech}\left(\tilde{z}\right)\right\} B_0
 \label{fieldsrotational}
\ea
where $\gamma_0 = 1/\sqrt{1-\beta_0^2} $ parametrized the value of the initial  charge (for $\gamma_0=1$ the initial state is charge-neutral).

 The following parameterization describes charged (relativistically sheared) rotational   current layer, 
\ba &&
   n_\pm= n_0 (1\pm f_p)
   \nn &&
   f_p = \tanh \left(\tilde{z}\right) {\rm sech}\left(\tilde{z}\right)  \frac{B_0 \sqrt{\gamma _0^2-1}}{2 L n_0}
   \nn &&
   \rho_e =2 n_0 f_p
      \nn &&
      {\bf v}_\pm  =  {\bf v} _{EM}   \pm v_d {\bf e}_B 
      \nn &&
      v_d = \frac{1}{ \sqrt{2}  \sqrt{2 \gamma _0^2+\cosh \left({2 \tilde{z}}\right)-1}} 
       \frac{B_0 \gamma _0}{   L n_0 } 
\nn &&
 {\bf v} _{EM} =
   \left\{\frac{2 \gamma_0 \sqrt{\gamma_0^2-1}}{2{\gamma_
   0}^2+\cosh \left({2 \tilde{z}}\right)-1},-\frac{2 \sqrt{\gamma_0^2-1} \sinh
   \left(\tilde{z}\right)}{2 \gamma_0^2+\cosh \left({2  \tilde{z}}\right)-1},0\right\}
            \nn &&
               \J  = \left\{\gamma _0 \tanh \left(\tilde{z}\right)
   {\rm sech}\left(\tilde{z}\right),{\rm sech}^2\left(\tilde{z}\right),0\right\}  \frac{ B_0}{L}= \J_\rho + \J_\parallel
   \nn &&
    \J_\rho = \rho_e {\bf v} _{EM} 
    \nn && 
    \J_\parallel = \J - \J_\rho =\left\{\frac{\sinh \left(\tilde{z}\right)}{2 \gamma _0^2+\cosh \left(\frac{2
   z}{L}\right)-1},\frac{\gamma _0}{2 \gamma _0^2+\cosh \left(\frac{2
   z}{L}\right)-1},0\right\}  \frac{2 B_0 \gamma _0}{ L}
   \label{setuprotational}
   \ea
  Two contributions to current are conduction current along \Bf\  $\J_\parallel$ along ${\bf e}_B = \B/B$,  and advection currents  $ \J_\rho = \rho_e  {\bf v} _{EM} $.    For $\gamma_0 =1$ the above relations reduce to a rotational switch of the \Bf, with no charge density.

   The current is always space-like,
   \be
   \rho_e^2 - j^2 = -\frac{B_0^2 {\rm sech}^2\left(\tilde{z}\right)}{ L^2}
   \ee
   Total charge  in each $\pm z$ region is 
   \be
   Q= {B_0 \sqrt{\gamma _0^2-1}}
   \ee
   
   Electric potential 
   \be
  \Phi  = \left(  1-2 H (z)+\frac{2 \cot ^{-1}(\sinh (z))}{\pi } \right) \sqrt{\gamma_0^2-1} B_0
  \ee
   ($H(z)$ is Heaviside function). 
   
\begin{figure}[h!]
\includegraphics[width=.38\linewidth]{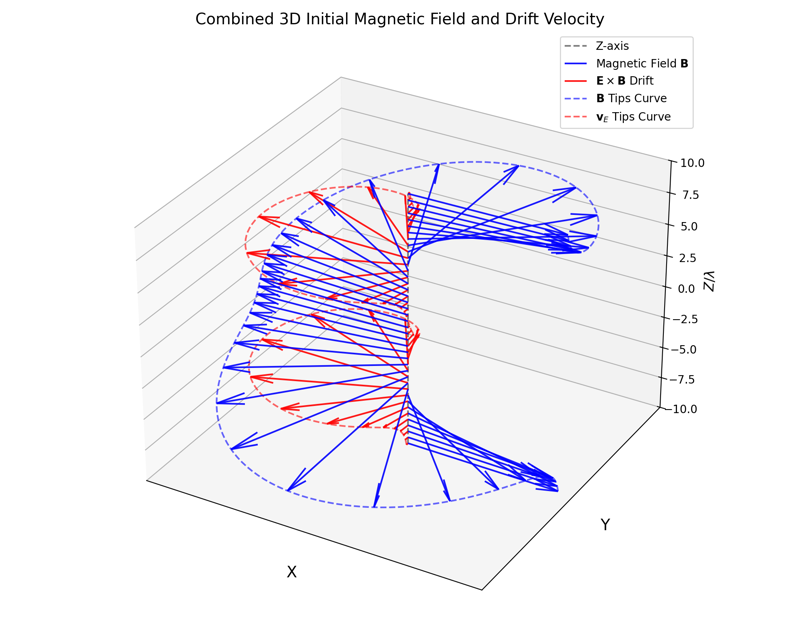}
\includegraphics[width=.33\linewidth]{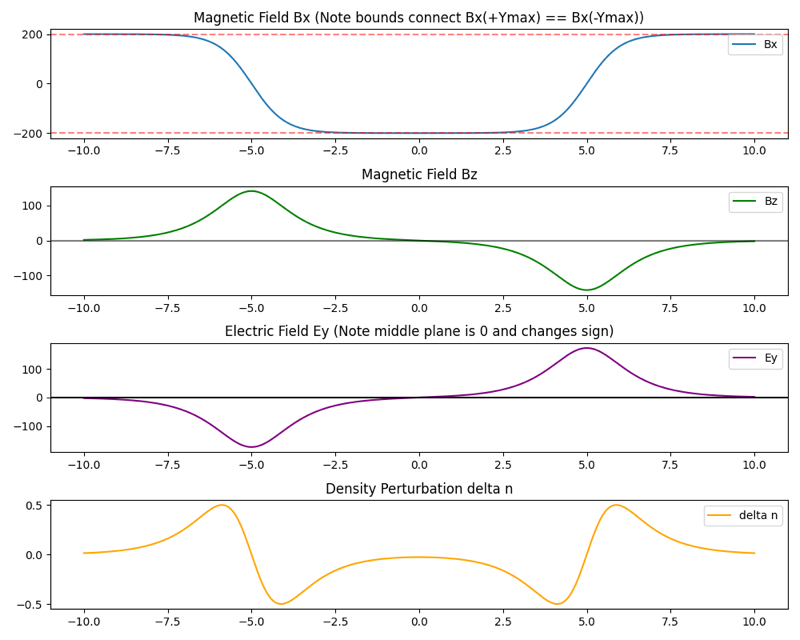}
\includegraphics[width=.25\linewidth]{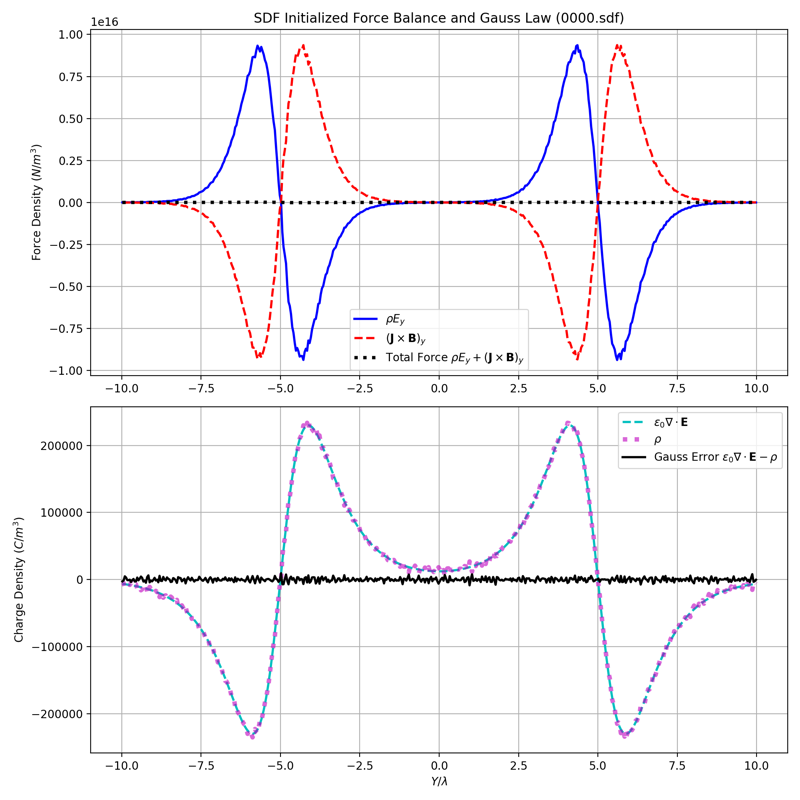}
\caption{Rotaional-CRCS. Left panel:   \Bf\ and velocity structure across the rotational-CRC. Middle and right panels: , analytical  profiles  of forces, and  a check of numerical initialization for double-Rotaional-CRC configuration.}
\label{CRCS10}
\end{figure}

   Thus, importantly,  there is potential different between $z \pm \infty$. This is the main  reasons why in  actually simulations we use double current layer, with zero potential different between $z \pm \infty$, Fig. \ref{CRCS10}. (Otherwise, in the nonlinear stage  of tearing, large global charge flows develop.)

\subsection{Simulations: charged rotational current sheet  - 2D}

\begin{figure}[h!]
\includegraphics[width=.34\linewidth]{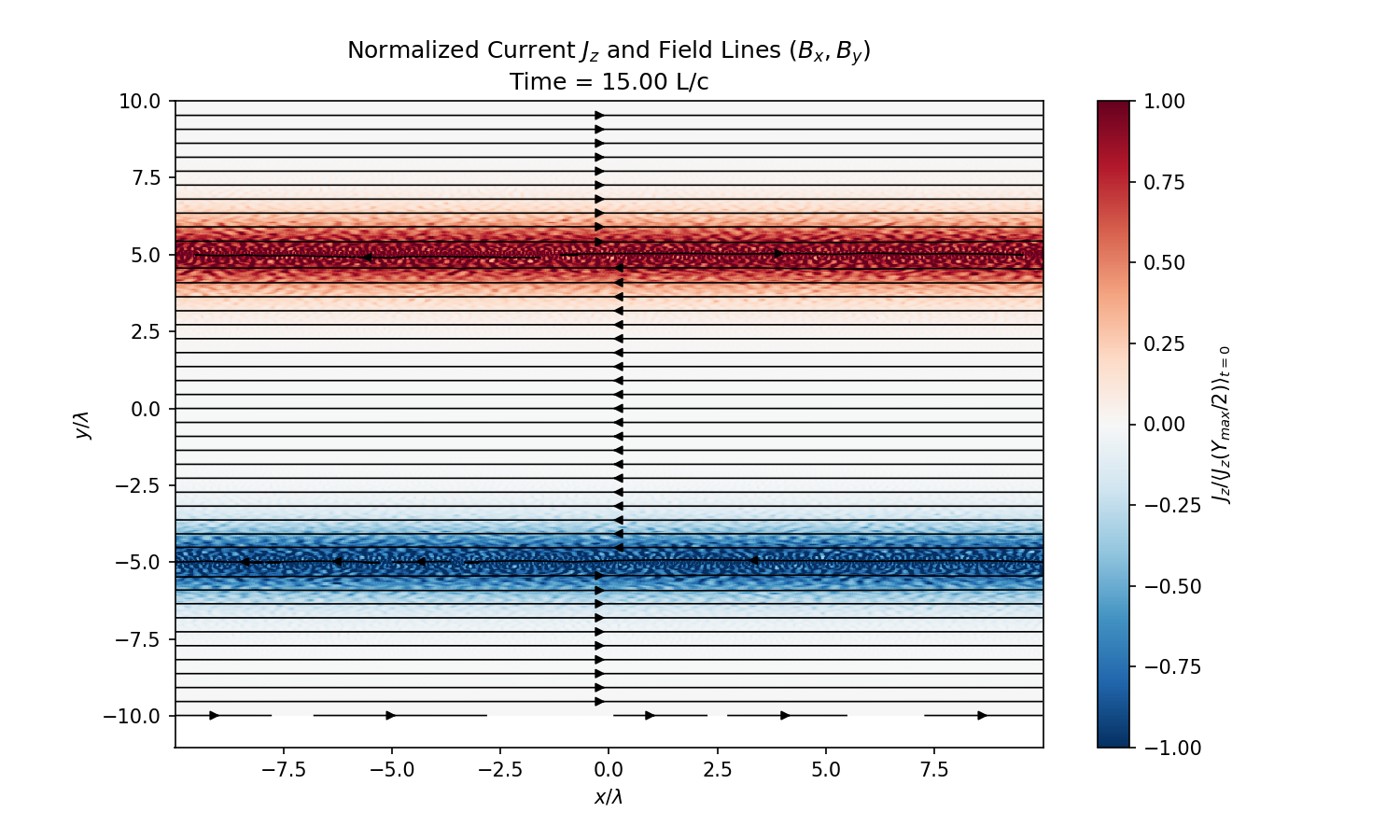}\vline
\includegraphics[width=.34\linewidth]{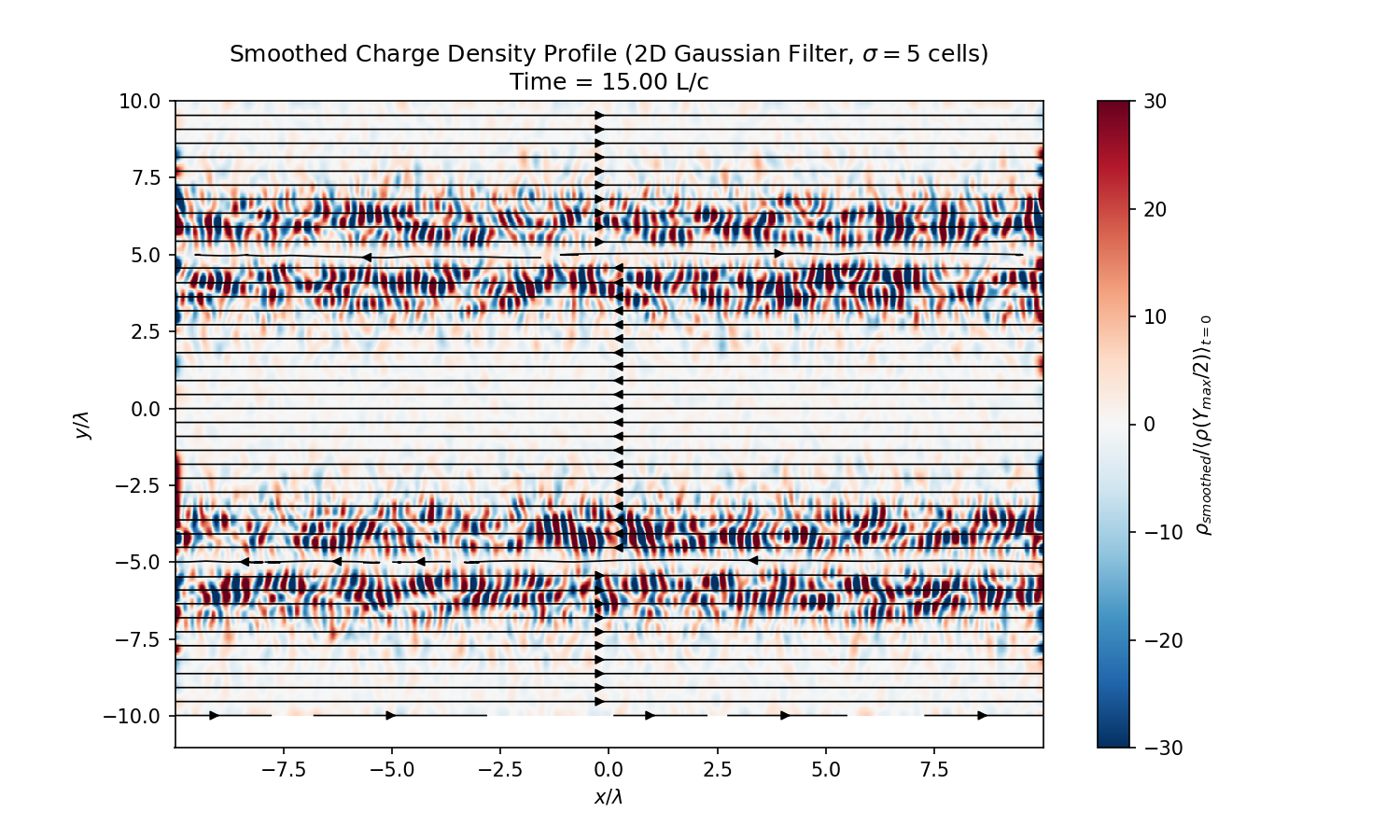}\vline
\includegraphics[width=.34\linewidth]{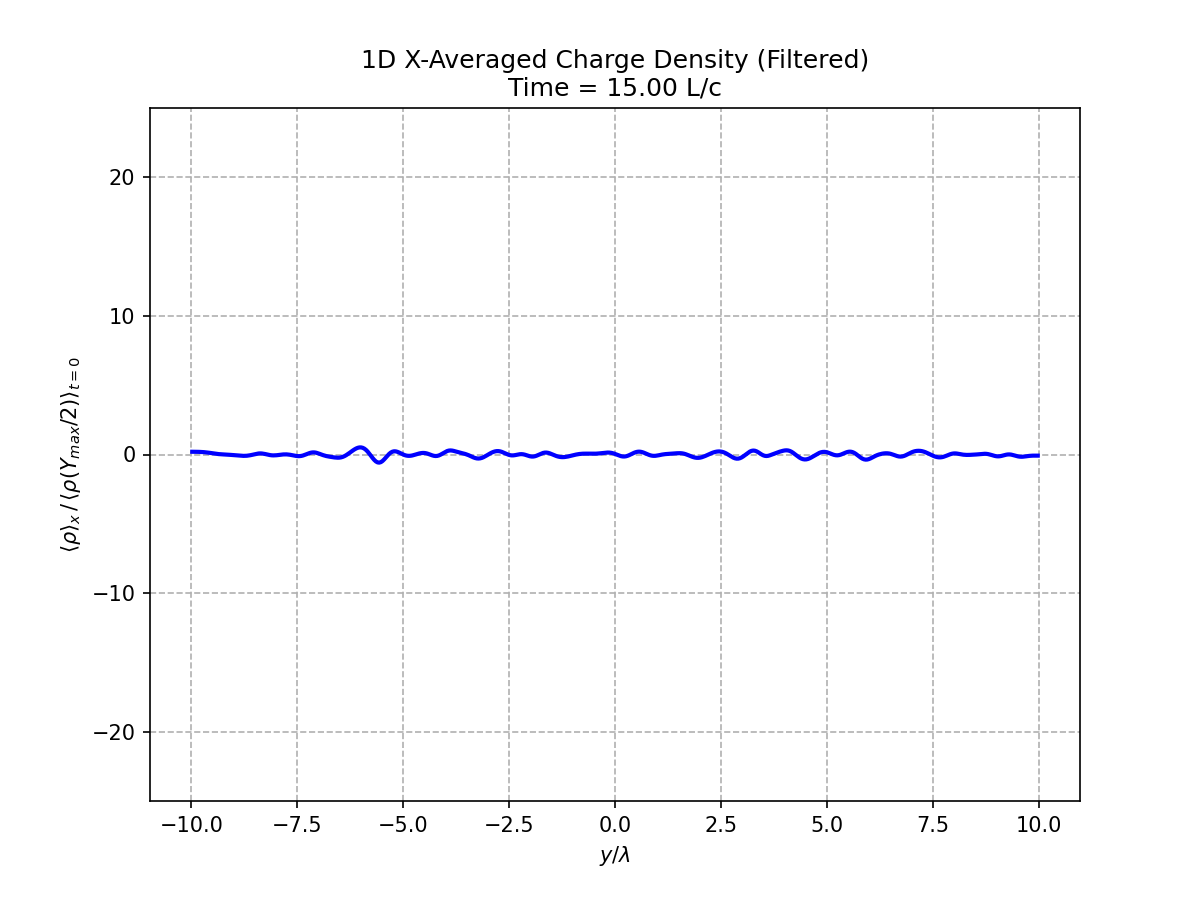}\\
\includegraphics[width=.34\linewidth]{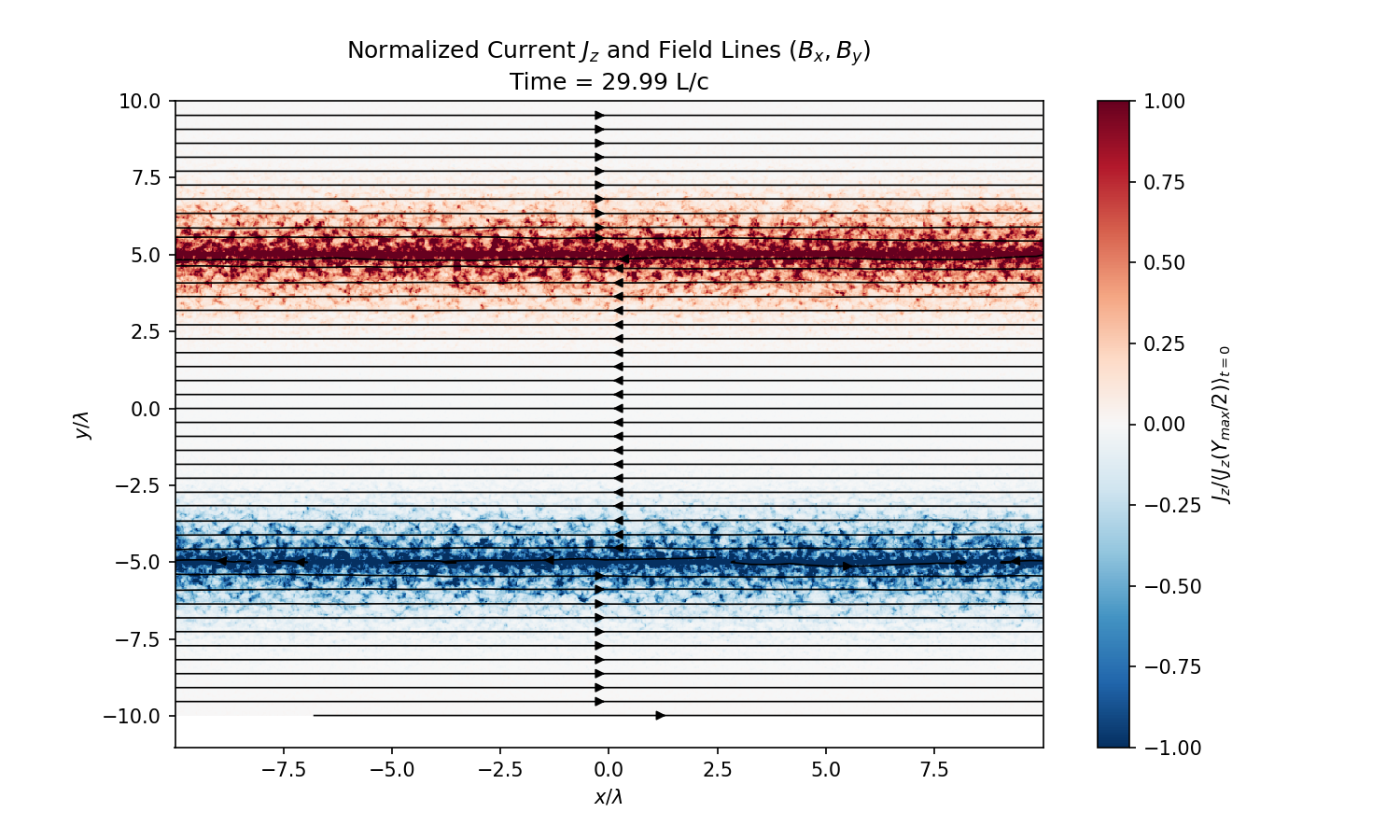}\vline
\includegraphics[width=.34\linewidth]{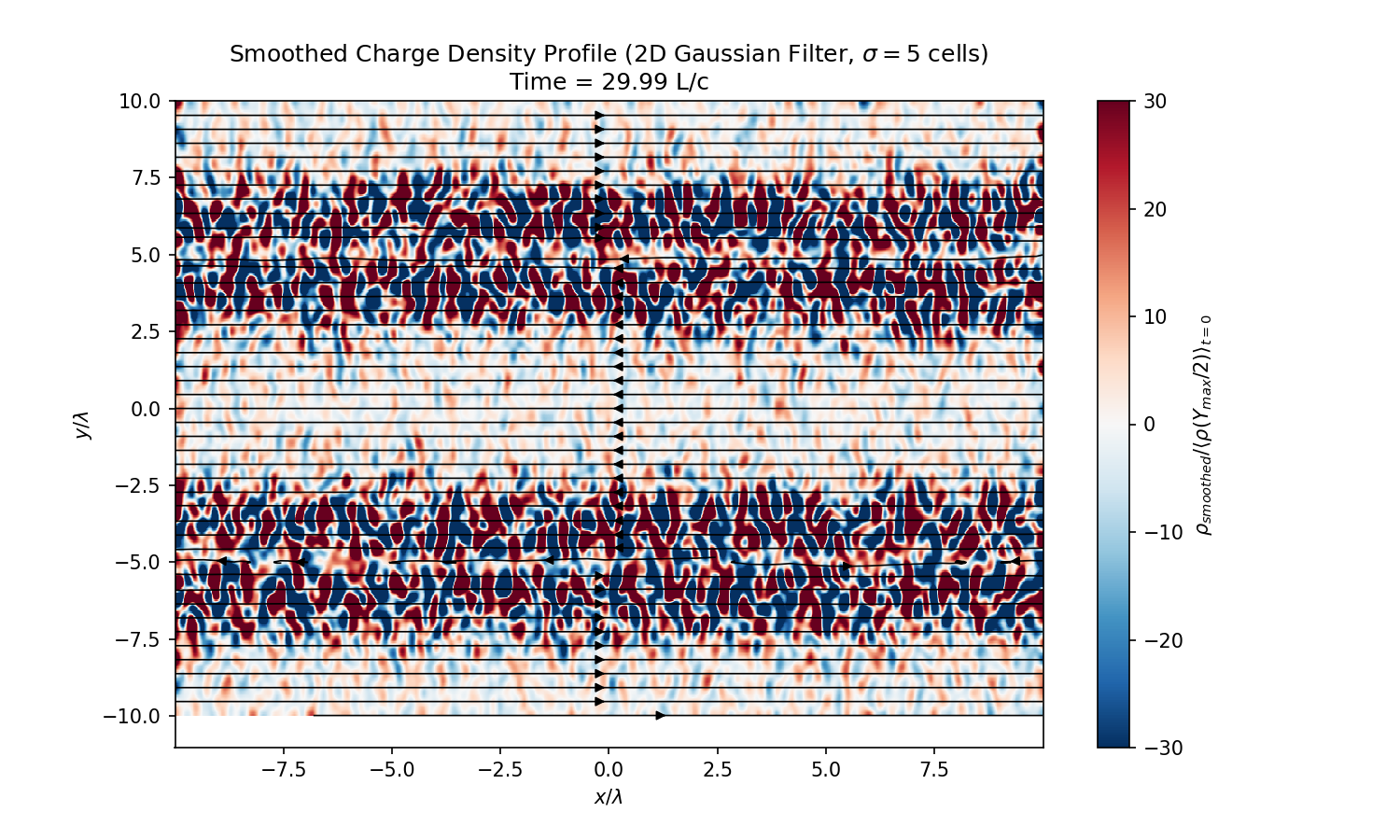}\vline
\includegraphics[width=.34\linewidth]{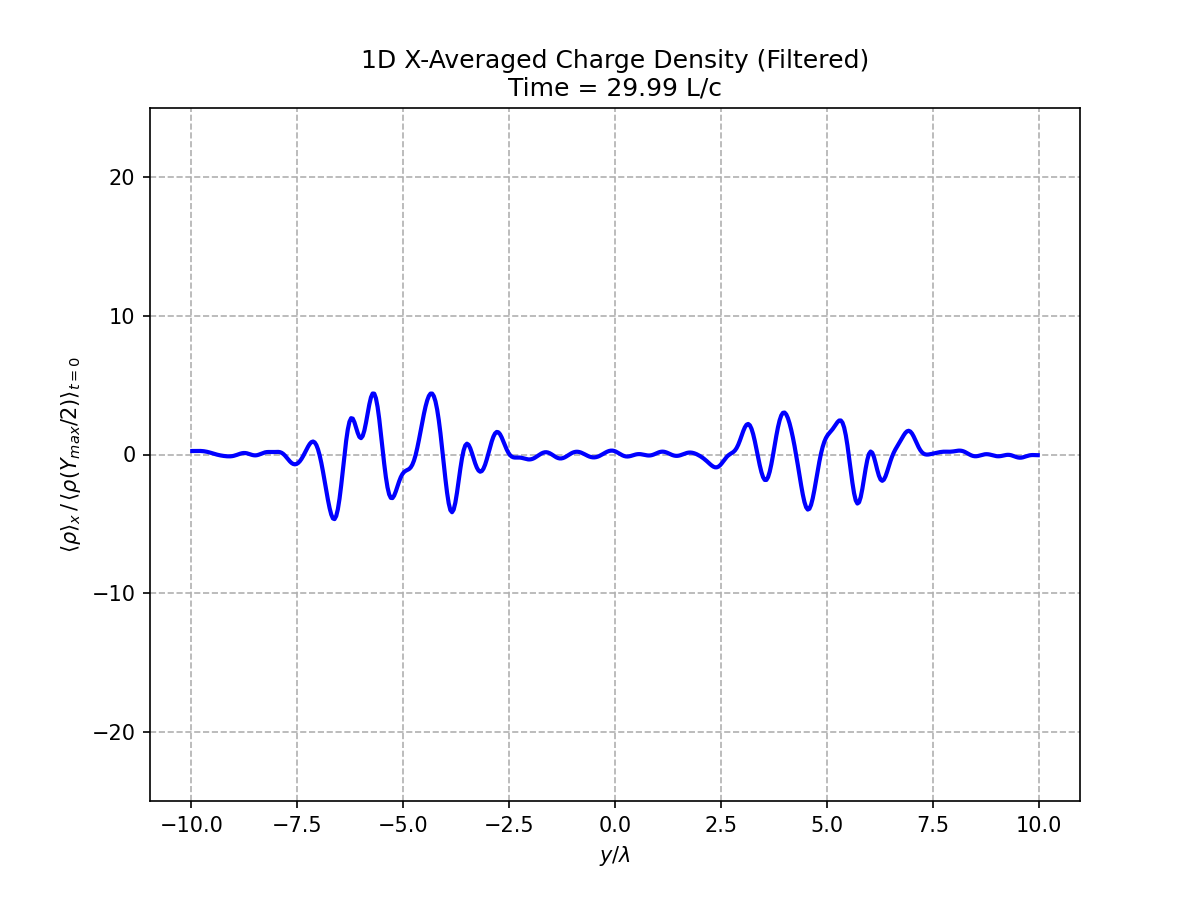}\\
\includegraphics[width=.34\linewidth]{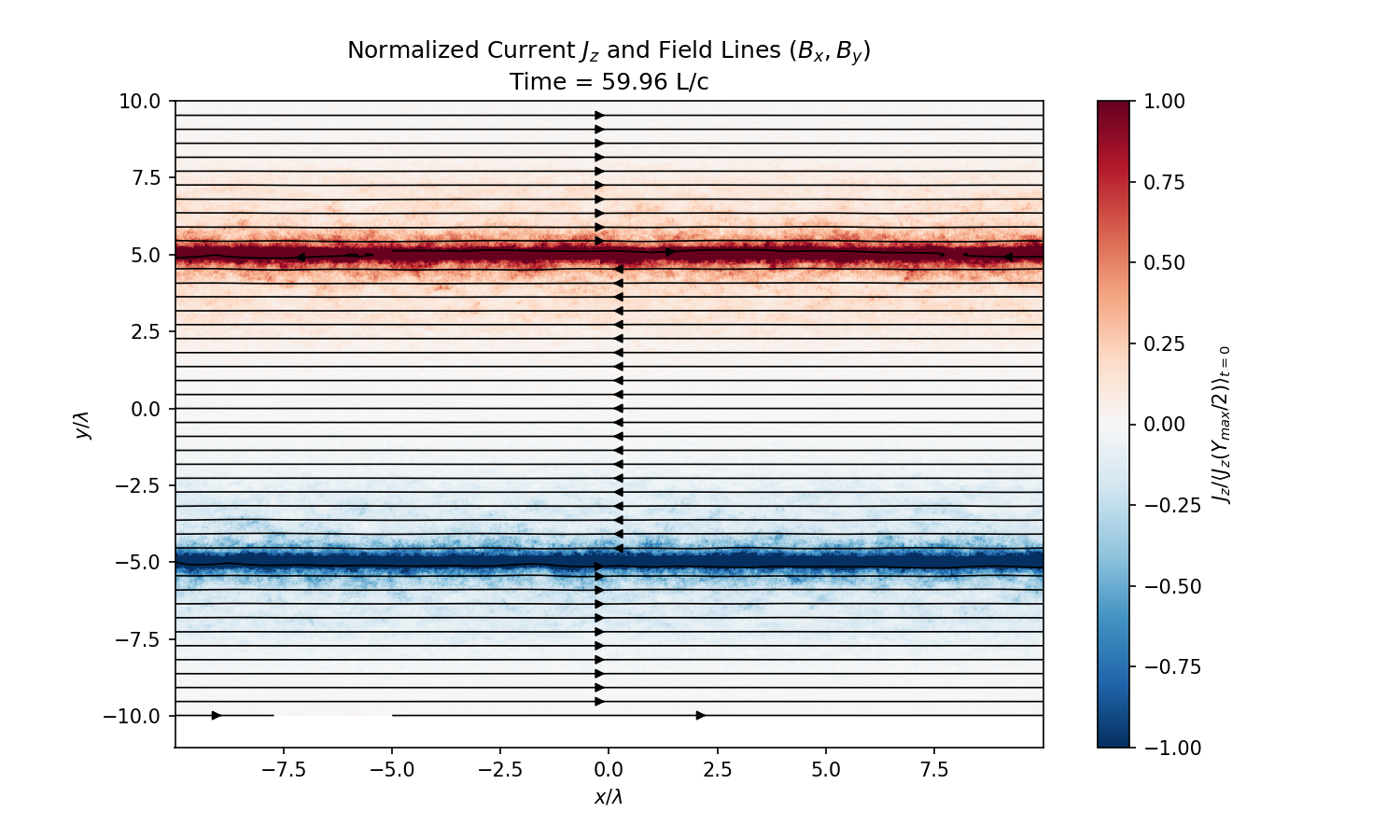}\vline
\includegraphics[width=.34\linewidth]{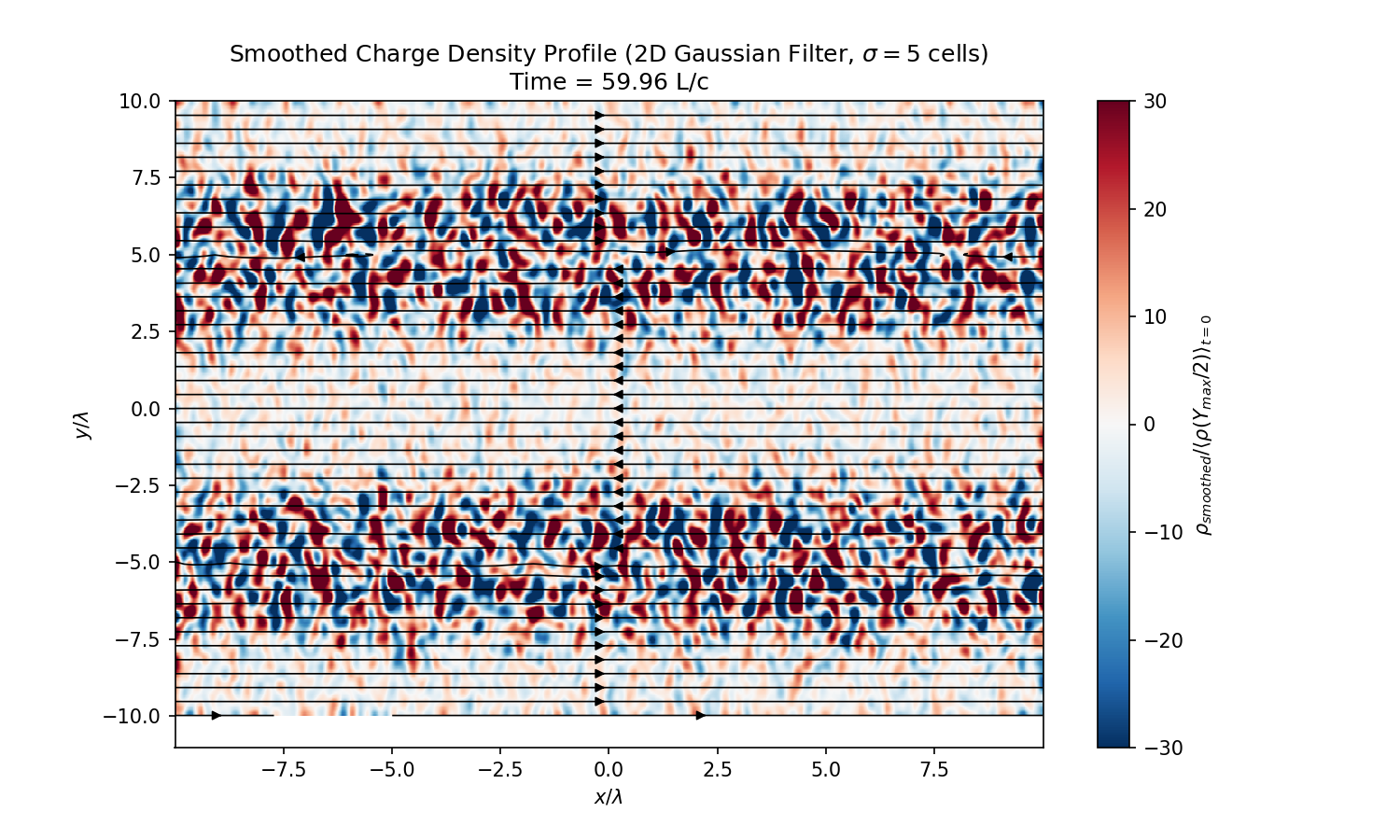}\vline
\includegraphics[width=.34\linewidth]{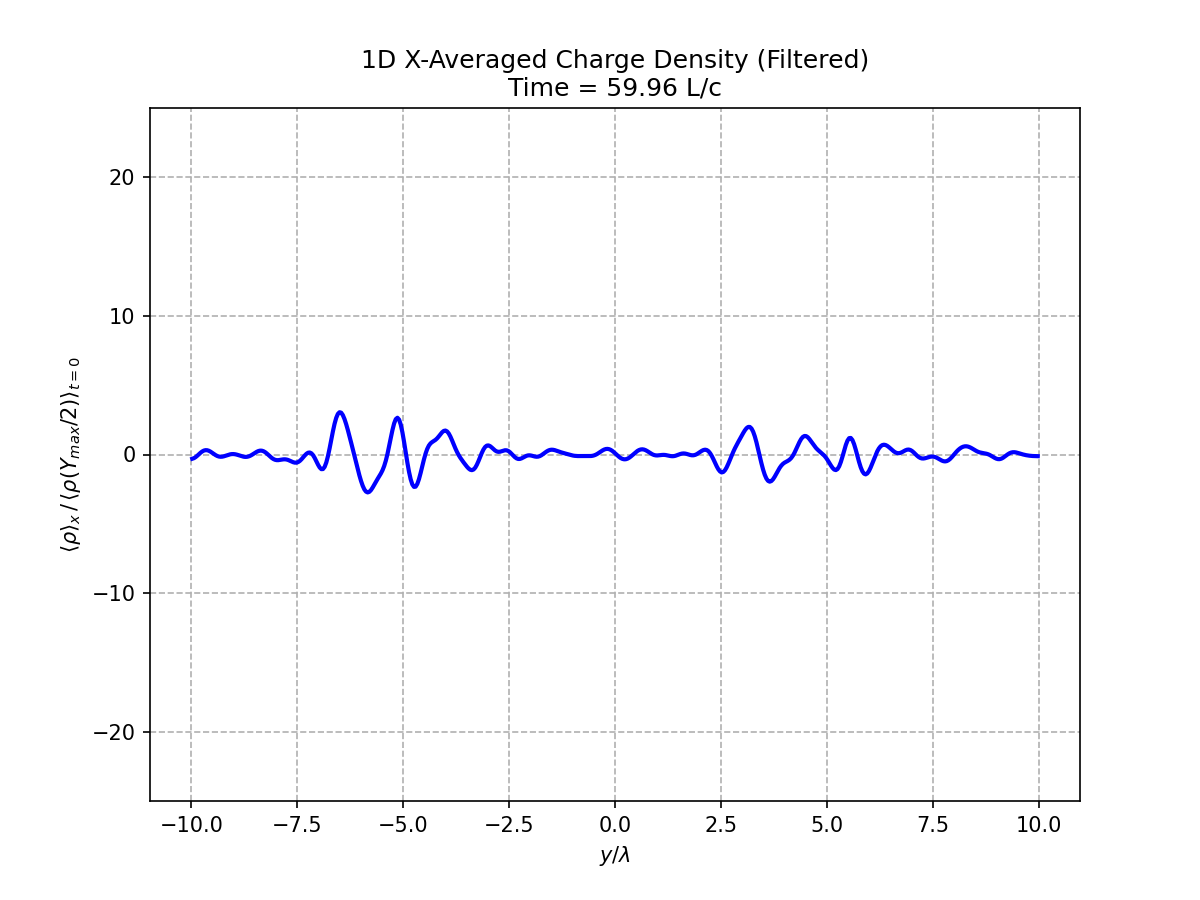}\\
\includegraphics[width=.34\linewidth]{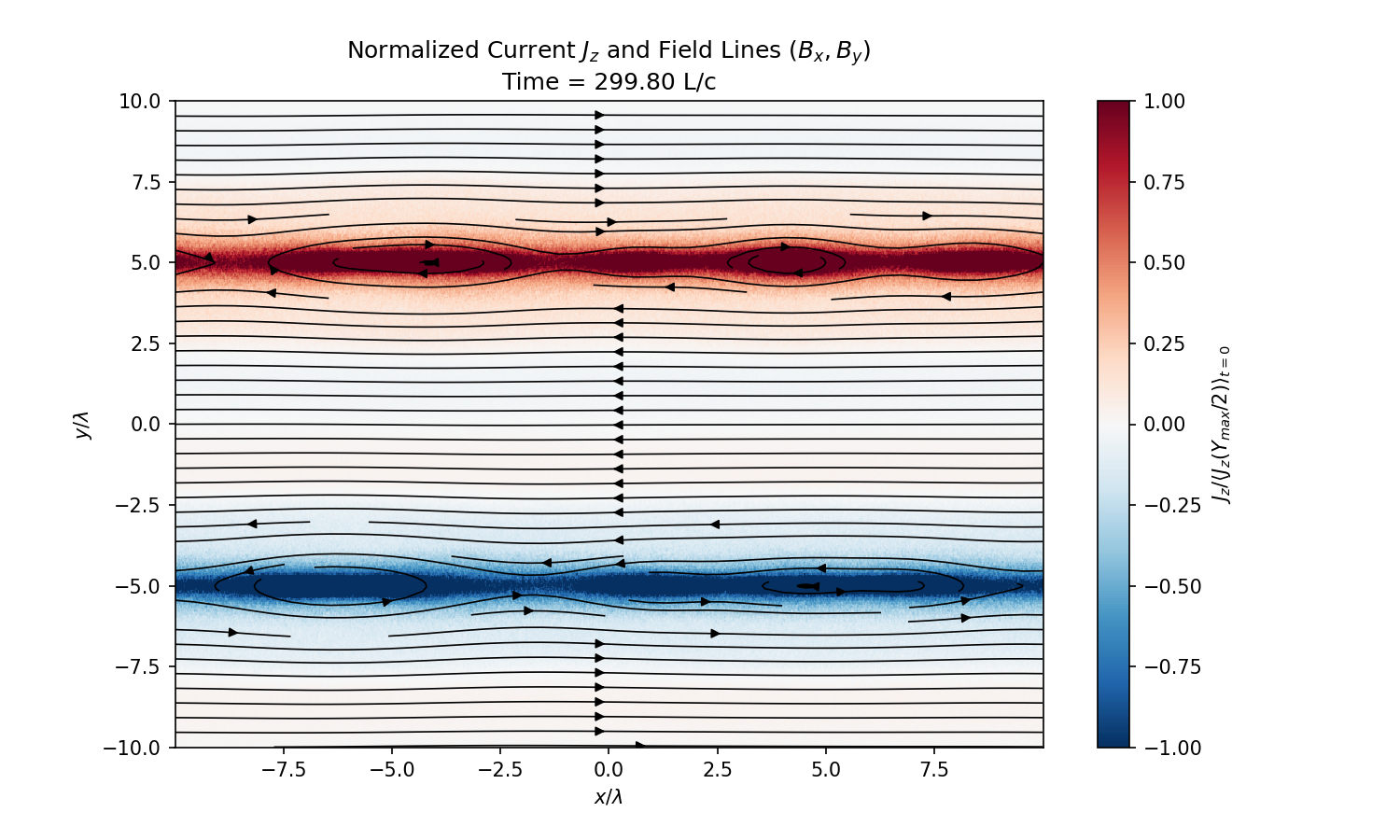}\vline
\includegraphics[width=.34\linewidth]{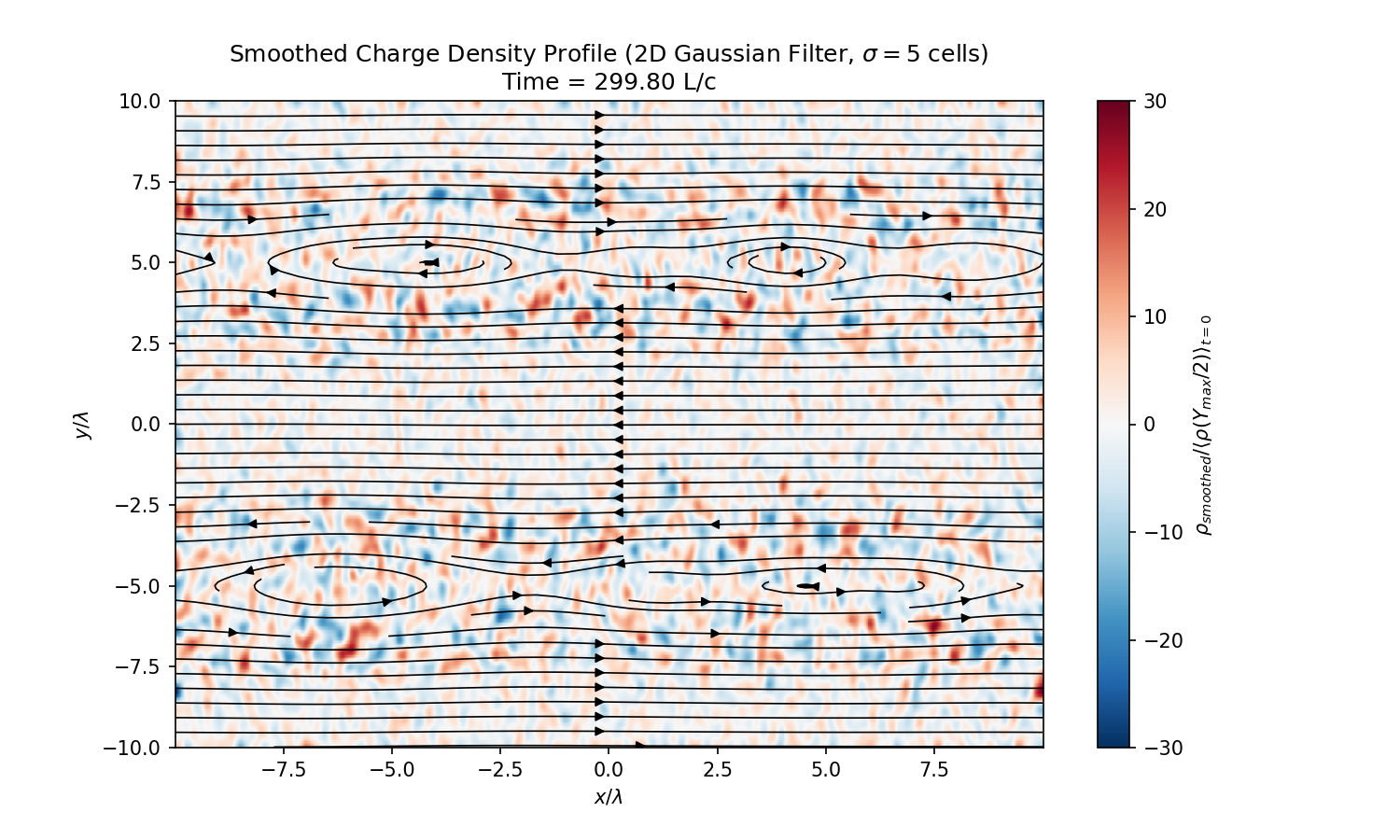}\vline
\includegraphics[width=.34\linewidth]{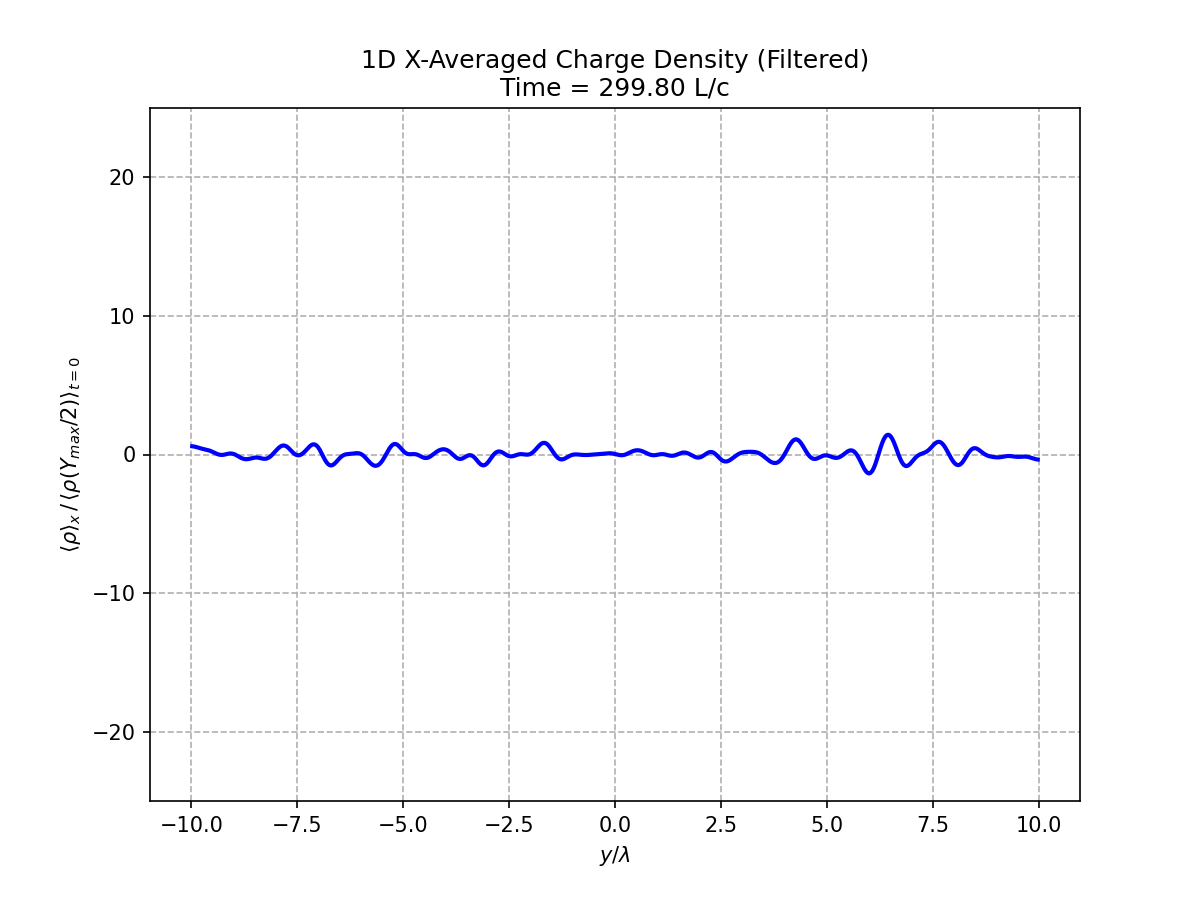}
\caption{Basic uncharged, cold force-free current layer; times 15, 30, 60, 300 $L/c$, parameters Cold-2, Table \ref{table},  with $\gamma_0=1$.  Notice temporary  appearance of large fluctuations of charge density, \eg,   at time = 30. }
\label{cold-uncharged}
\end{figure}

First, even basic uncharged  rotational current layer  develops large charge density fluctuations, Fig \ref{cold-uncharged}.   \citep[see][Fig. 3]{2025ApJ...979..104D}

\begin{figure}[h!]
\includegraphics[width=.34\linewidth]{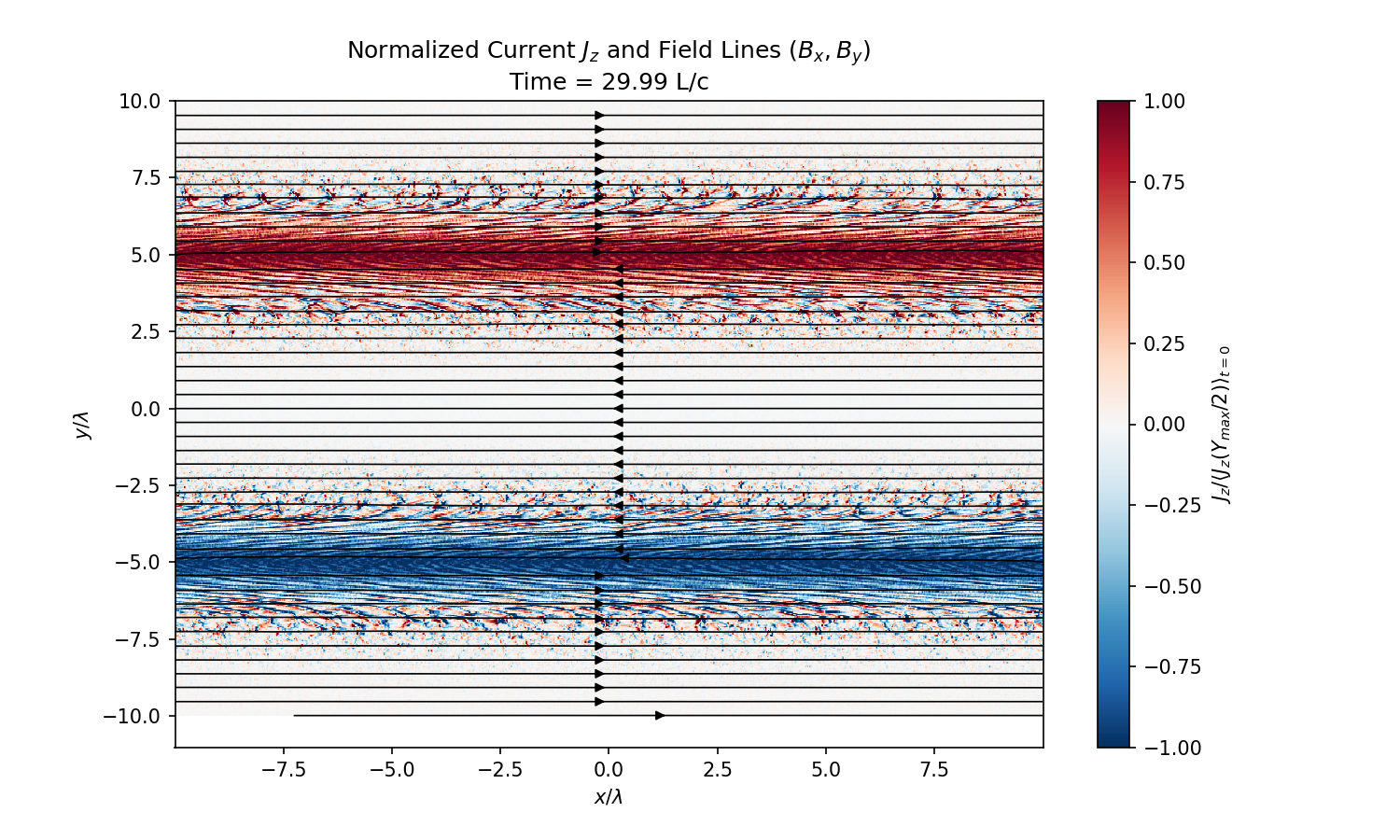}\vline
\includegraphics[width=.34\linewidth]{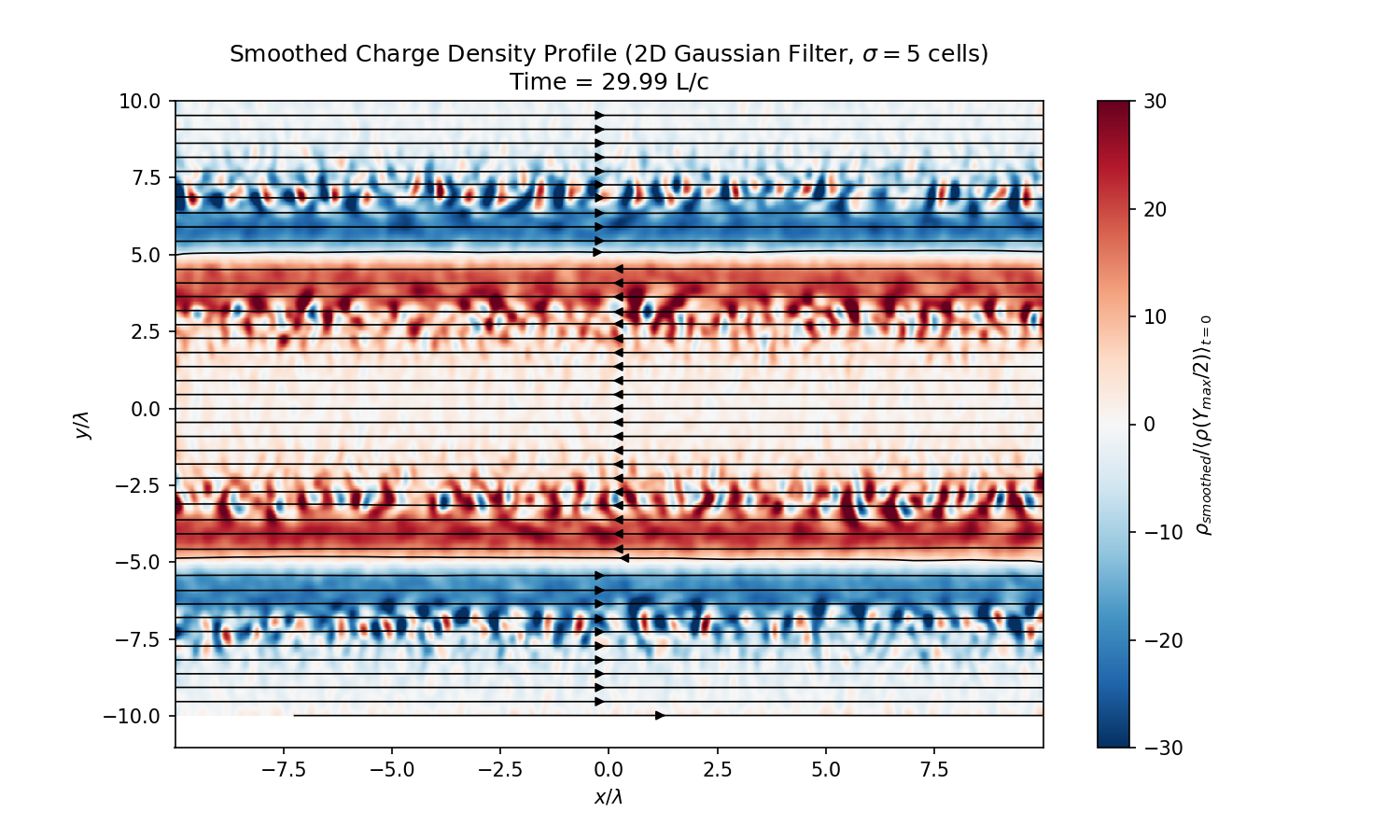}\vline
\includegraphics[width=.3\linewidth]{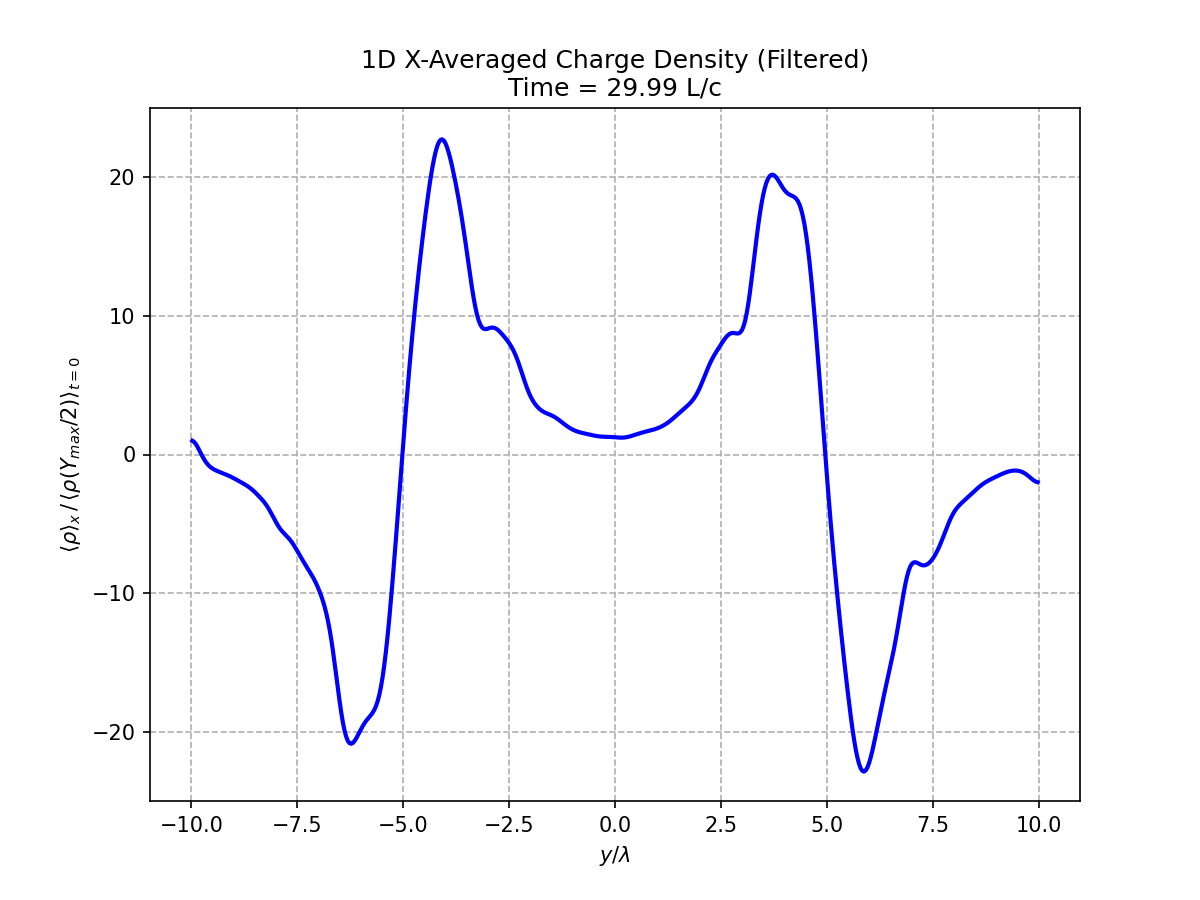}\\
\includegraphics[width=.34\linewidth]{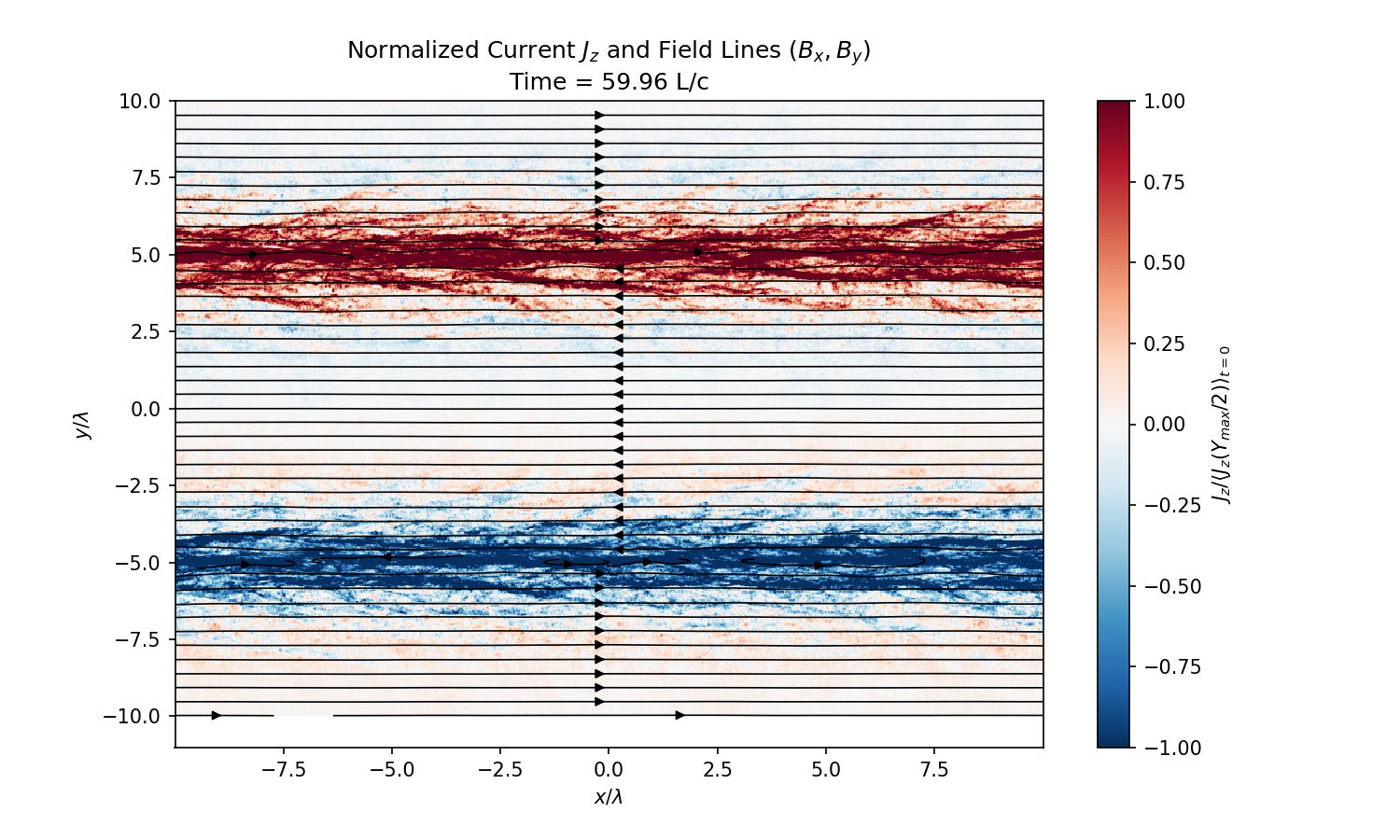}\vline
\includegraphics[width=.34\linewidth]{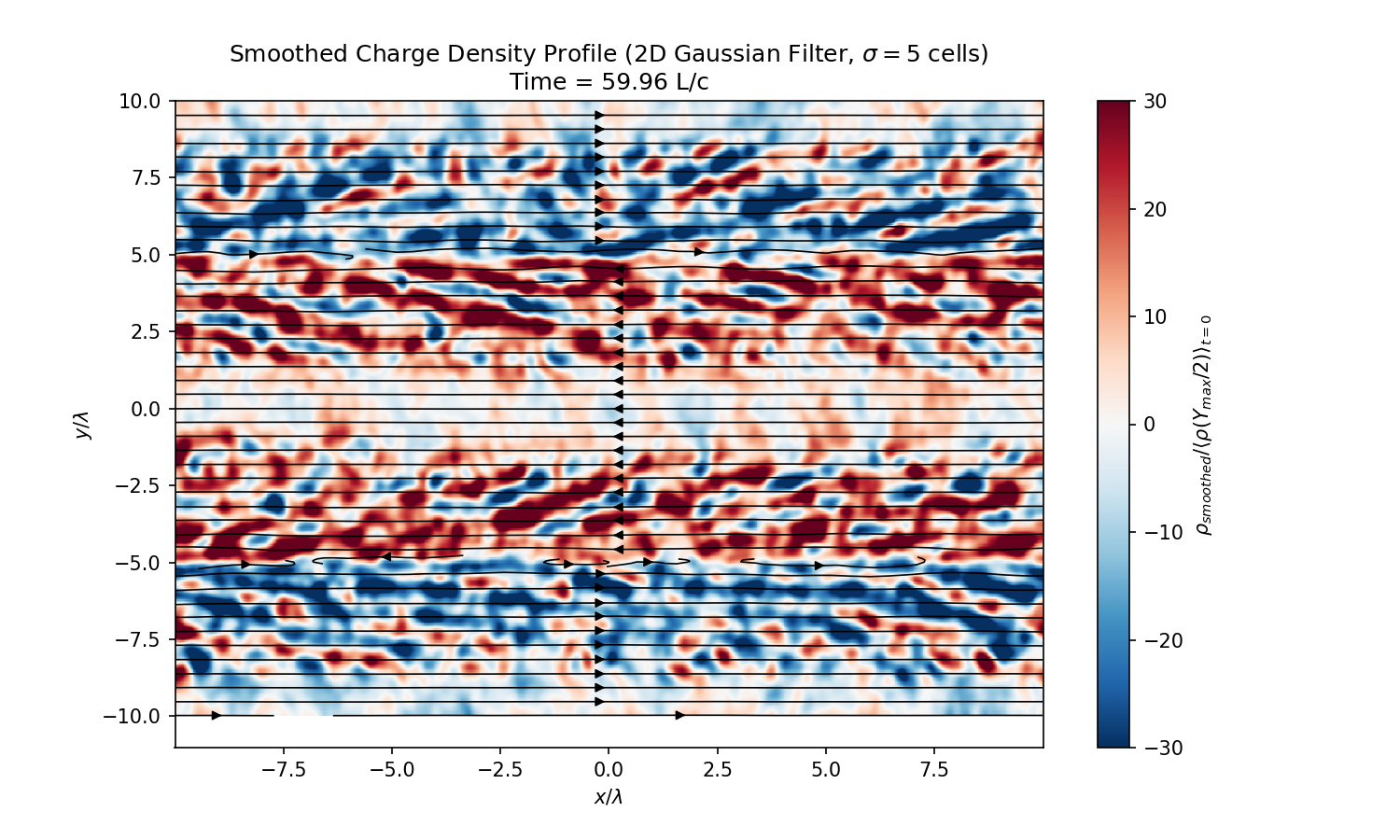}\vline
\includegraphics[width=.3\linewidth]{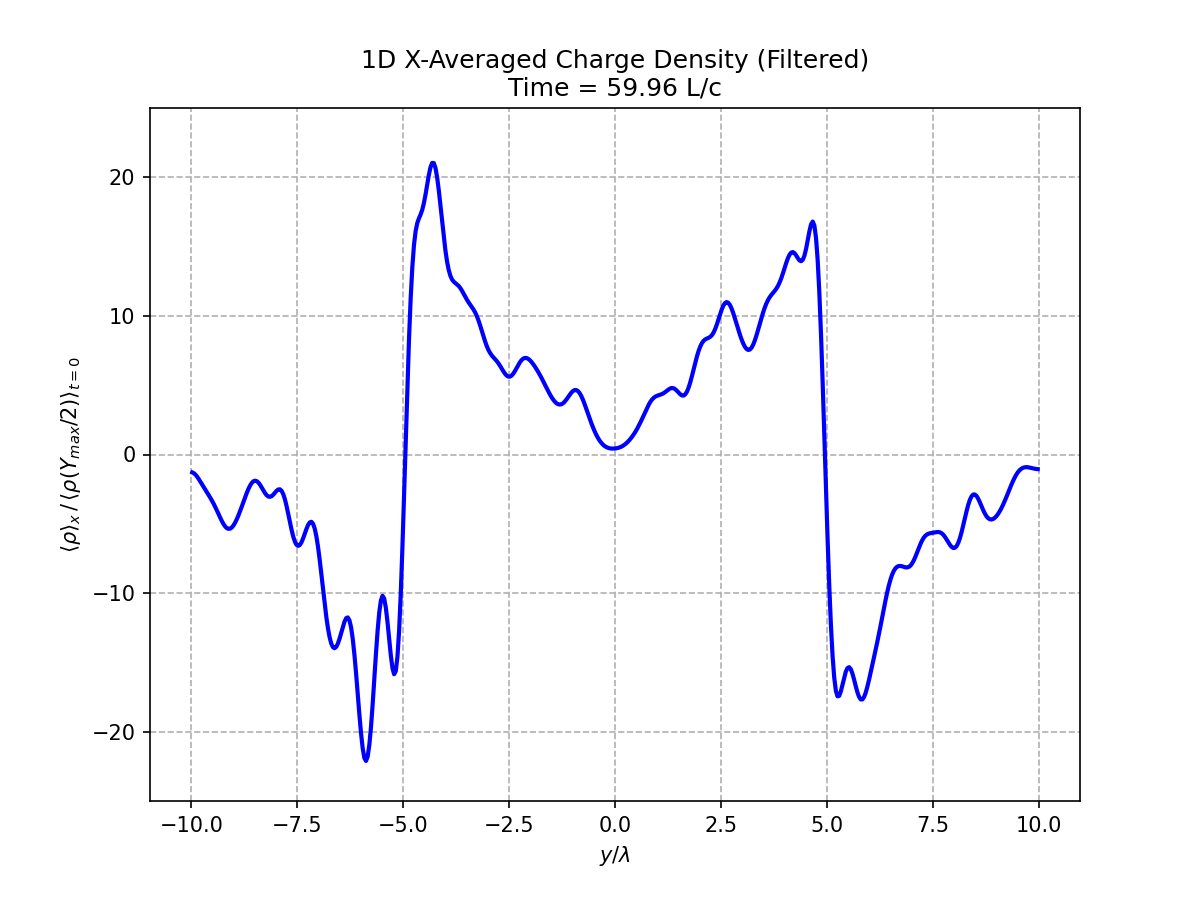}\\
\includegraphics[width=.34\linewidth]{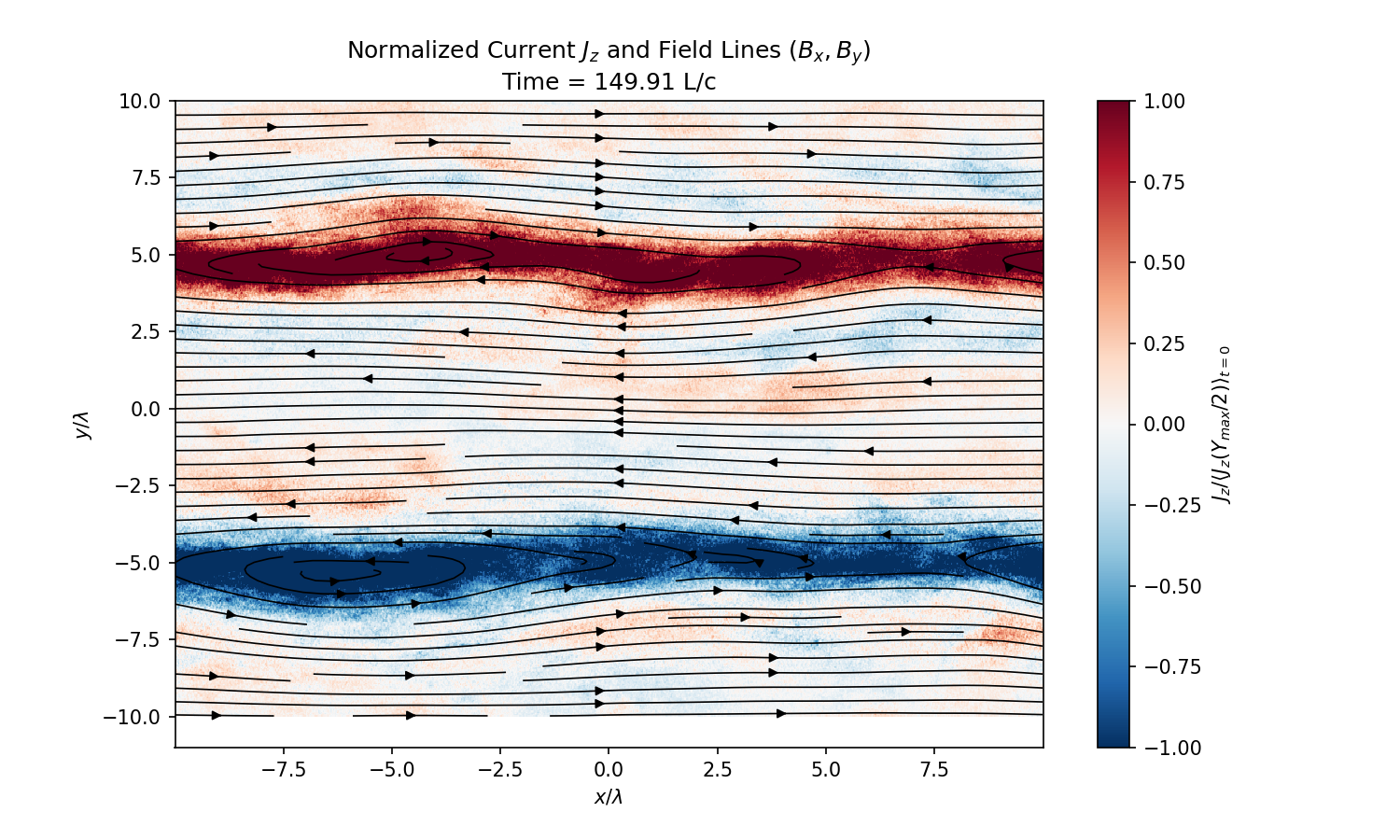}\vline
\includegraphics[width=.34\linewidth]{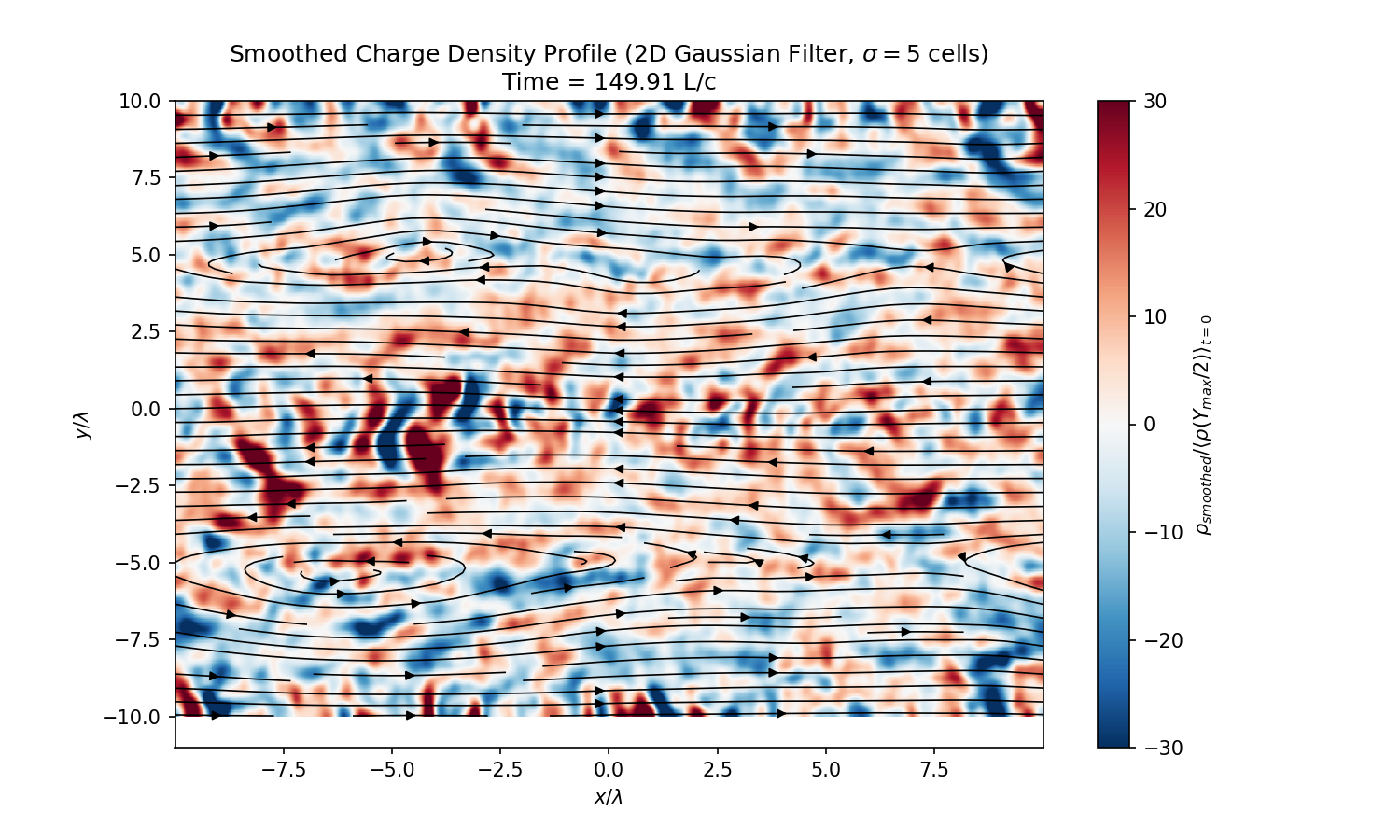}\vline
\includegraphics[width=.3\linewidth]{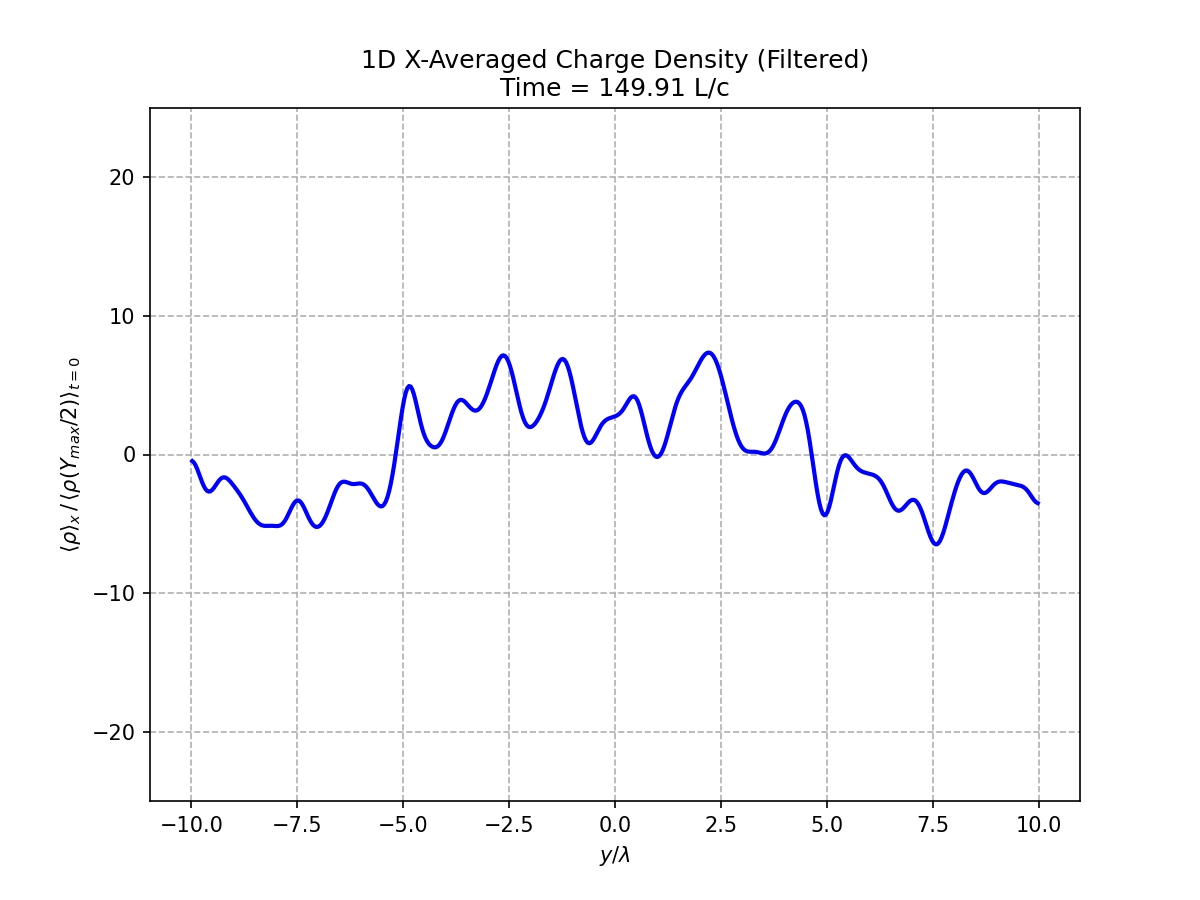}\\
\includegraphics[width=.34\linewidth]{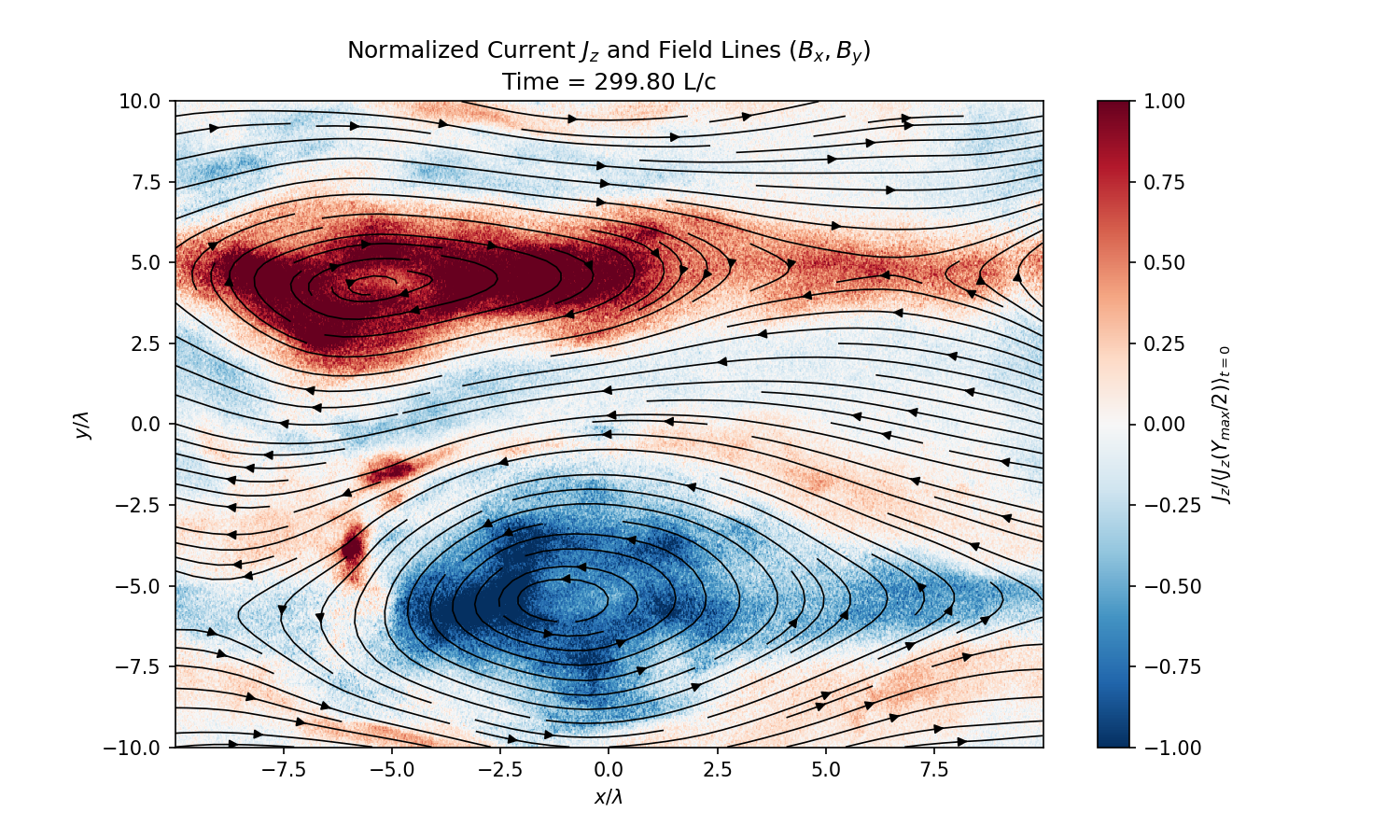}\vline
\includegraphics[width=.34\linewidth]{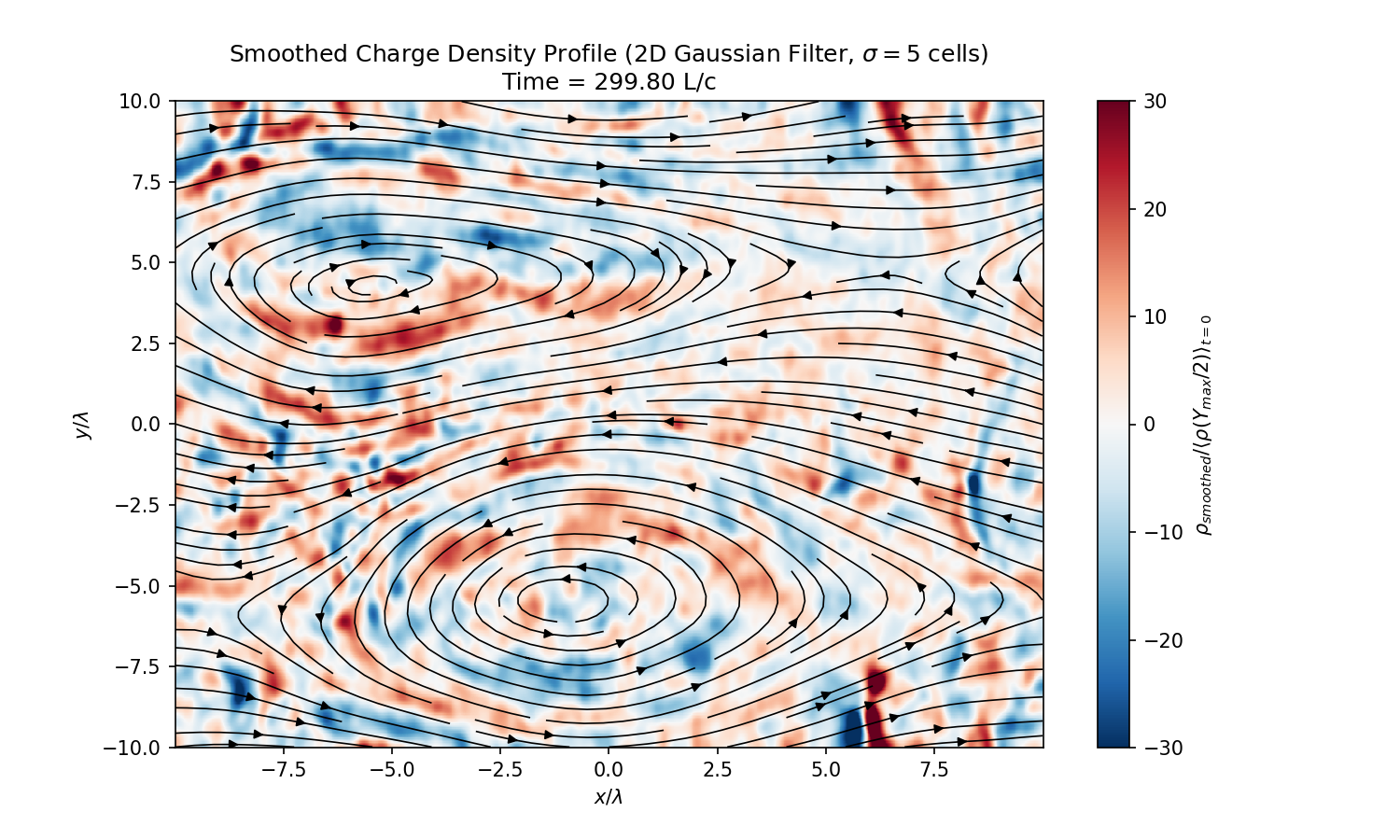}\vline
\includegraphics[width=.3\linewidth]{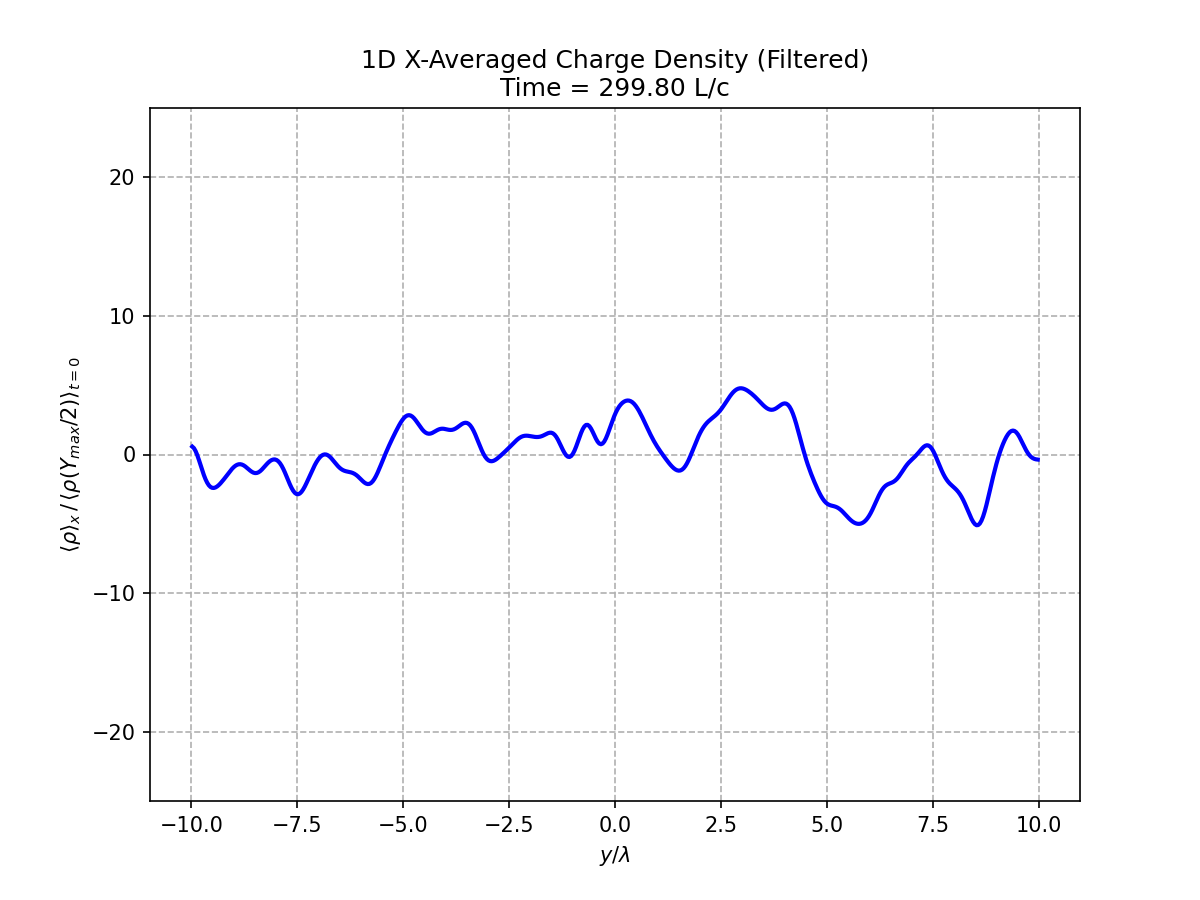}\\
\caption{Same as Fig. \ref{cold-uncharged} but with  $\gamma_0=2$ (charged: Rotaional-CRCS).}
\label{sdf-main00}
\end{figure}

Next, is one of our main unexpected results, Fig. \ref{sdf-main00}:  treating instability in initially charged  force-free current layers develops faster than in the uncharged case (compare last lines in Figs. \ref{cold-uncharged} and \ref{sdf-main00}.

\begin{figure}[h!]
\includegraphics[width=.34\linewidth]{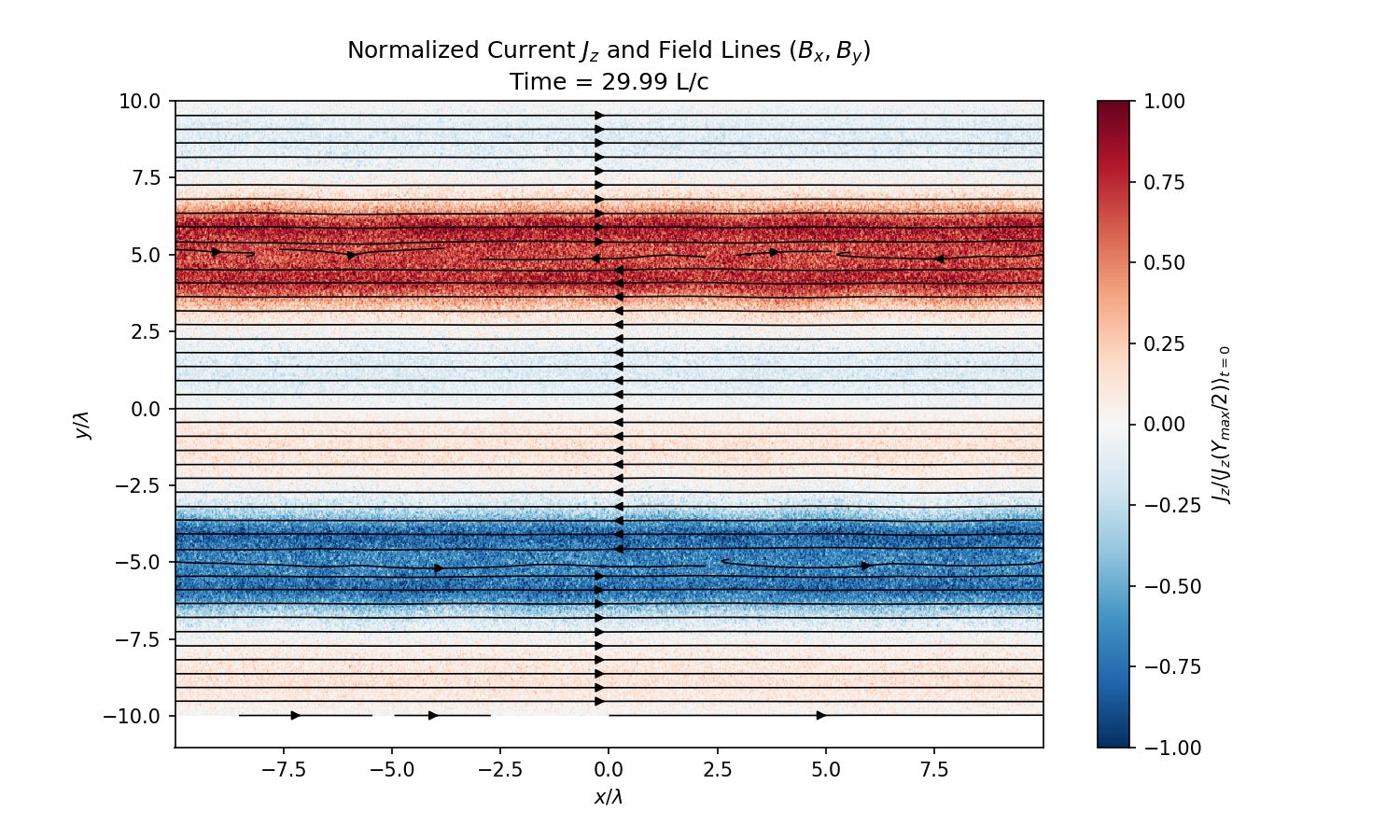}\vline
\includegraphics[width=.34\linewidth]{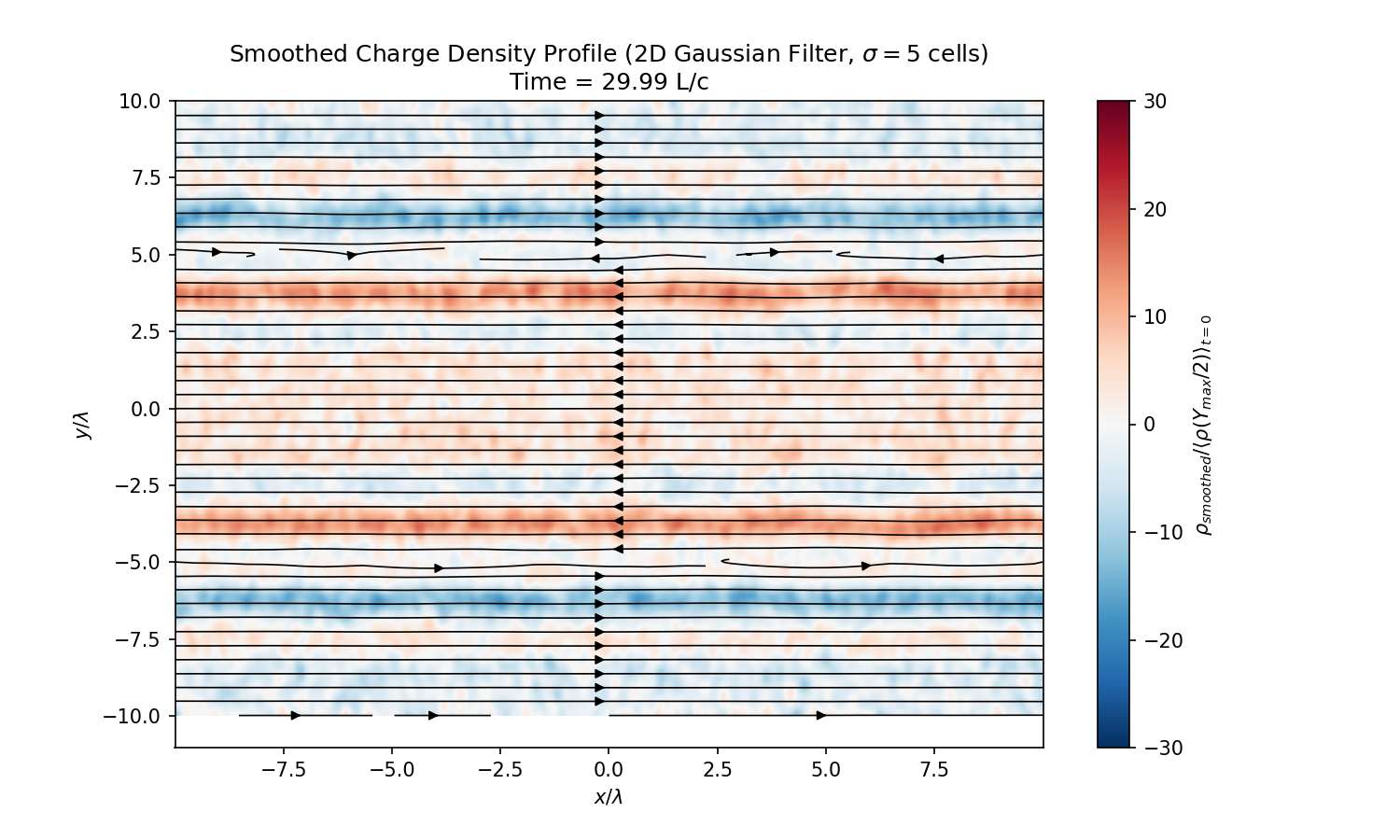}\vline
\includegraphics[width=.3\linewidth]{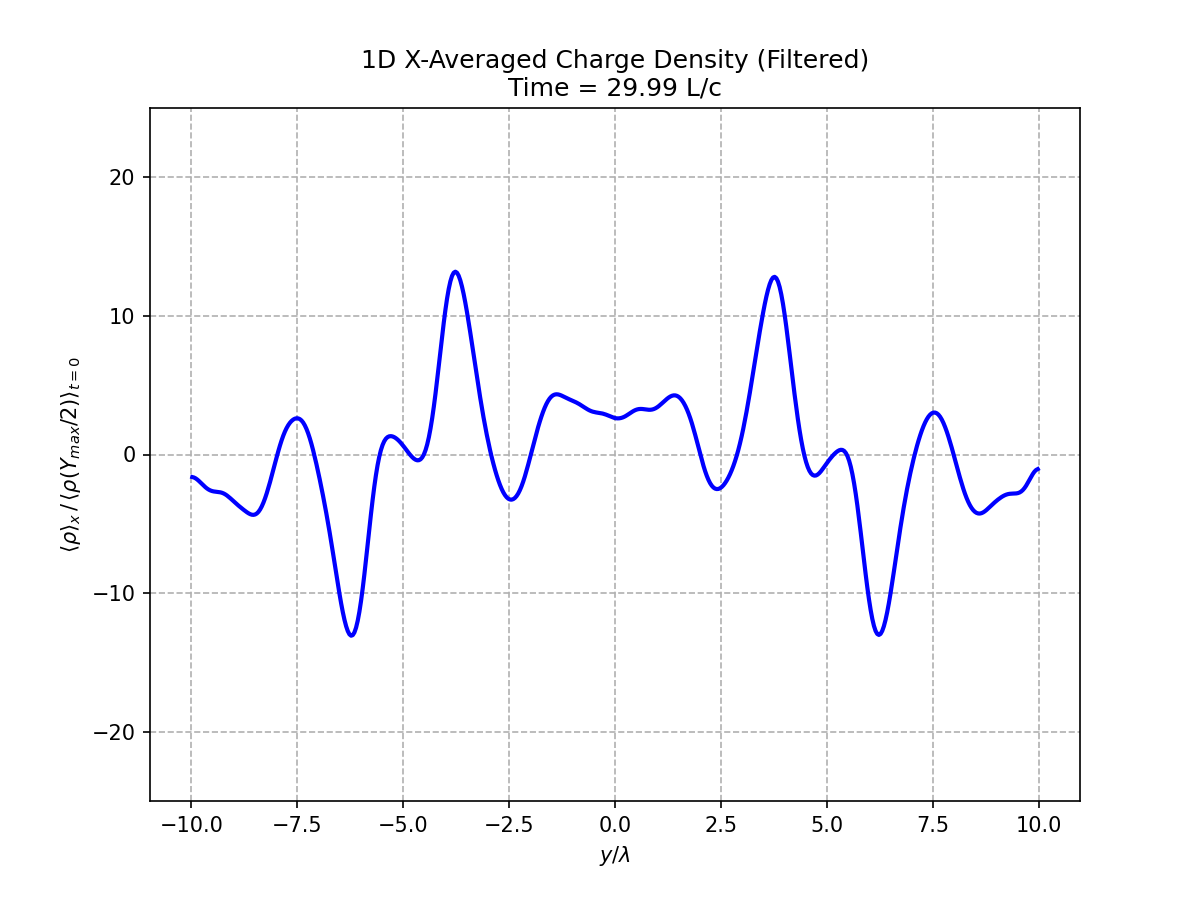}\\
\includegraphics[width=.34\linewidth]{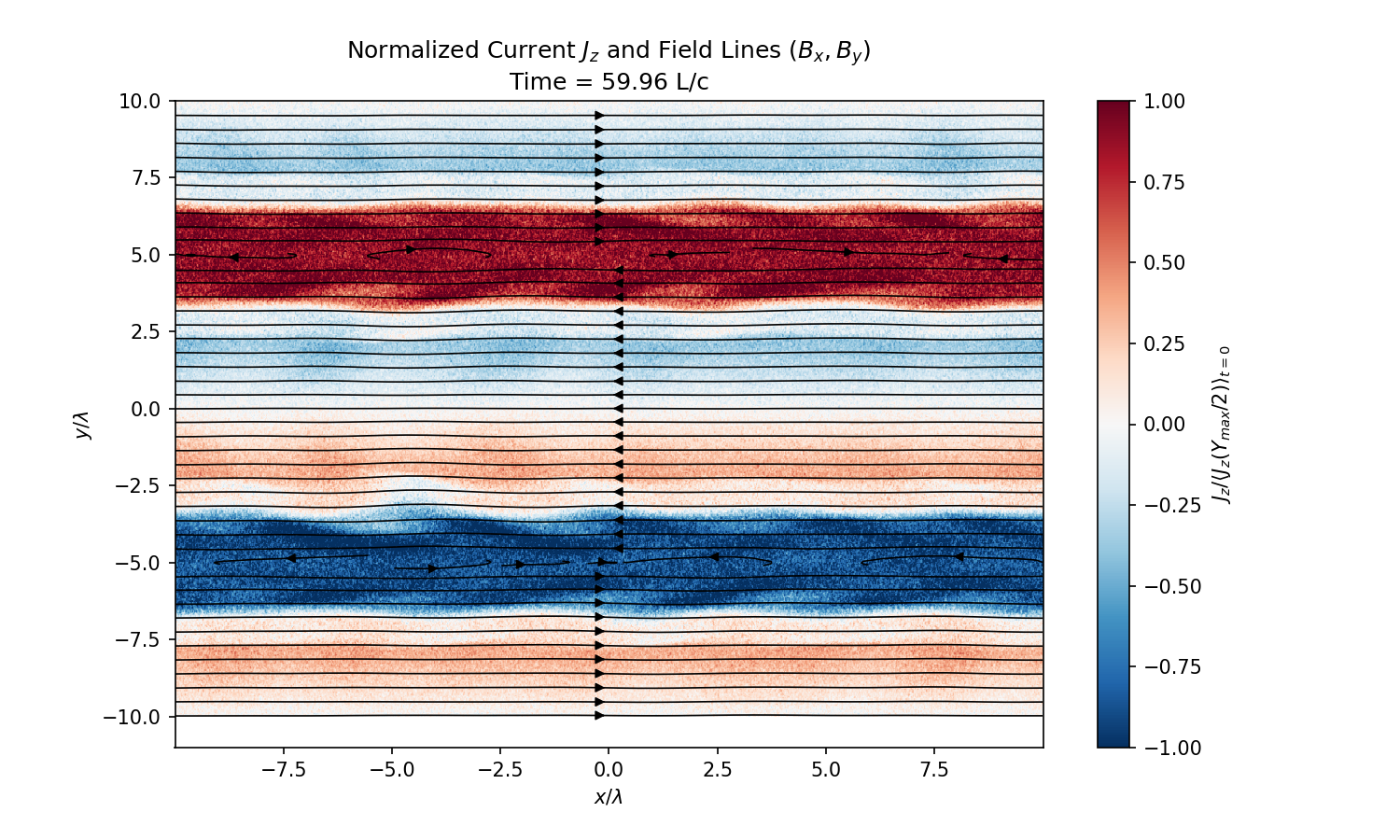}\vline
\includegraphics[width=.34\linewidth]{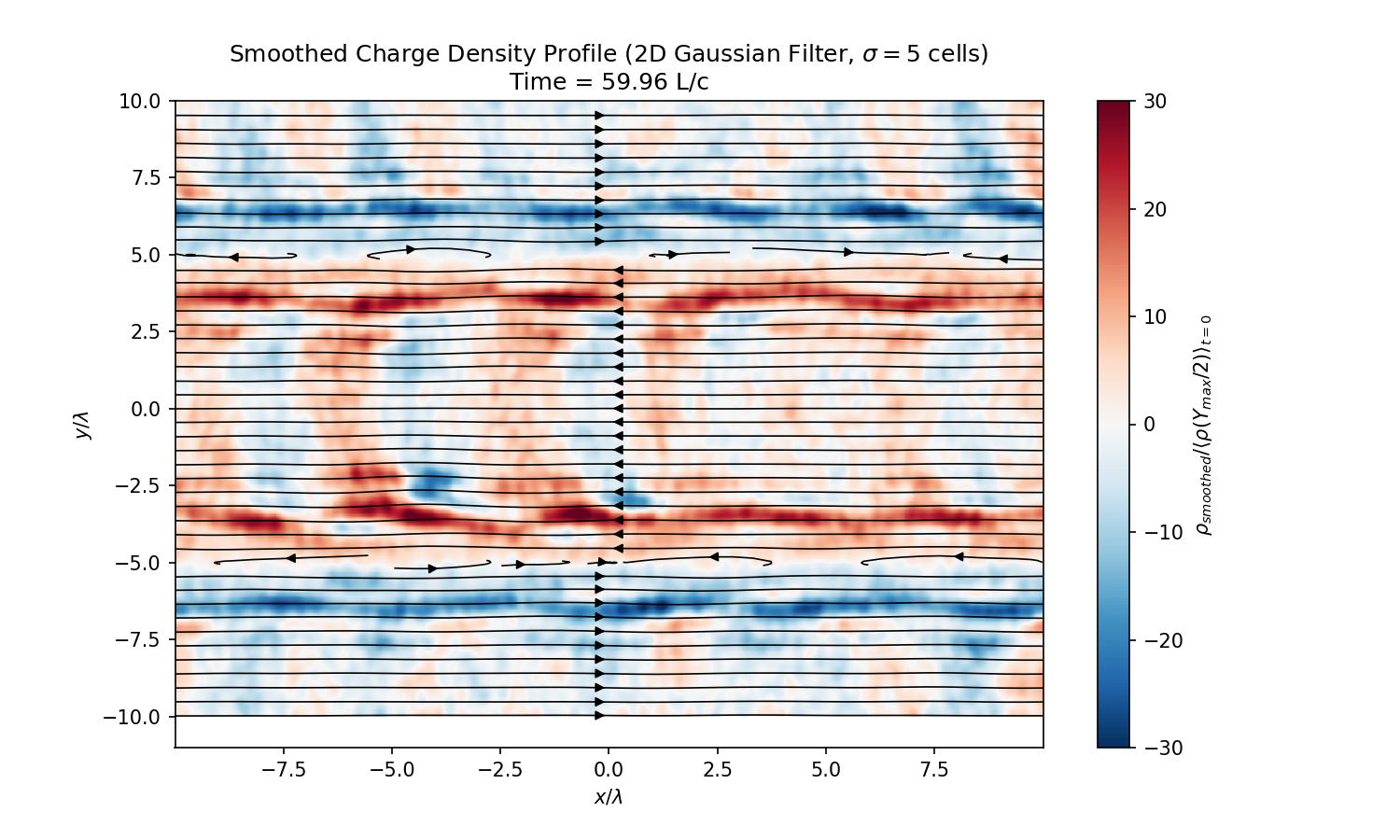}\vline
\includegraphics[width=.3\linewidth]{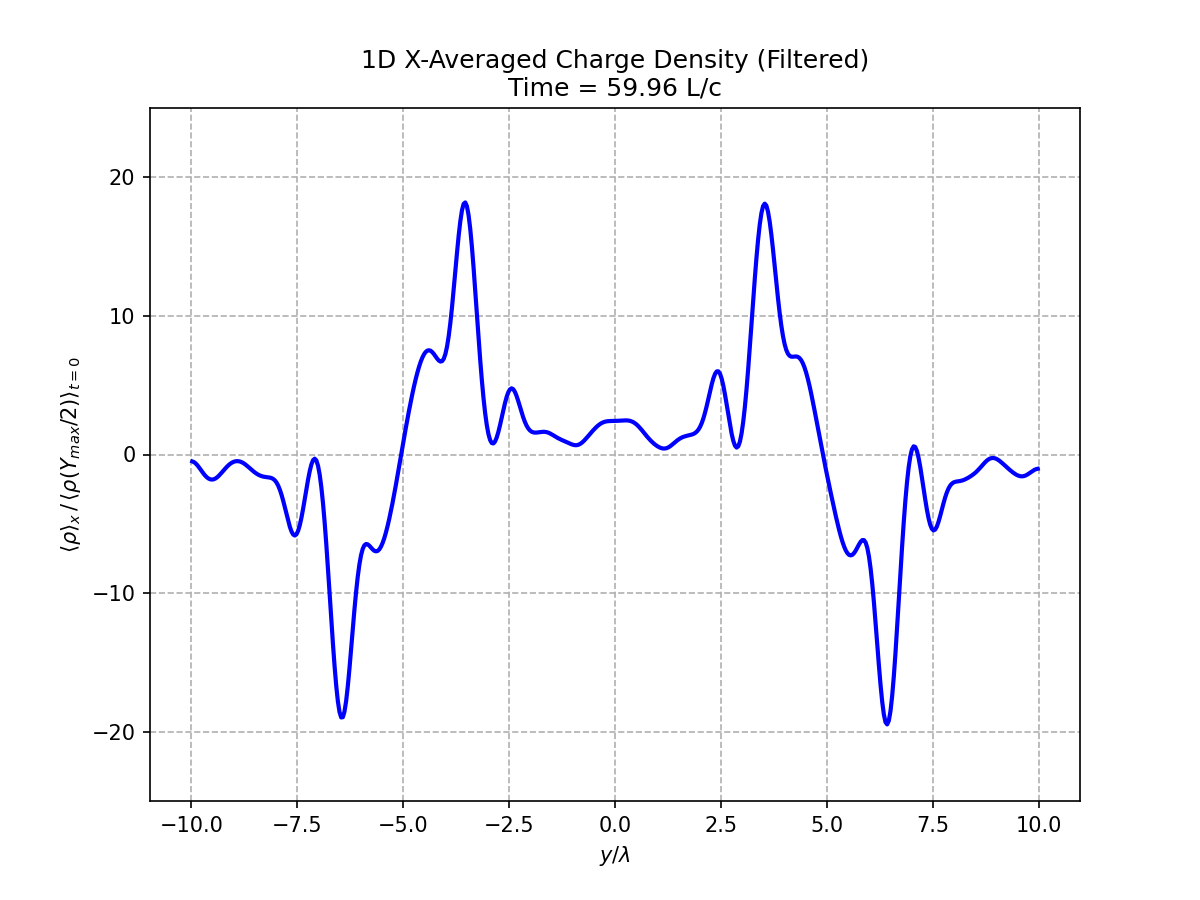}\\
\includegraphics[width=.34\linewidth]{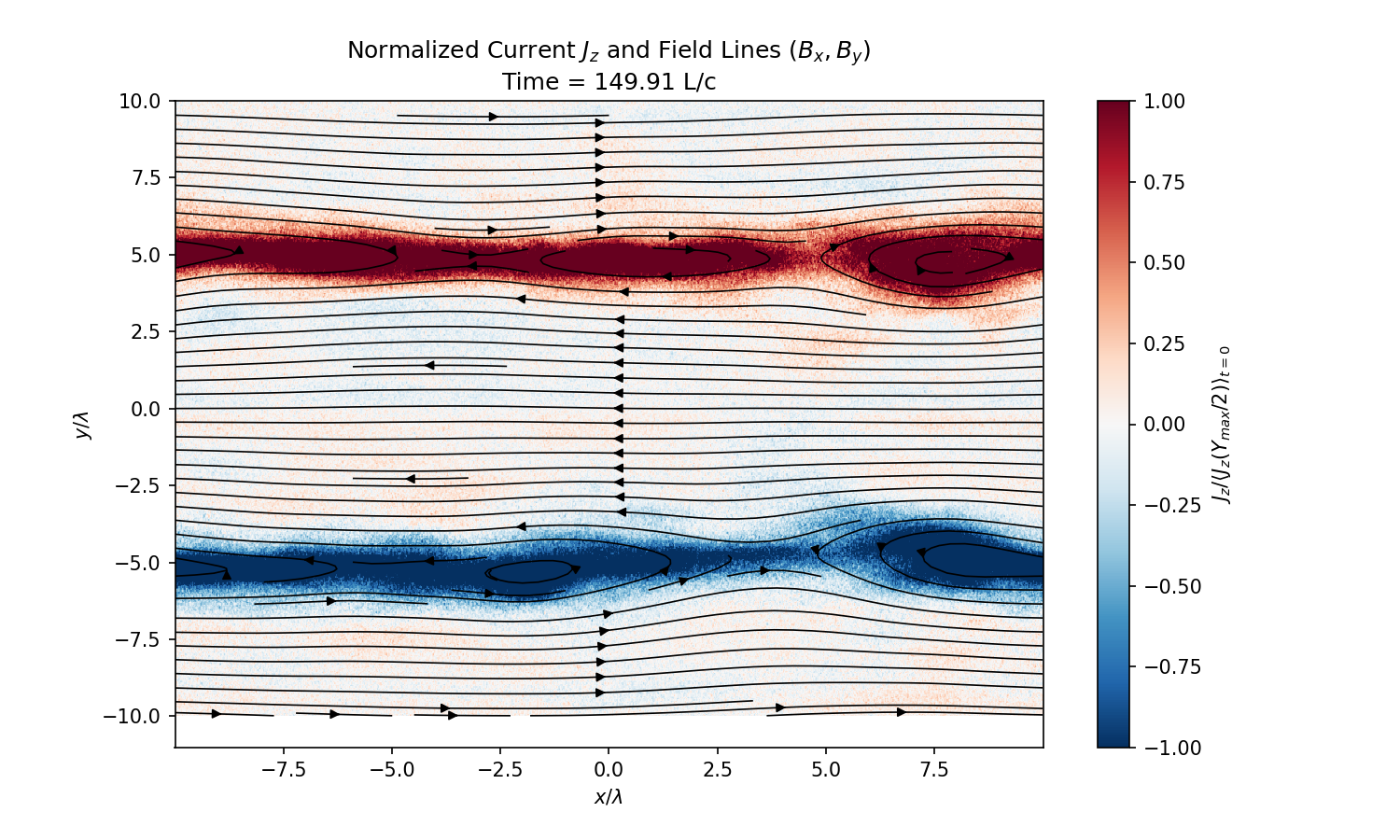}\vline
\includegraphics[width=.34\linewidth]{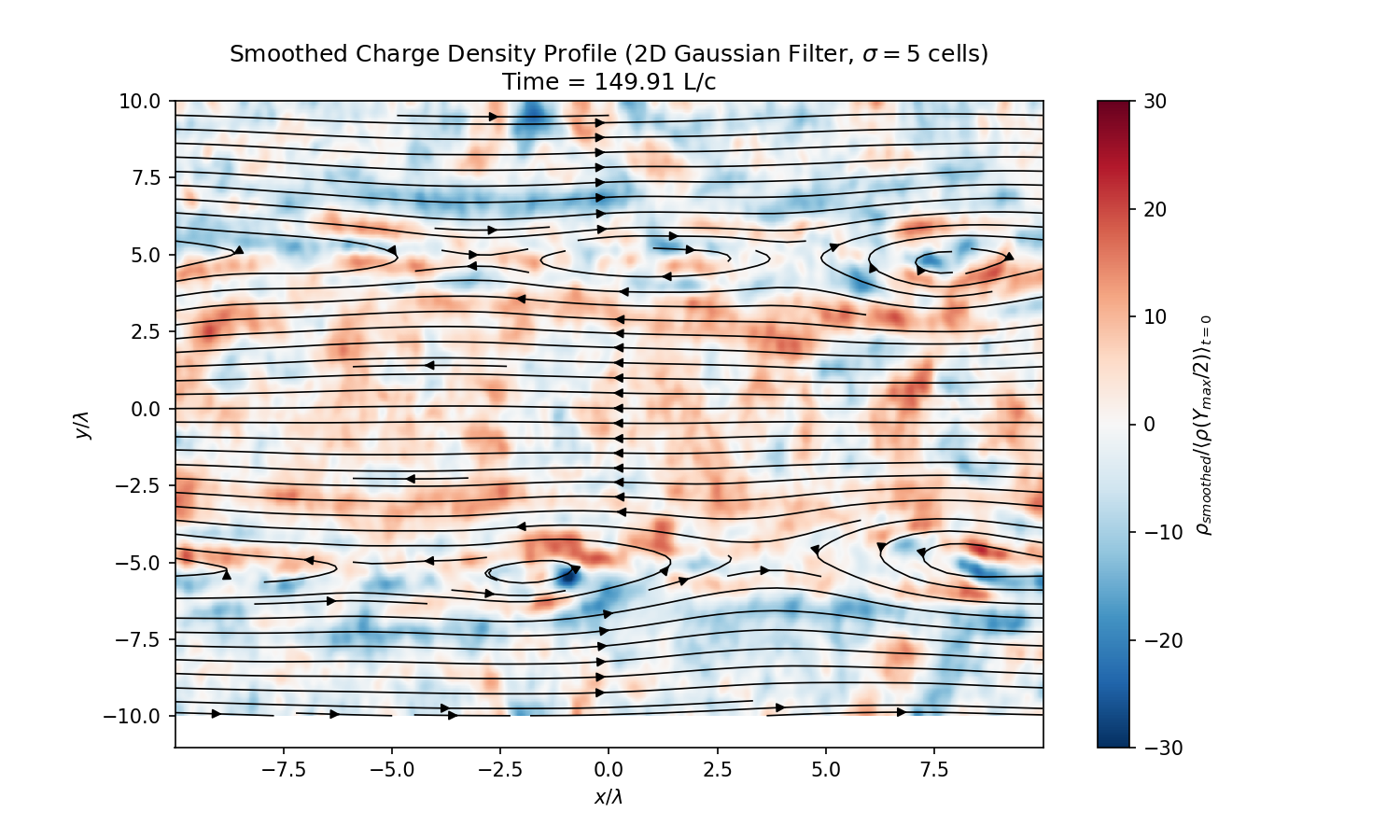}\vline
\includegraphics[width=.3\linewidth]{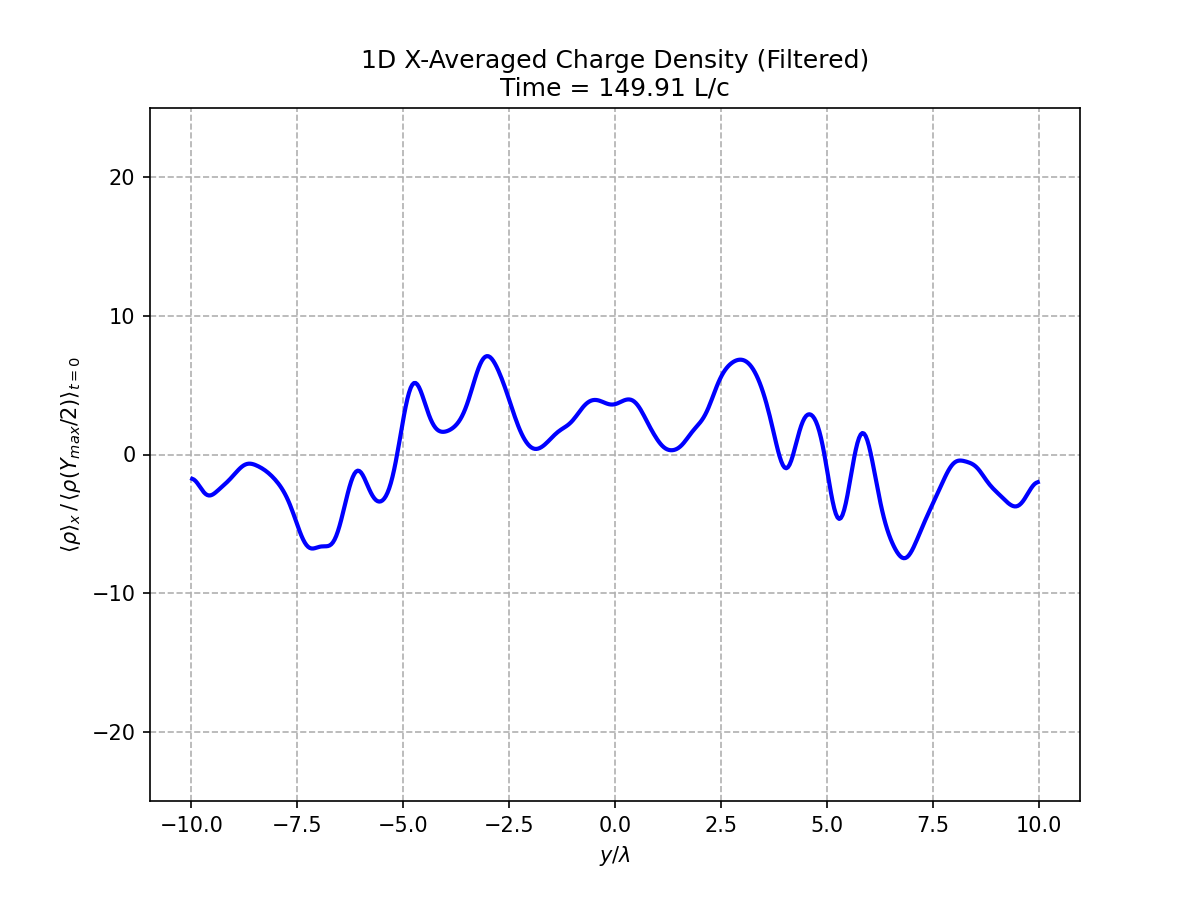}\\
\includegraphics[width=.34\linewidth]{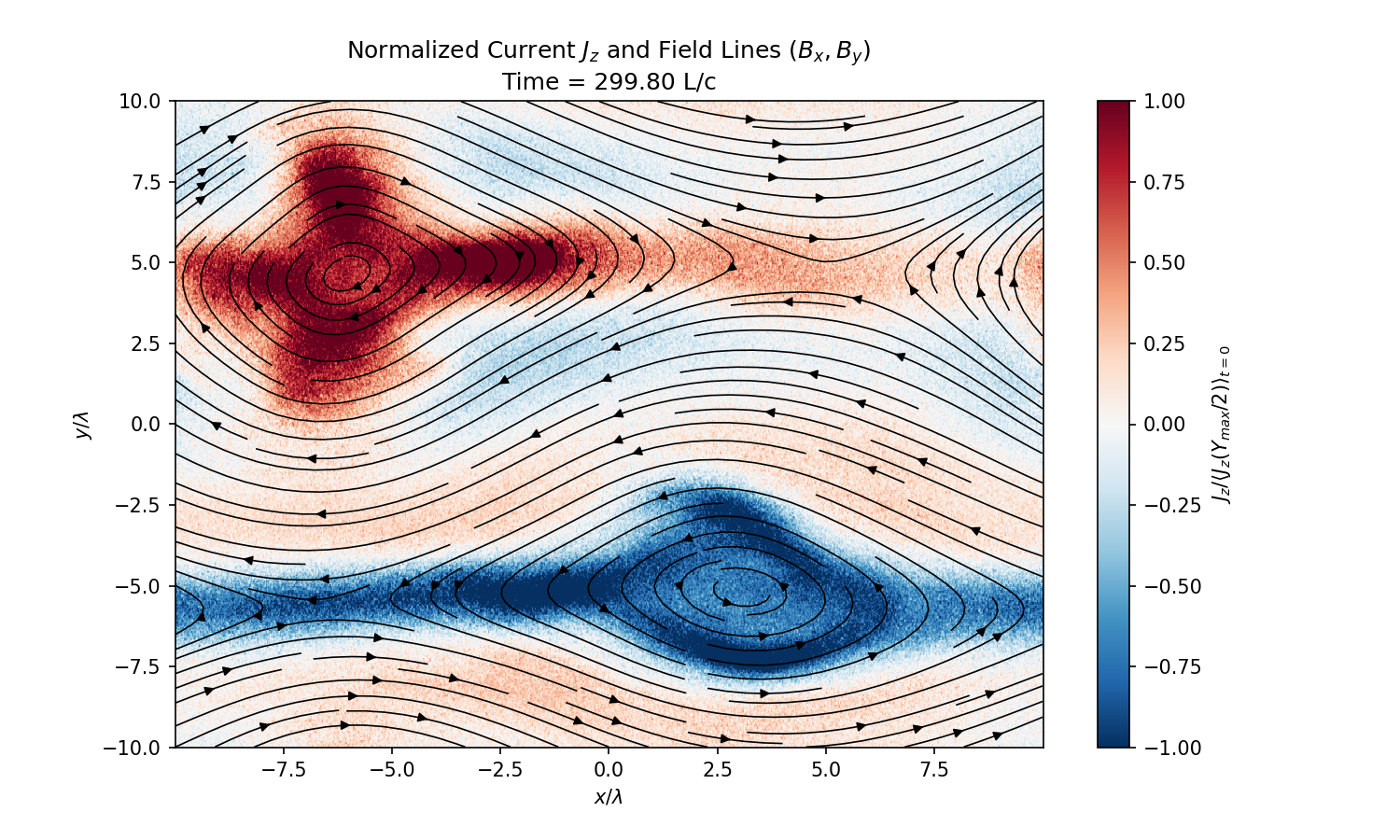}\vline
\includegraphics[width=.34\linewidth]{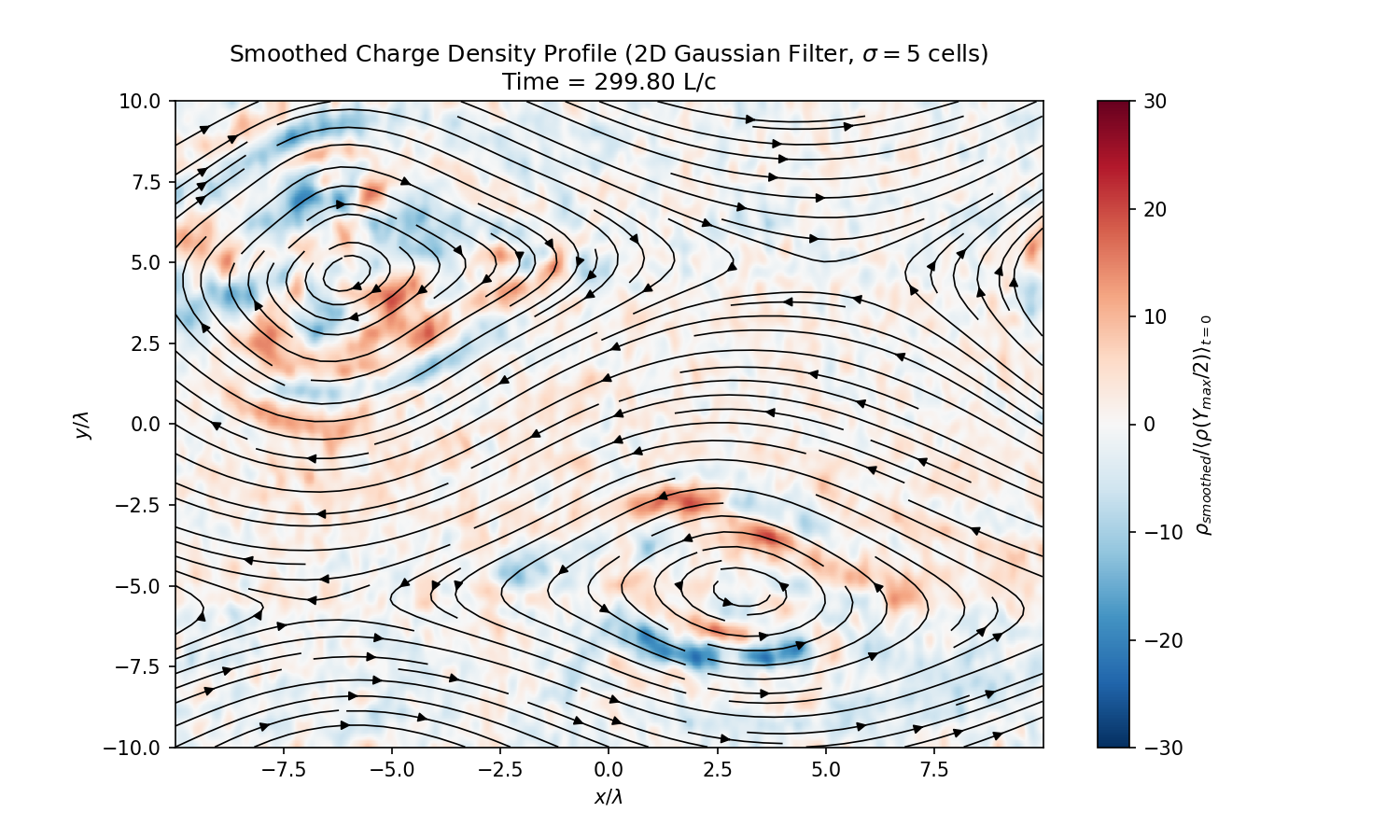}\vline
\includegraphics[width=.3\linewidth]{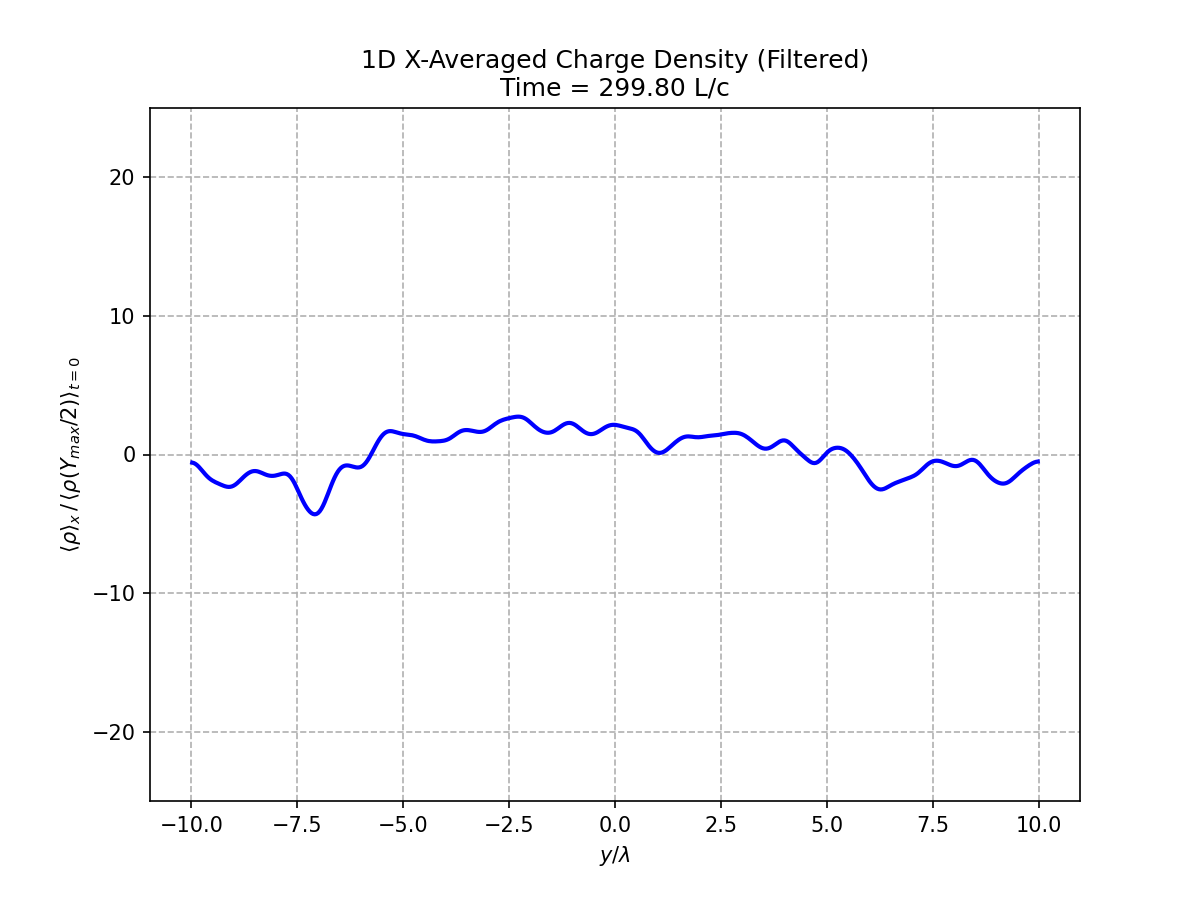}\\
\caption{Rotaional-CRCS, parameters Warm-1,  Table \ref{table}}
\label{sdf-main01}
\end{figure}

Making the initial plasma  warm, Fig. \ref{sdf-main01}, charge relaxation (right column) happens much faster than in the cold case Fig. \ref{sdf-main00},

\begin{figure}[h!]
\includegraphics[width=.3\linewidth]{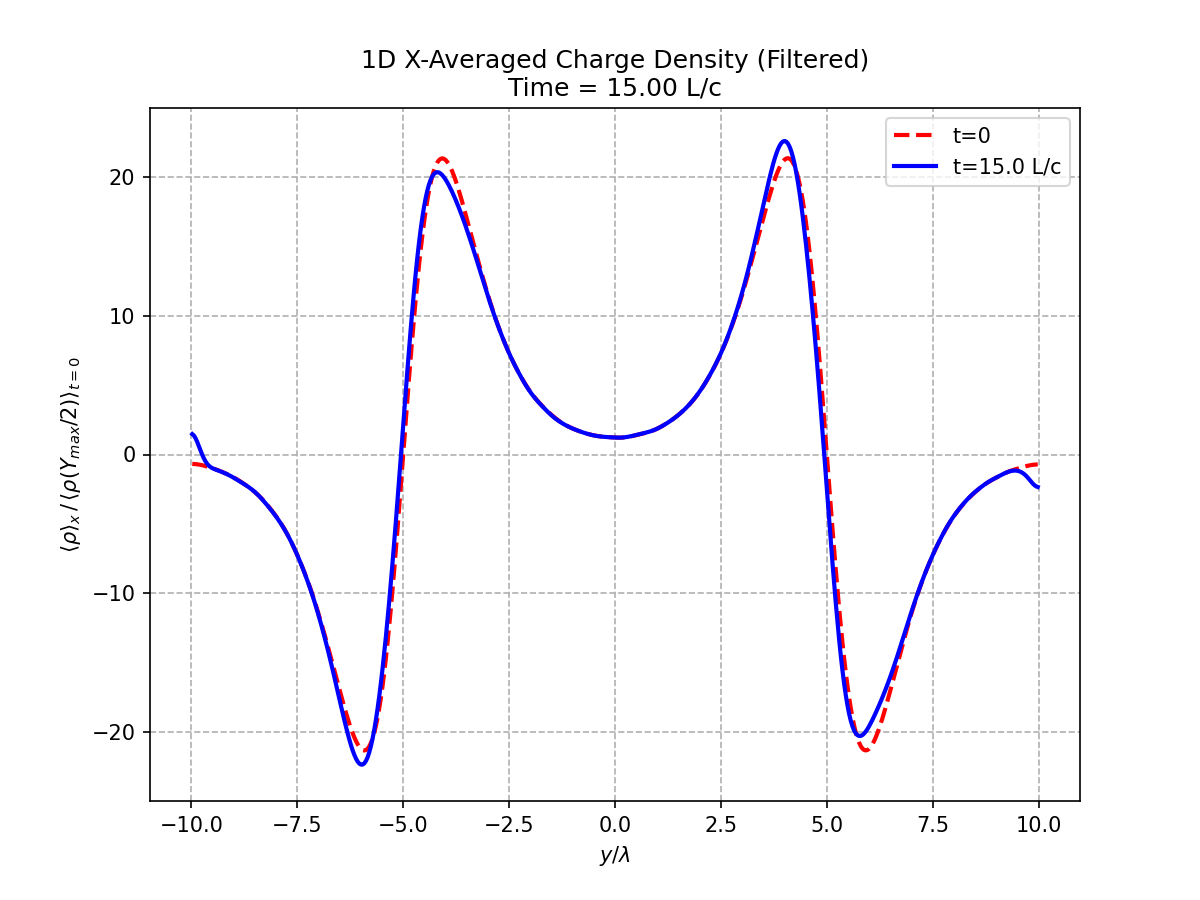}\vline
\includegraphics[width=.3\linewidth]{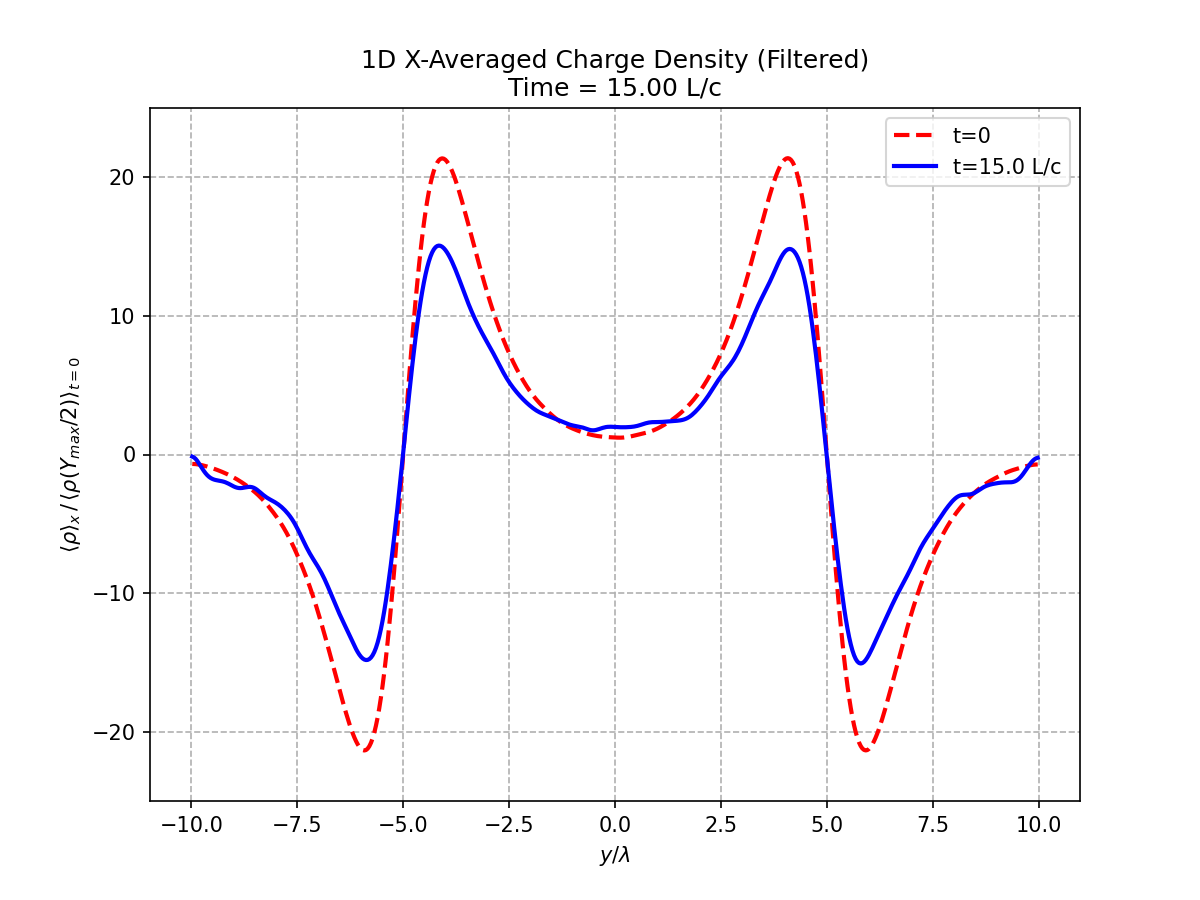}\vline
\includegraphics[width=.3\linewidth]{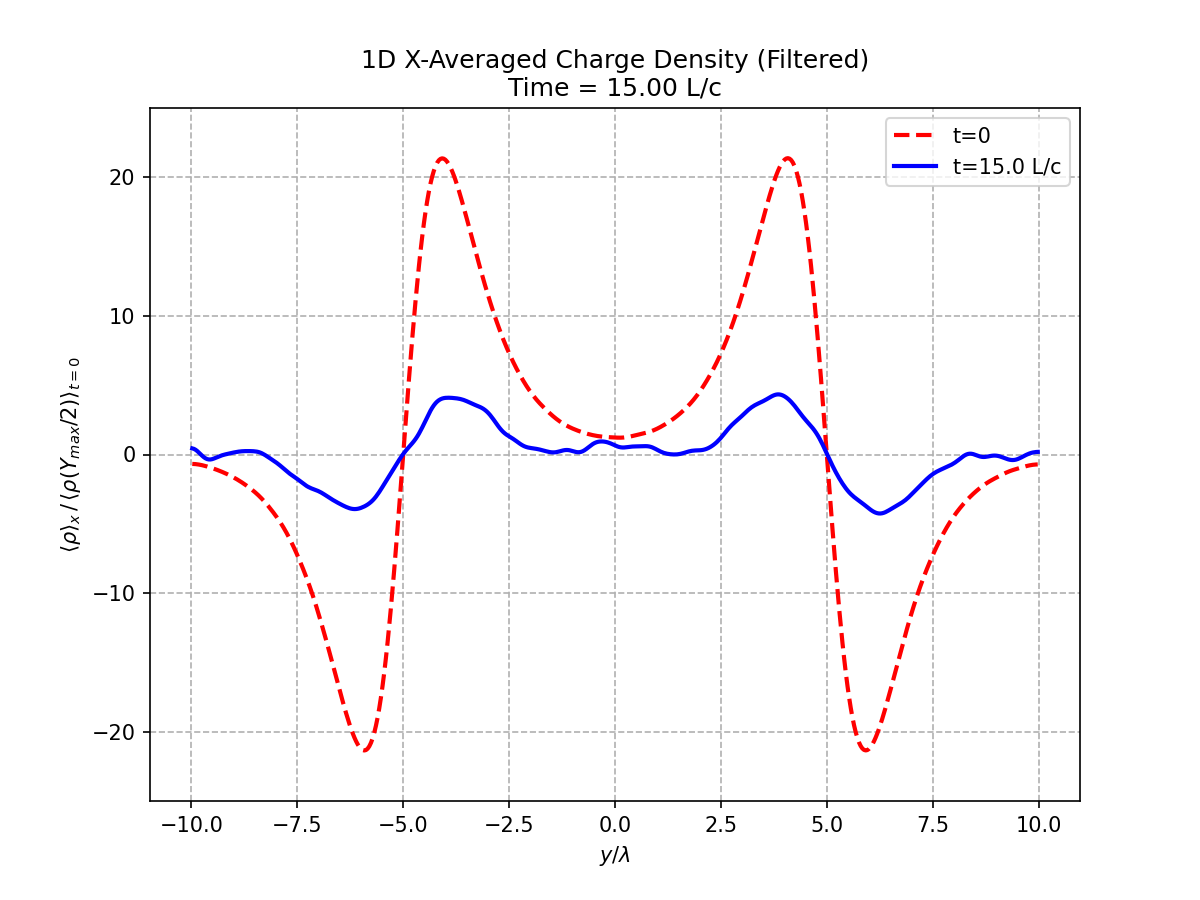}\\
\includegraphics[width=.3\linewidth]{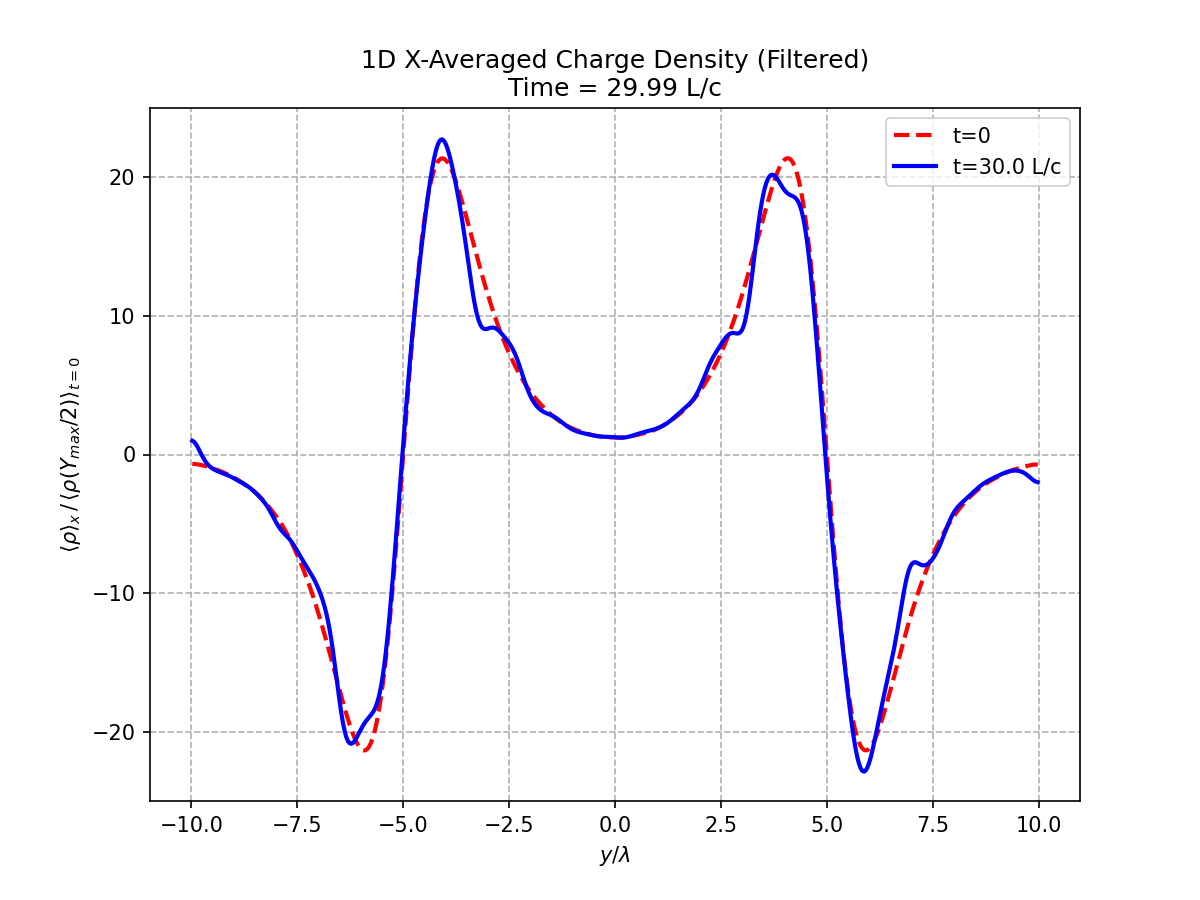}\vline
\includegraphics[width=.3\linewidth]{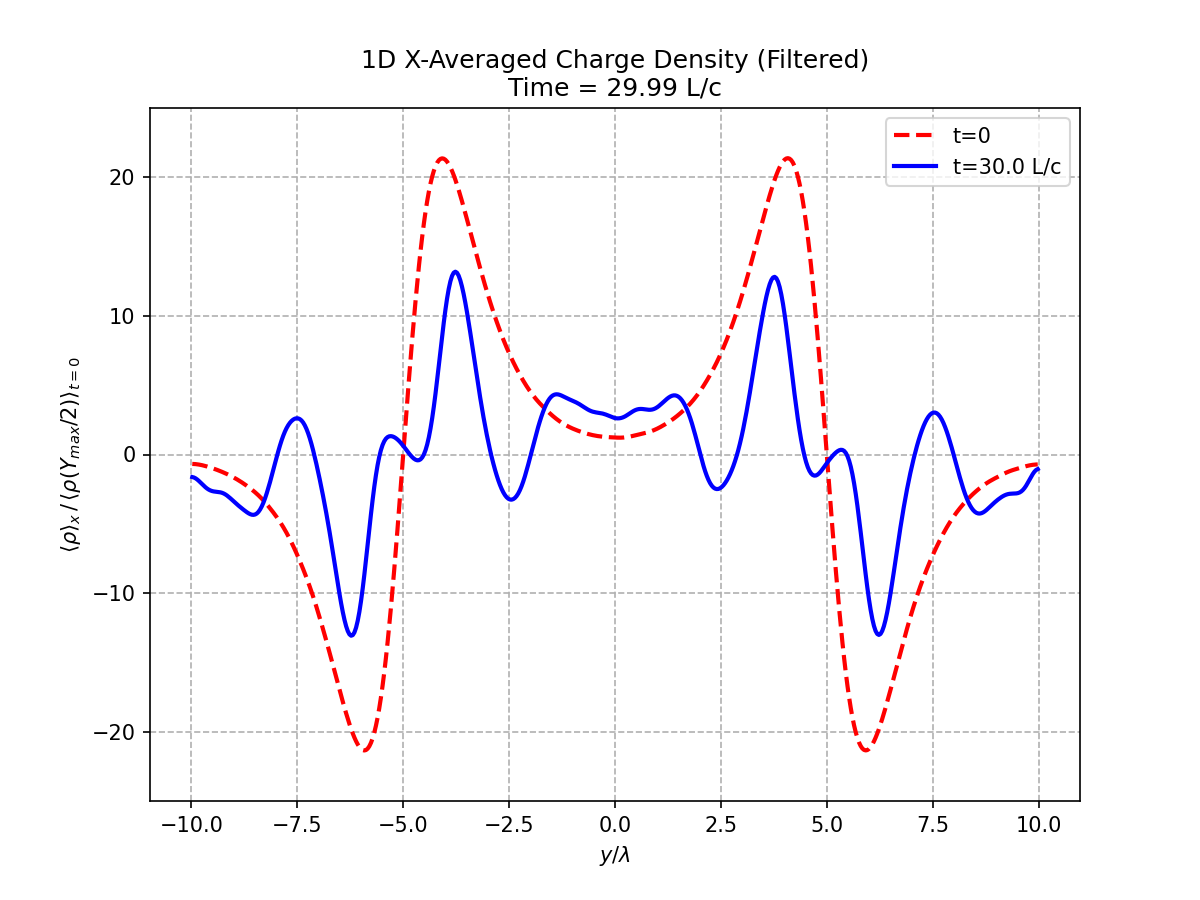}\vline
\includegraphics[width=.3\linewidth]{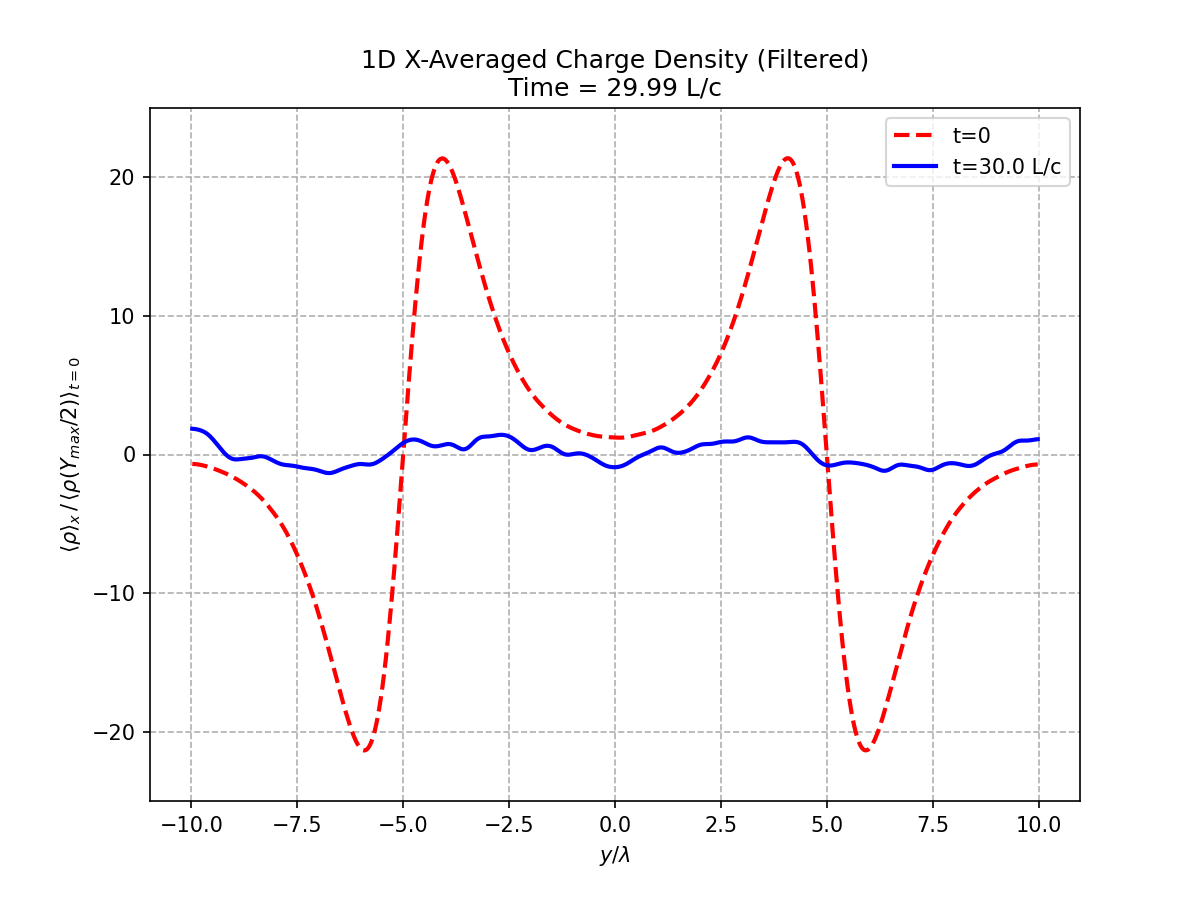}\\
\includegraphics[width=.3\linewidth]{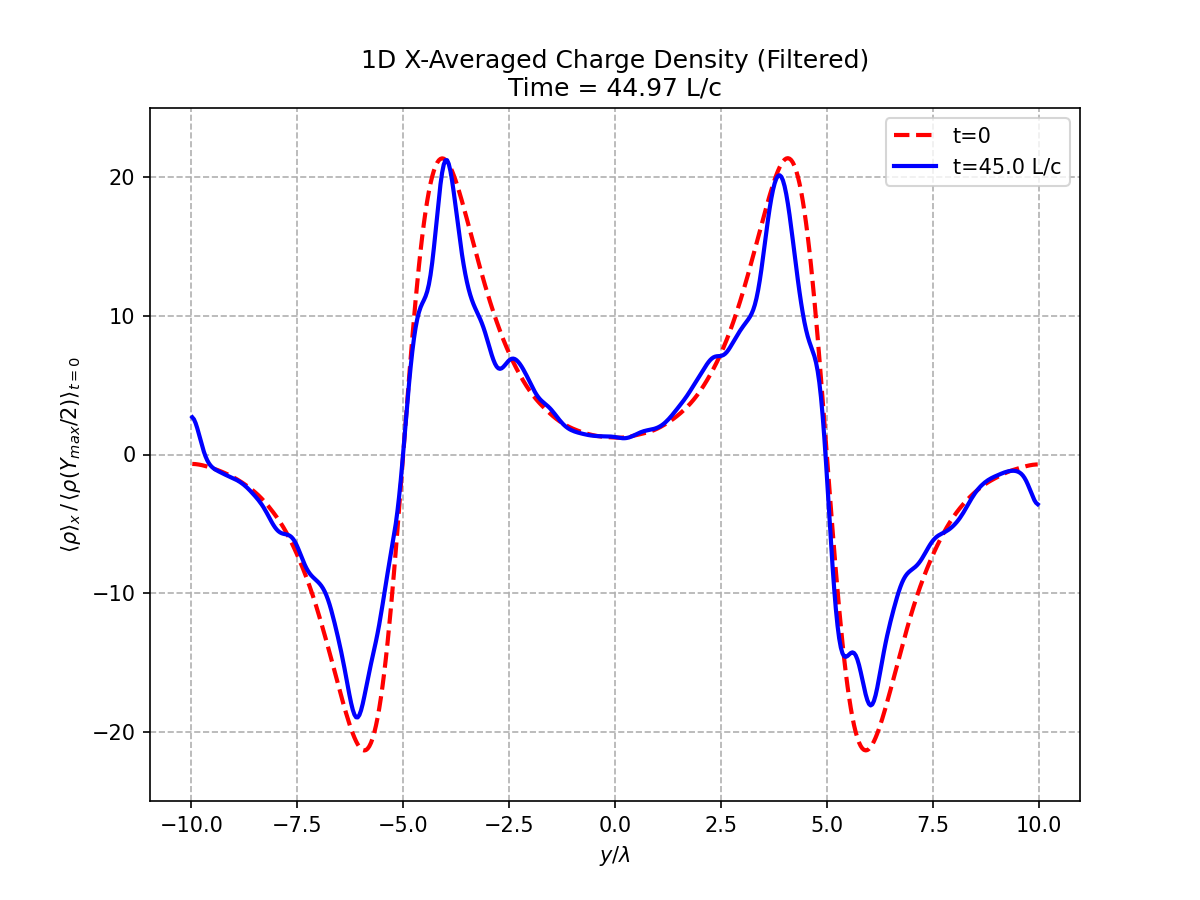}\vline
\includegraphics[width=.3\linewidth]{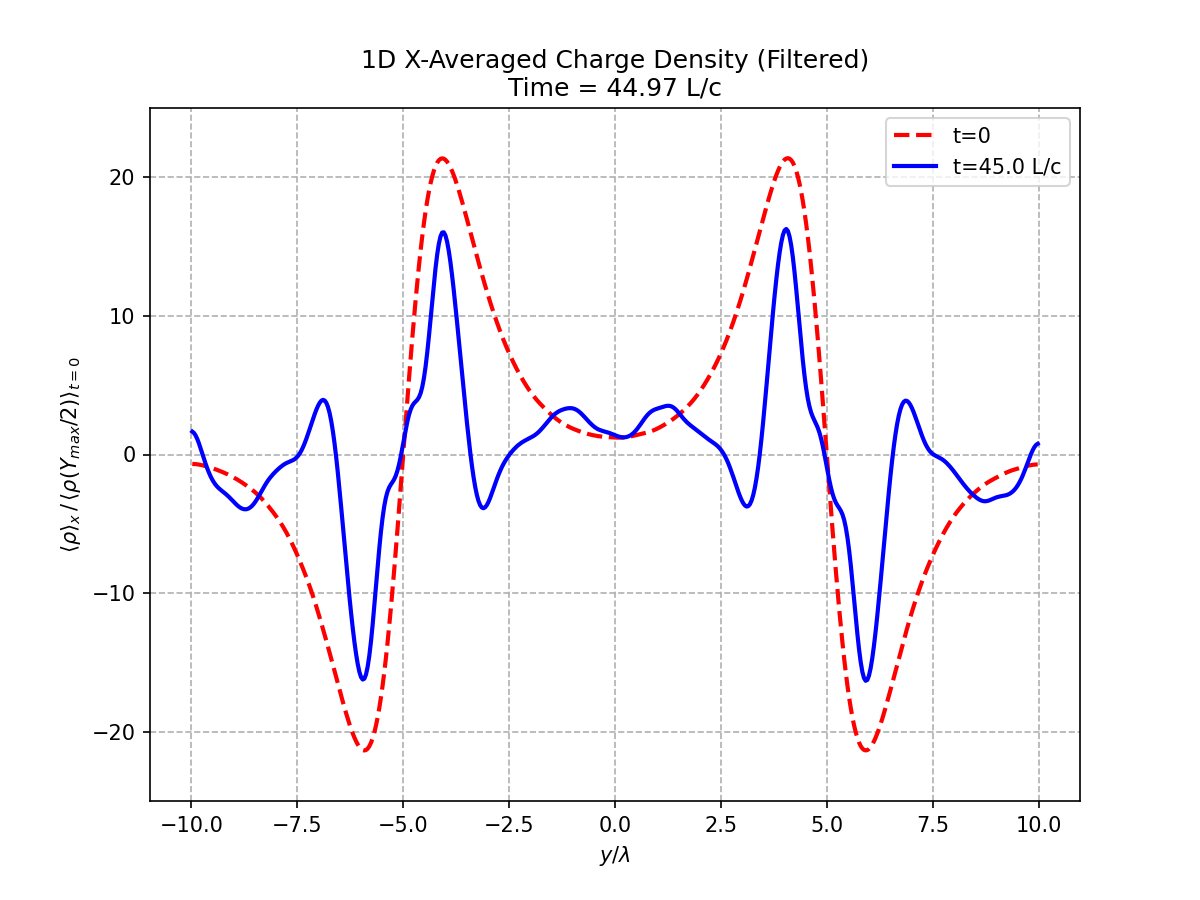}\vline
\includegraphics[width=.3\linewidth]{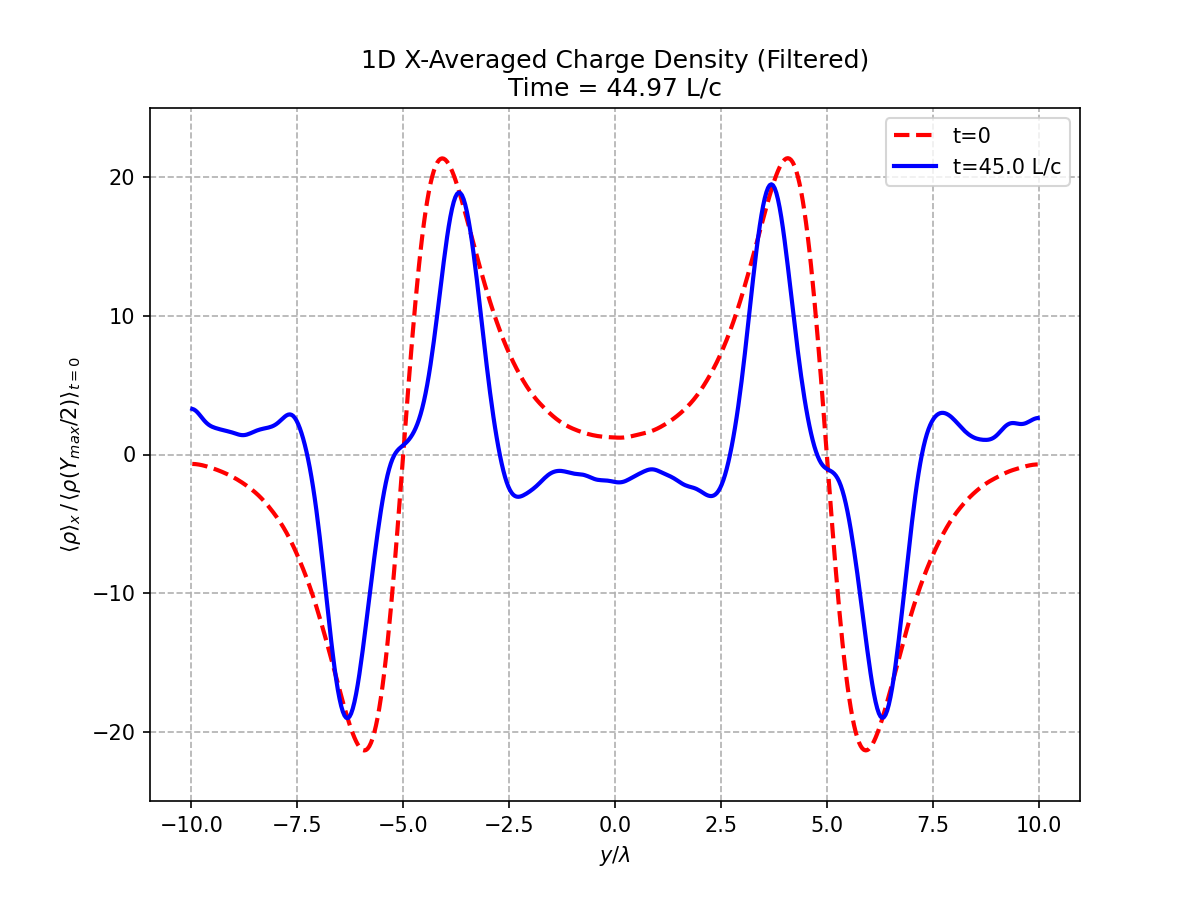}\\
\includegraphics[width=.3\linewidth]{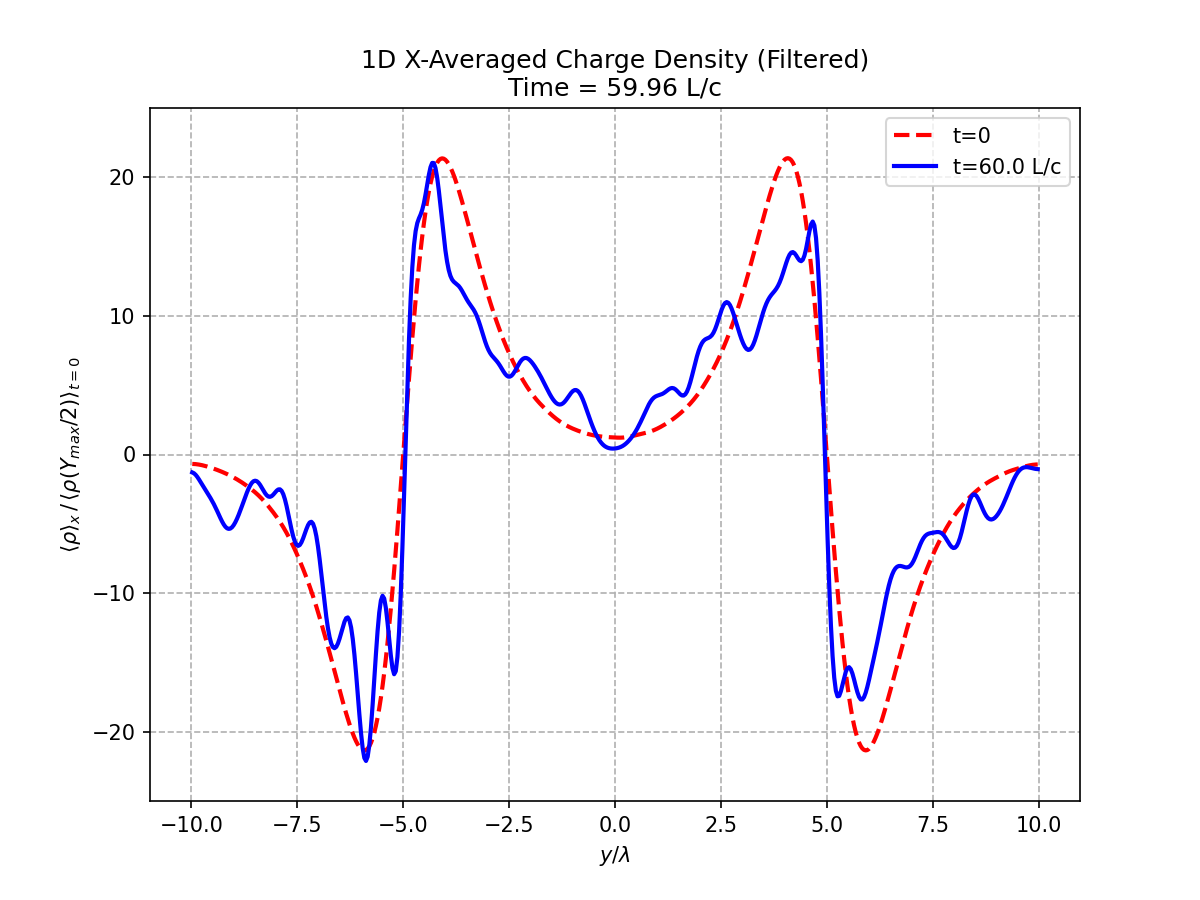}\vline
\includegraphics[width=.3\linewidth]{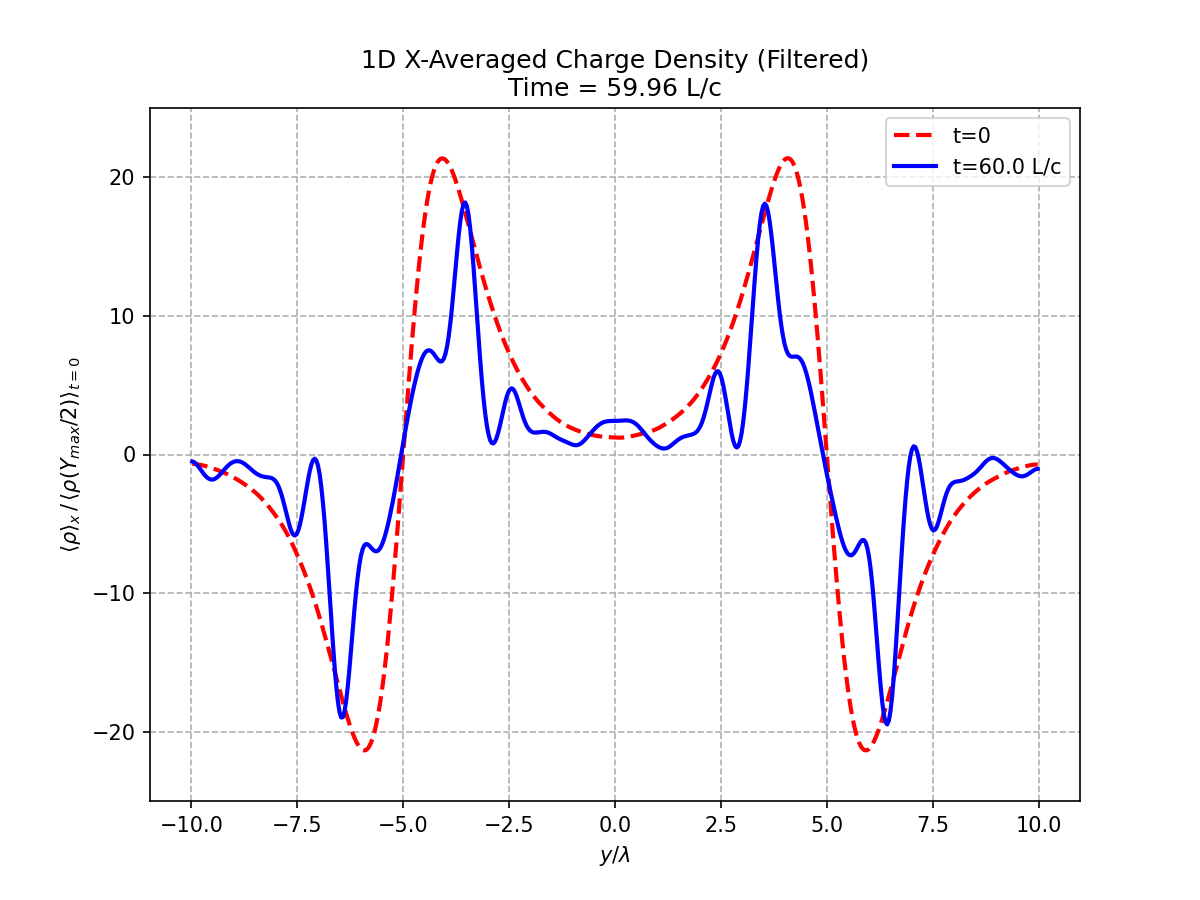}\vline
\includegraphics[width=.3\linewidth]{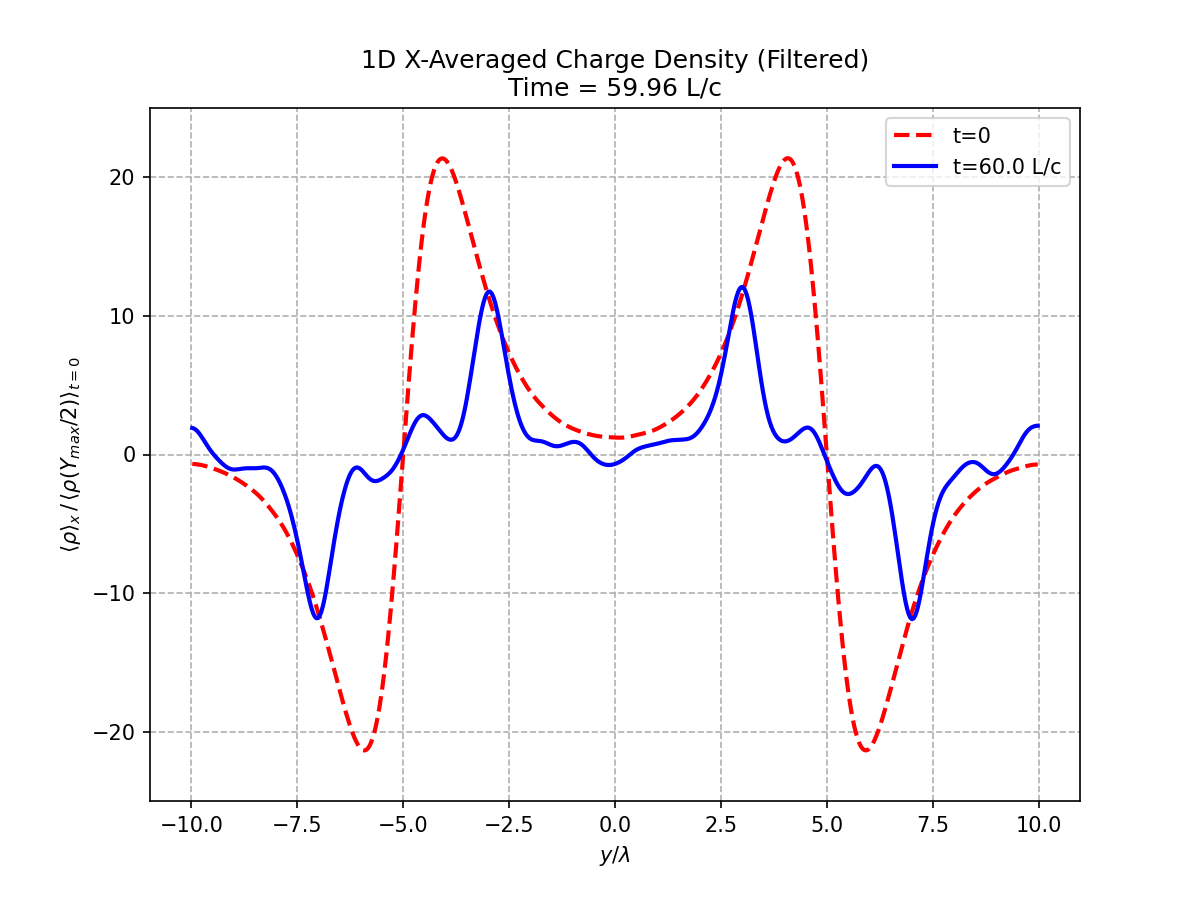}
\caption{BWs in Rotation-CRCS, short runs.  dependance on initial temperature: $\Theta=0$ (left column), $\Theta=0.01$ (middle column),  $\Theta=0.1$ (right column).  Initial state is  shown in dashed red. In warmer plasma BW oscillation occur faster. 
}
\label{short-t}
\end{figure}

In Fig. \ref{short-t} we compare averaged  dynamics of BW depending on temperature. As expected, in hotter  plasma BW oscillate/propagate faster.  BW form trapped oscillating structures; in hotter plasma trapped oscillations occur faster (since phase velocity of BWs $\propto $ sounds speed.)
In cold case (left column in Fig.  \ref{short-t})  even a single oscillation has not been completed, while in the hole case  (right column in Fig.  \ref{short-t}) a full global charge oscillation has completed by the third panel.

\begin{figure}[h!]
\includegraphics[width=.34\linewidth]{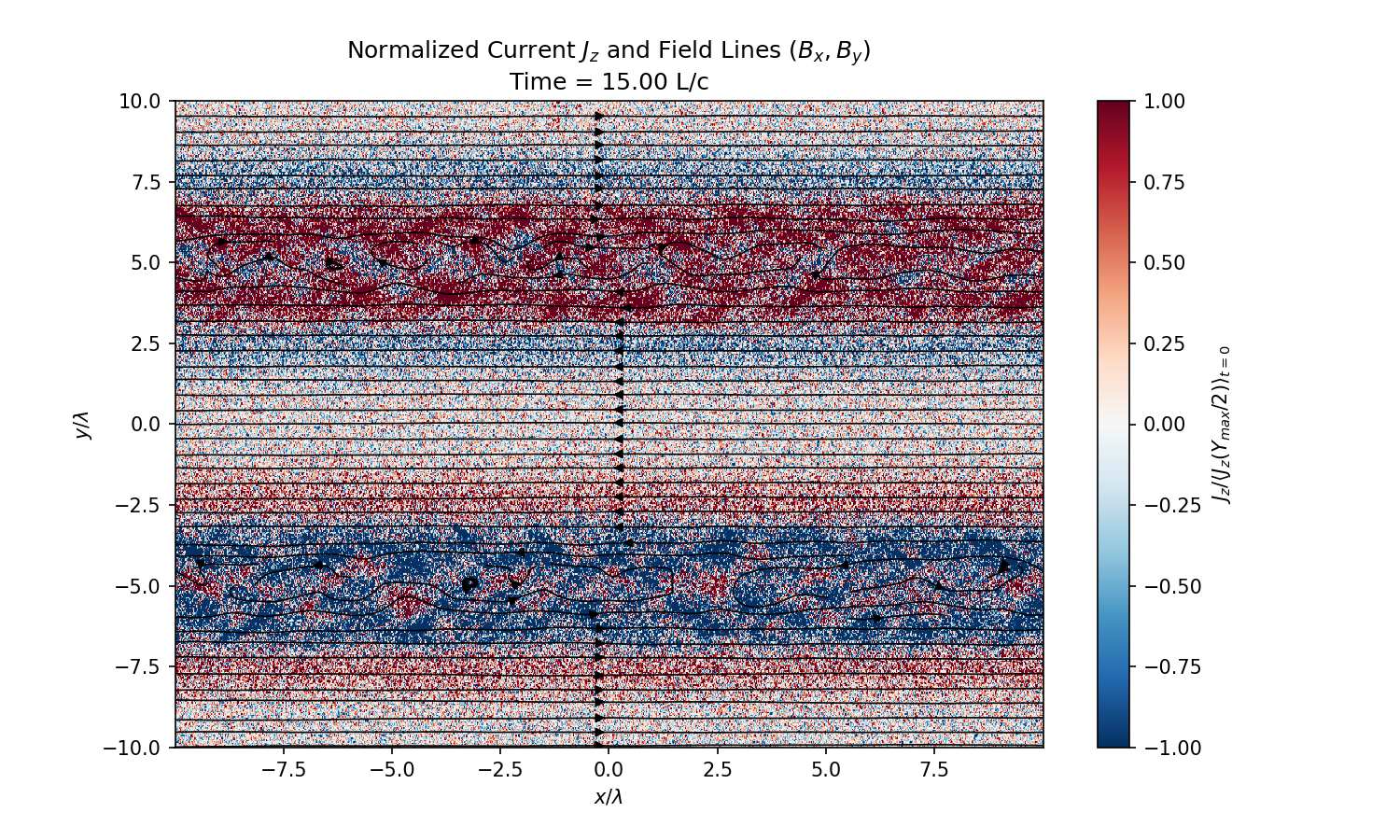}\vline
\includegraphics[width=.34\linewidth]{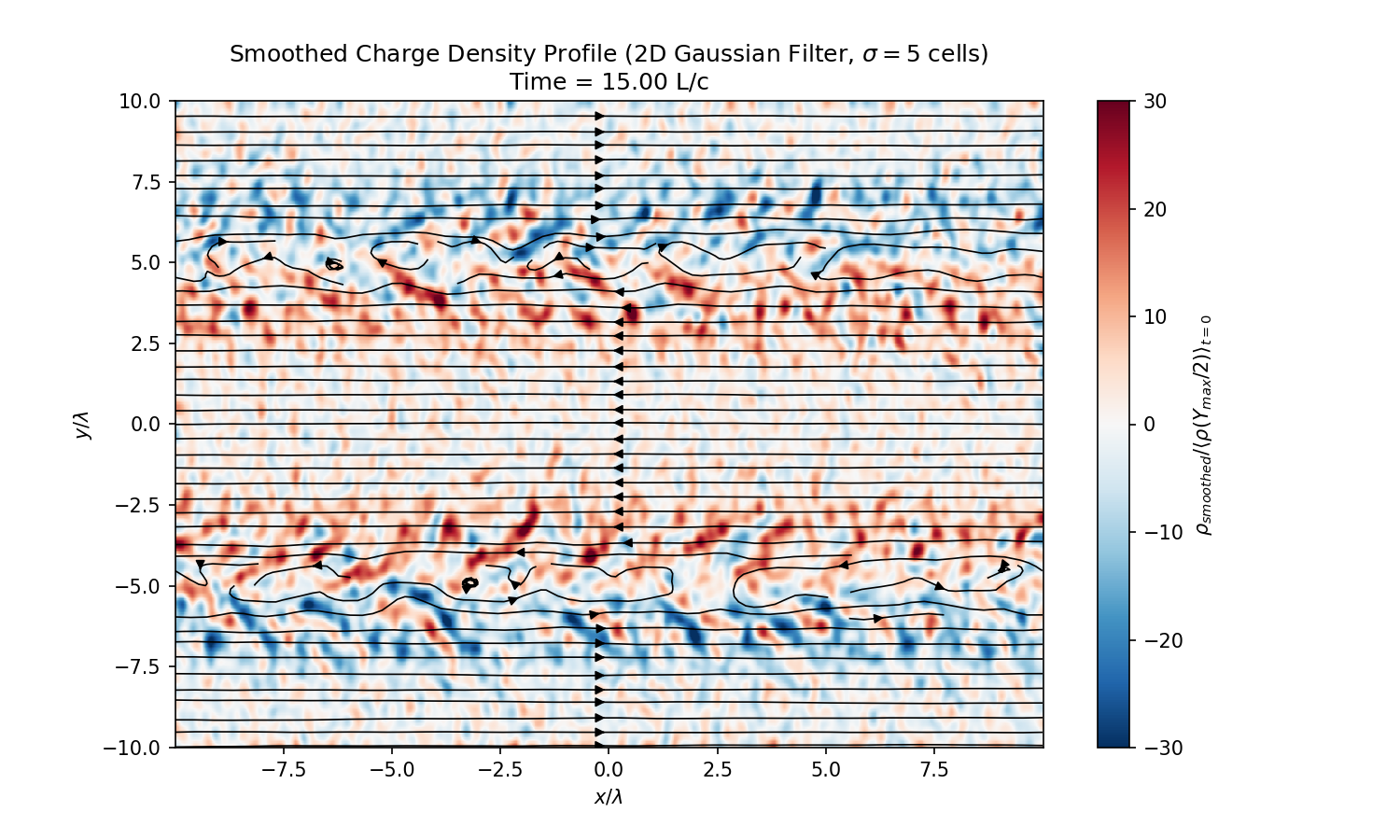}\vline
\includegraphics[width=.3\linewidth]{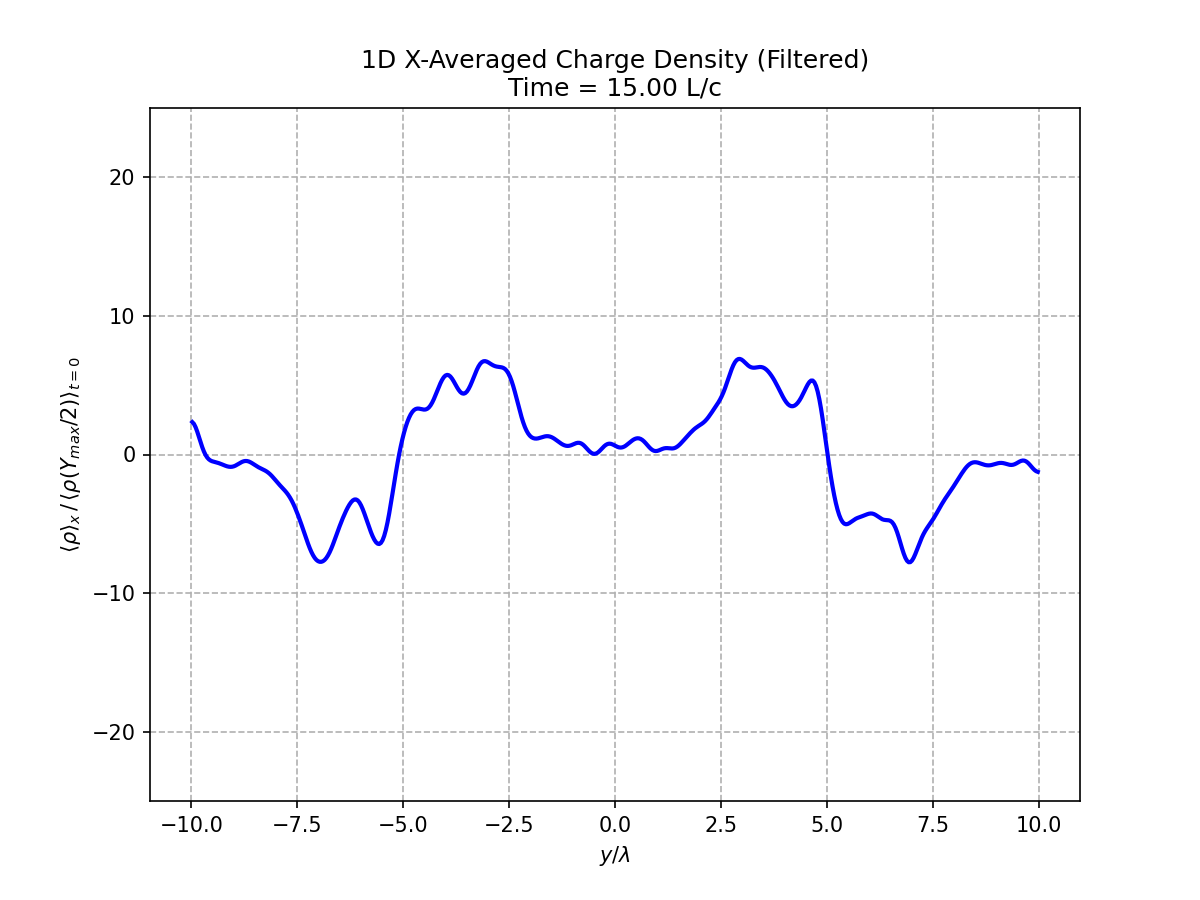}\\
\includegraphics[width=.34\linewidth]{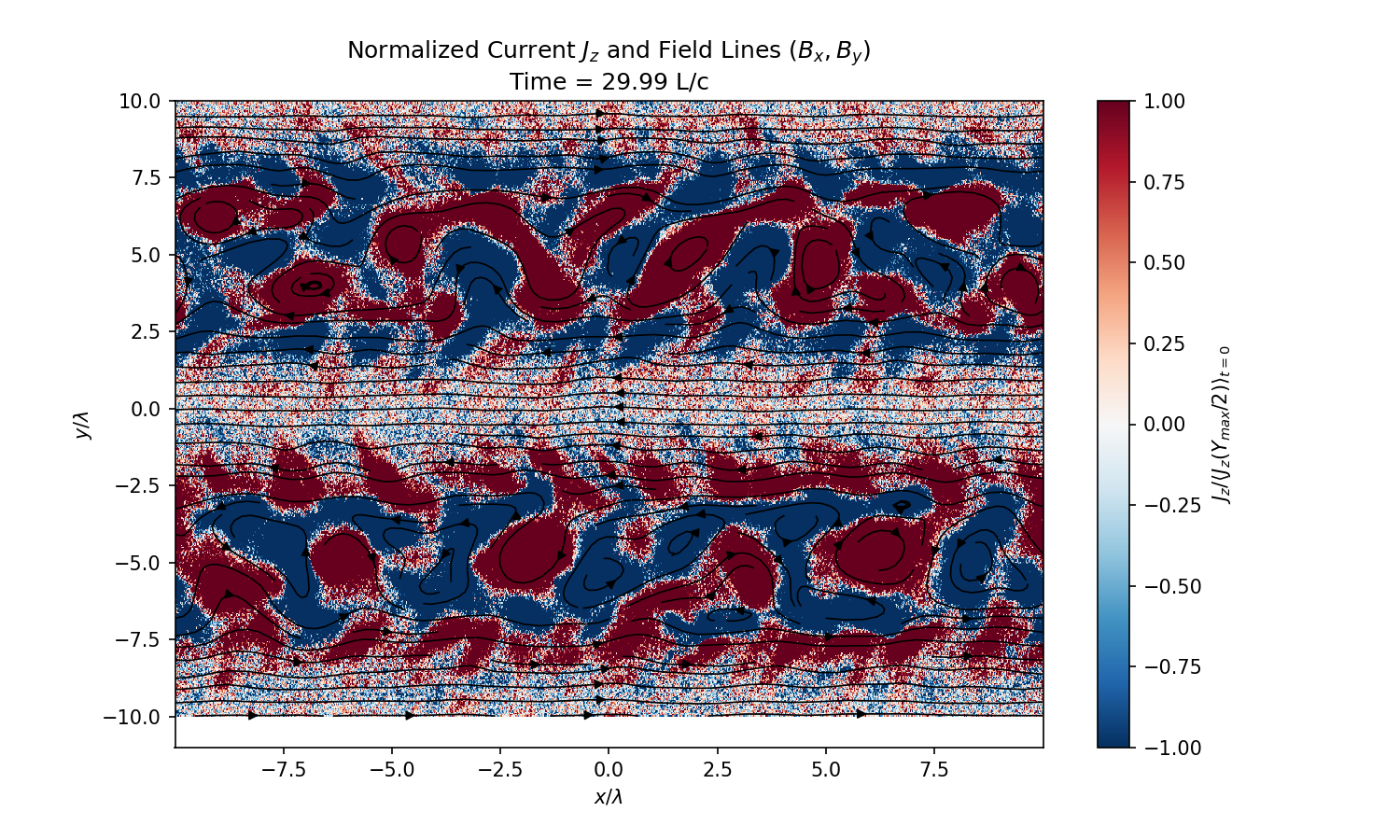}\vline
\includegraphics[width=.34\linewidth]{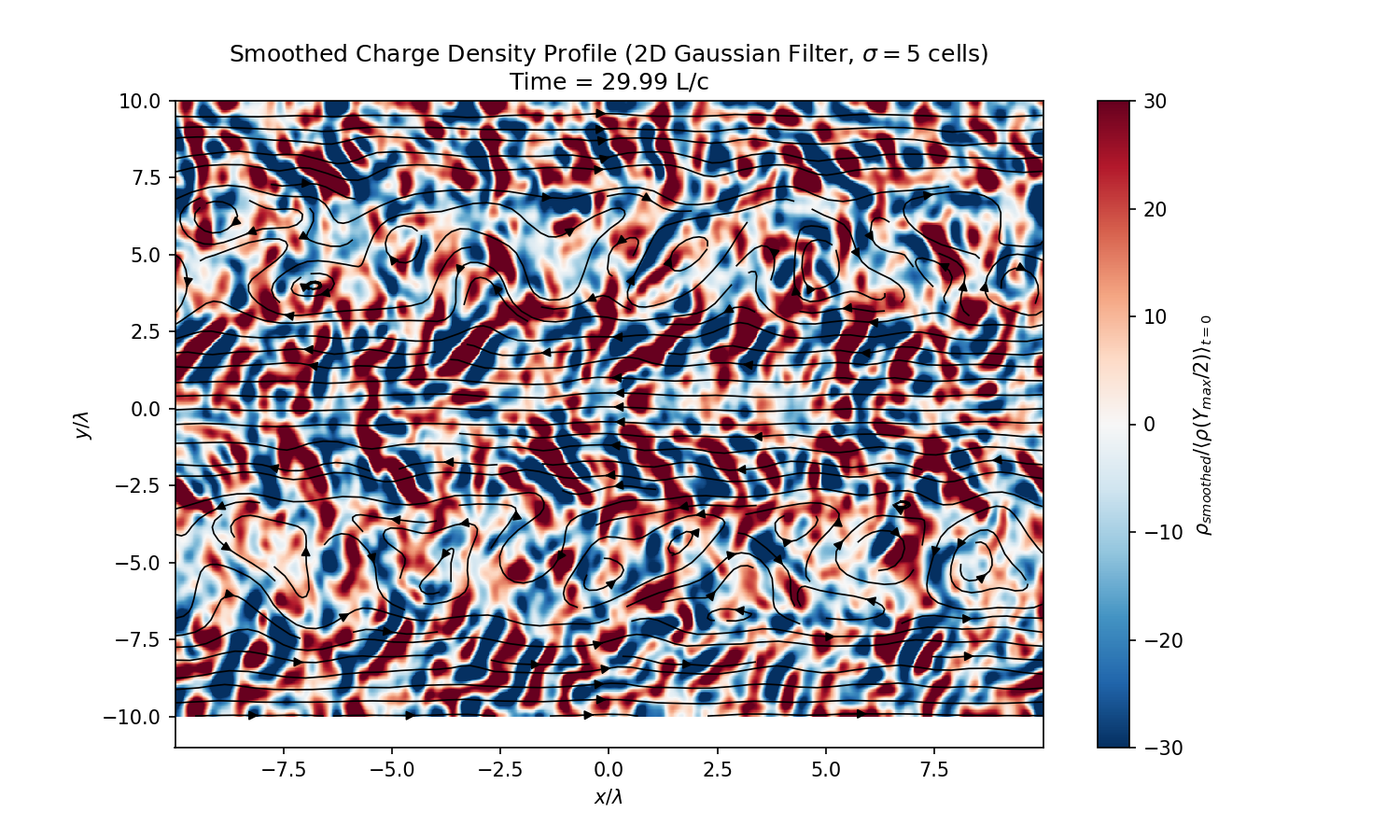}\vline
\includegraphics[width=.3\linewidth]{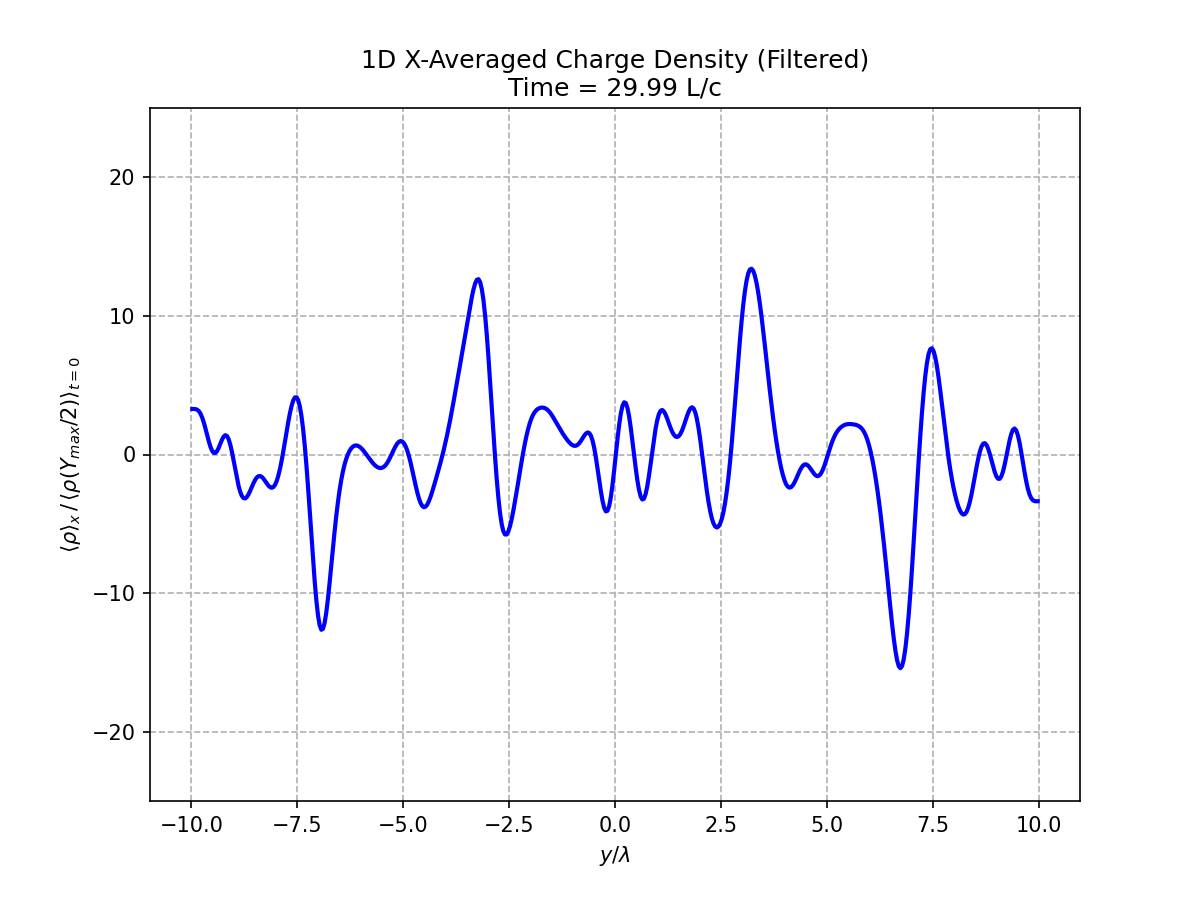}\\
\includegraphics[width=.34\linewidth]{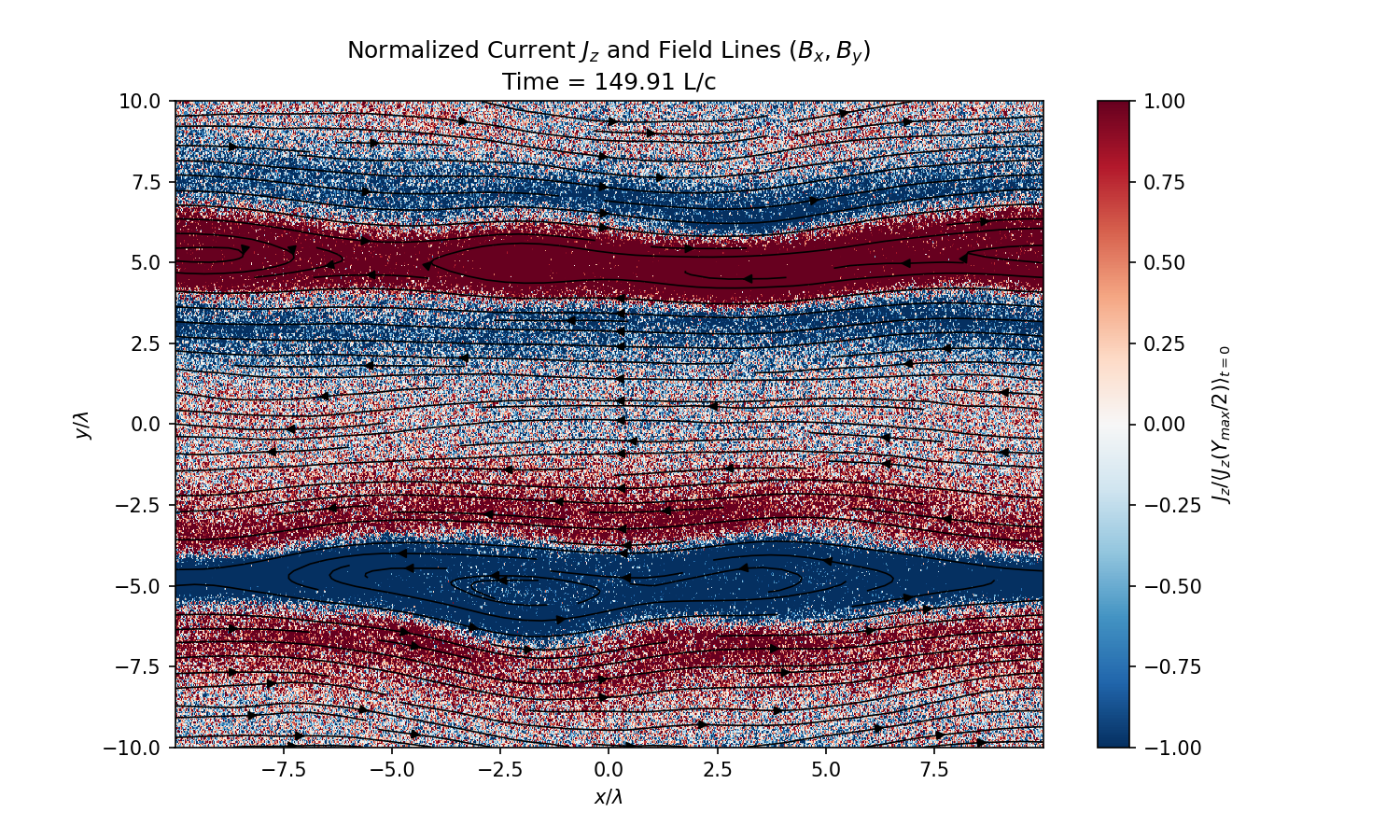}\vline
\includegraphics[width=.34\linewidth]{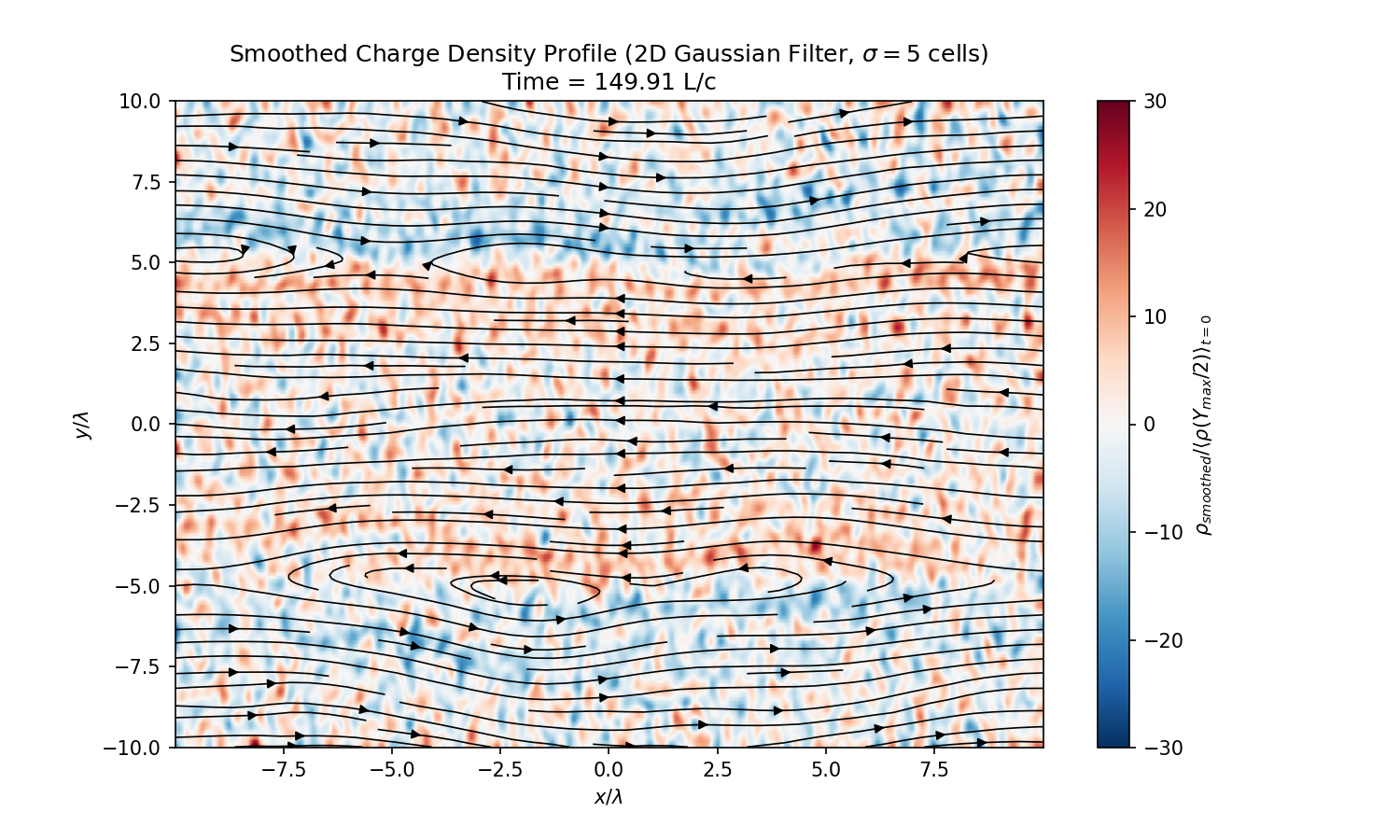}\vline
\includegraphics[width=.3\linewidth]{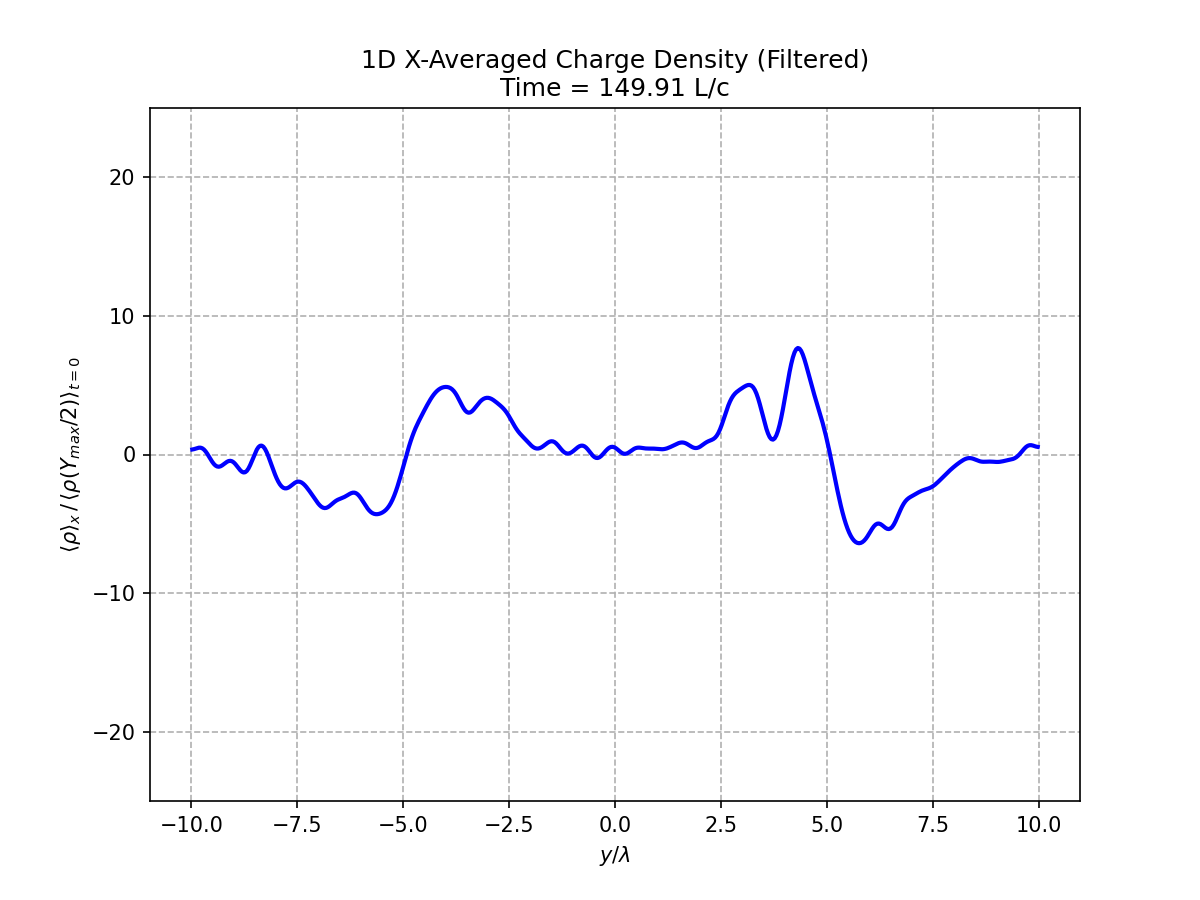}\\
\includegraphics[width=.34\linewidth]{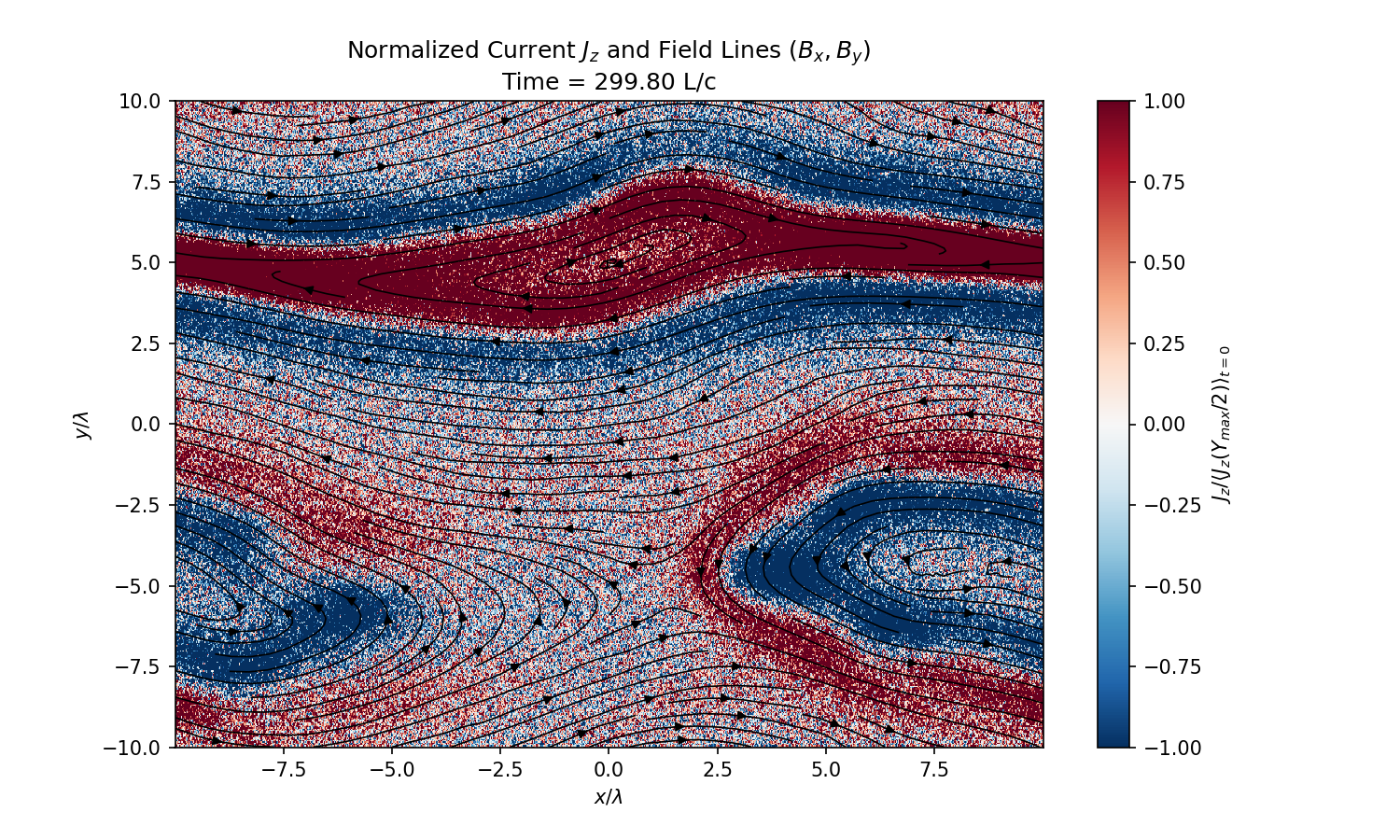}\vline
\includegraphics[width=.34\linewidth]{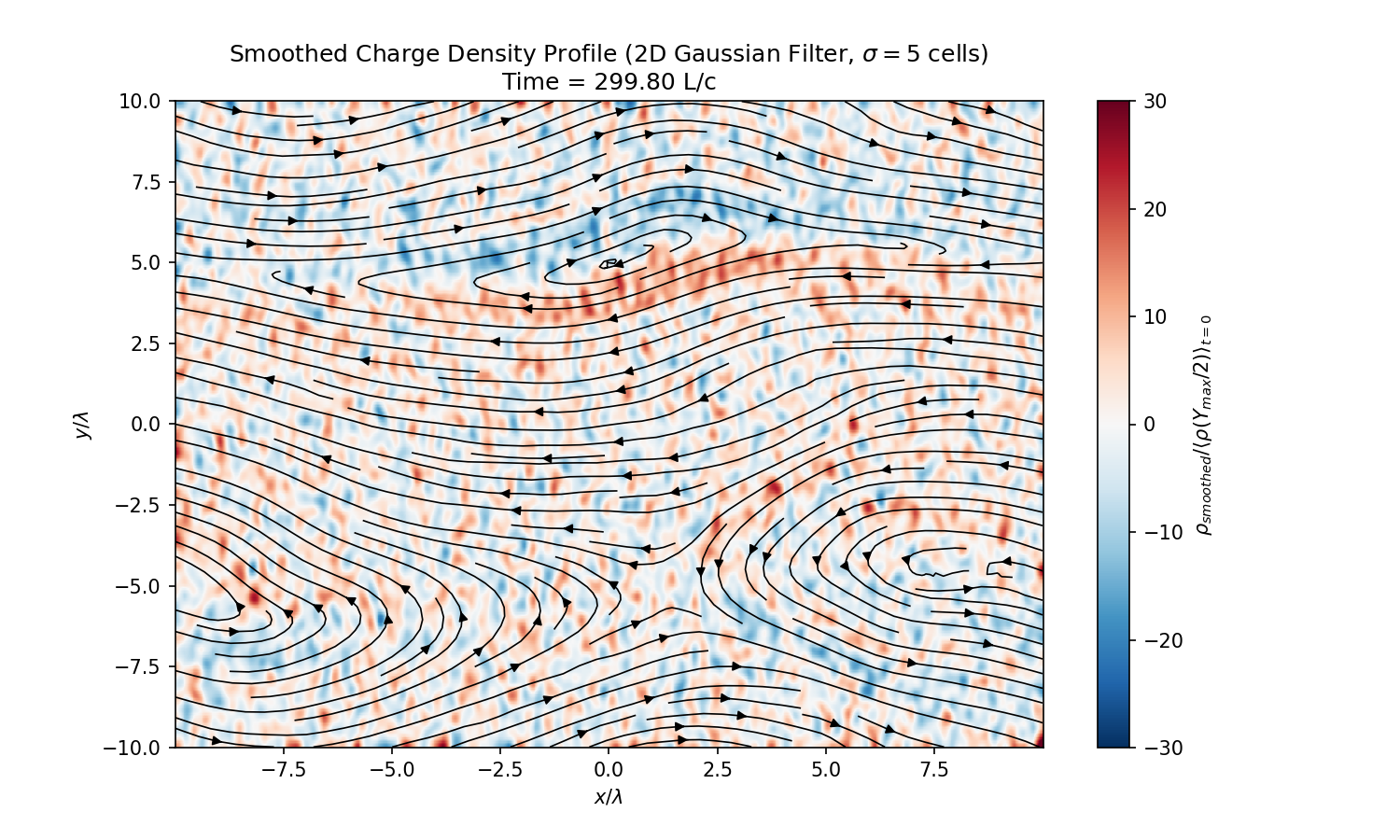}\vline
\includegraphics[width=.3\linewidth]{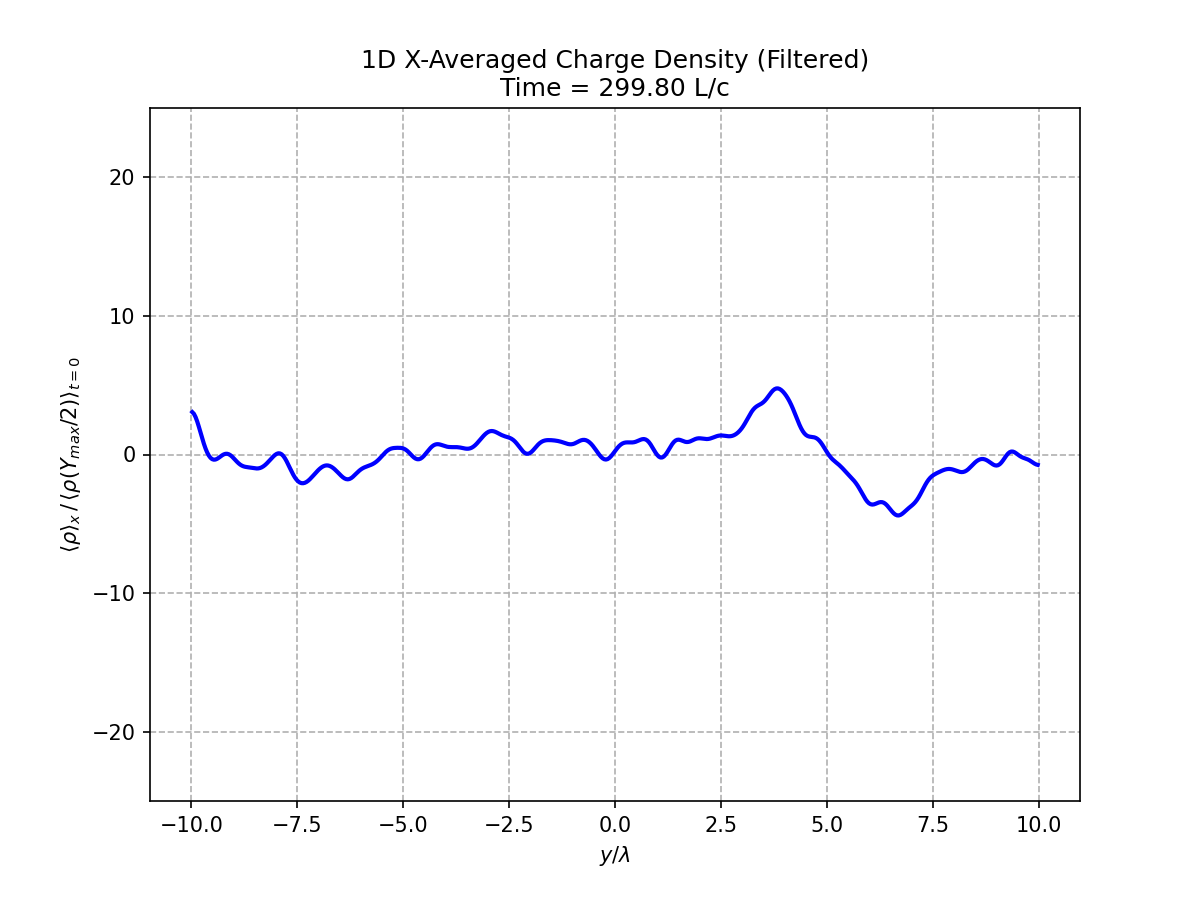}\\
\caption{BWs in Rotation-CRCS,  High-B-Cold run}
\label{Rotationn26}
\end{figure}

For higher \Bfs, $b_0=1000$  (Fig. \ref{Rotationn26}) evolution of BWs and tearing proceeds faster, naturally. 

\begin{figure}[h!]
\includegraphics[width=.24\linewidth]{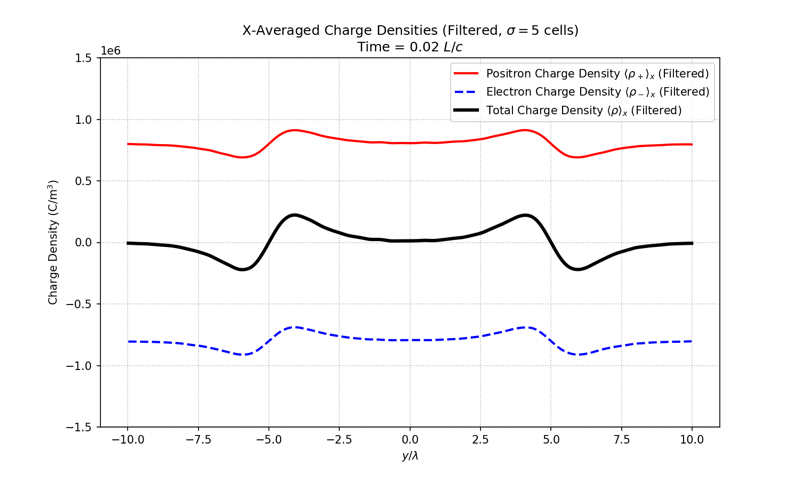}
\includegraphics[width=.24\linewidth]{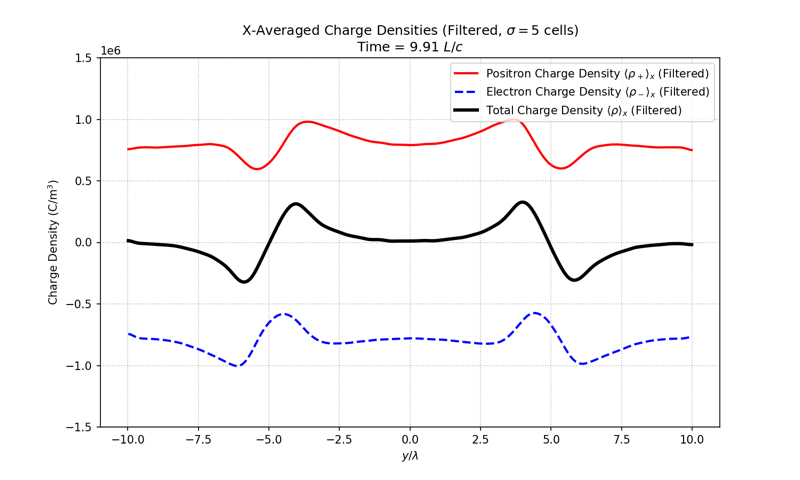}
\includegraphics[width=.24\linewidth]{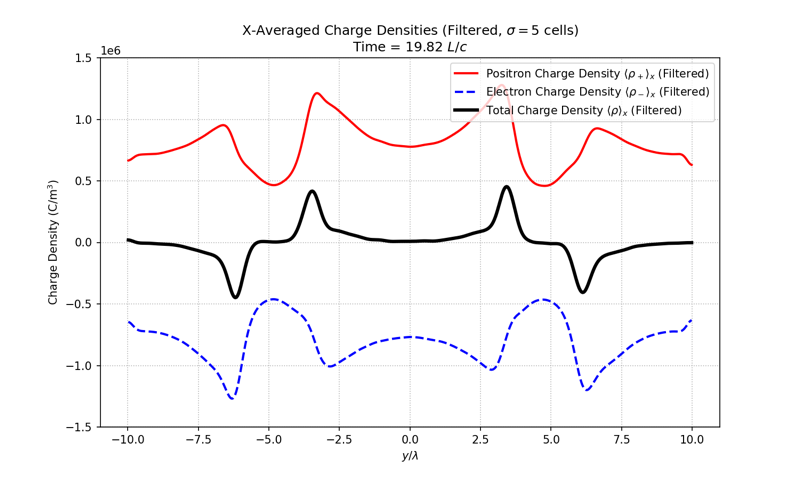}
\includegraphics[width=.24\linewidth]{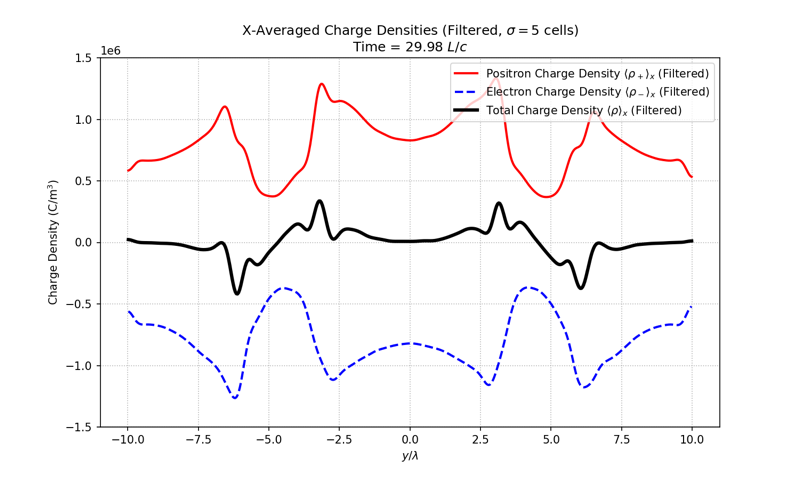}
\\
\includegraphics[width=.24\linewidth]{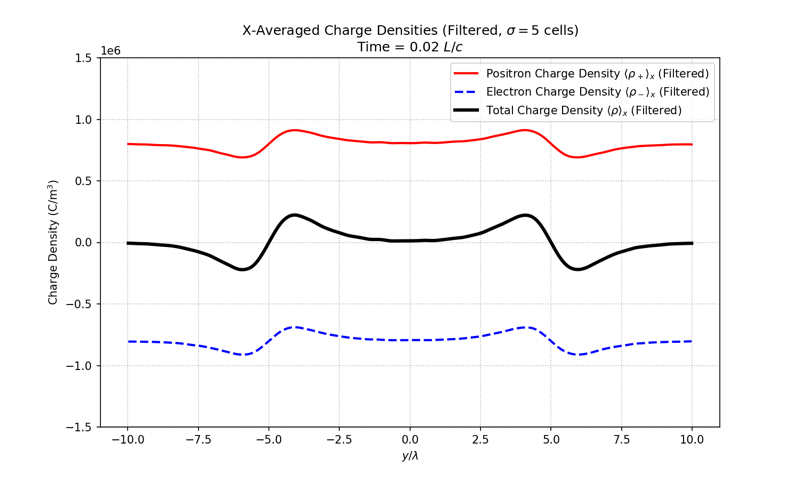}
\includegraphics[width=.24\linewidth]{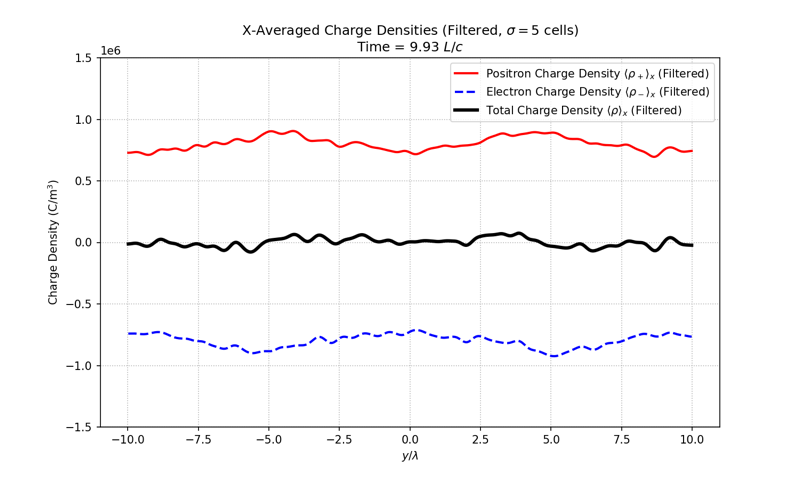}
\includegraphics[width=.24\linewidth]{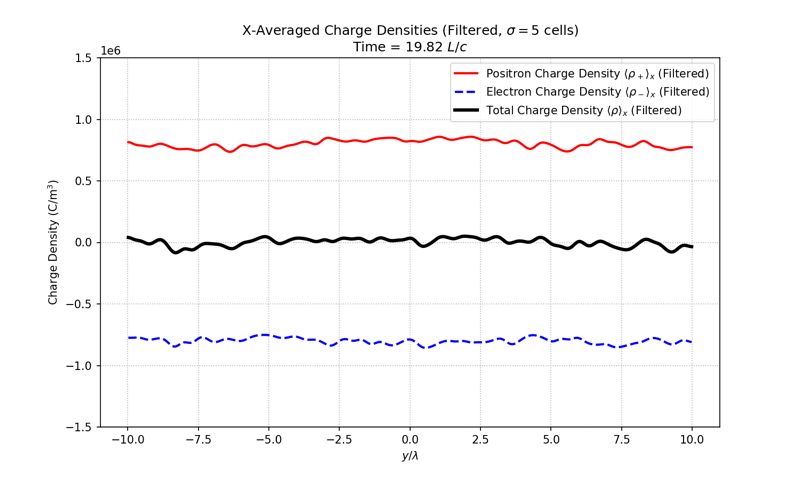}
\includegraphics[width=.24\linewidth]{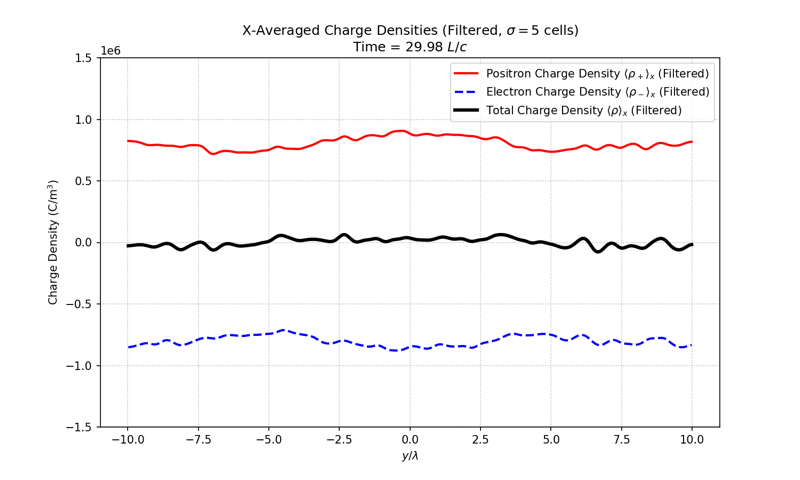}
\caption{Rotaional-CRCS. Evolution of charge densities, rotational current layer (constant temperature). Top: initially cold plasma. Relaxation is very slow.
Bottom row: hot plasm $\Theta=0.1$. In hotter plasma the relaxation is expected to occur faster since  phase speed of Bernstein modes 
is  of the order of the thermal velocity of the  species. We have also verified that  intermediate temperature produce intermediate time-scale relaxation. 
Quasi-1D in $y$ simulations.
}
\label{rho_avg-rot}
\end{figure}

 We  also performed a set of  quasi-1D simulations (minimal extension in $x$).  An important difference from the  Harris sheet is that in the rotation current sheet, temperature is a separate independent quantity. We can then test the rate of evolution of BWs as a function of temperature.   We indeed observe that hotter plasma relaxes faster. Fig. \ref{rho_avg-rot}.

\begin{figure}[h!]
\includegraphics[width=.33\linewidth]{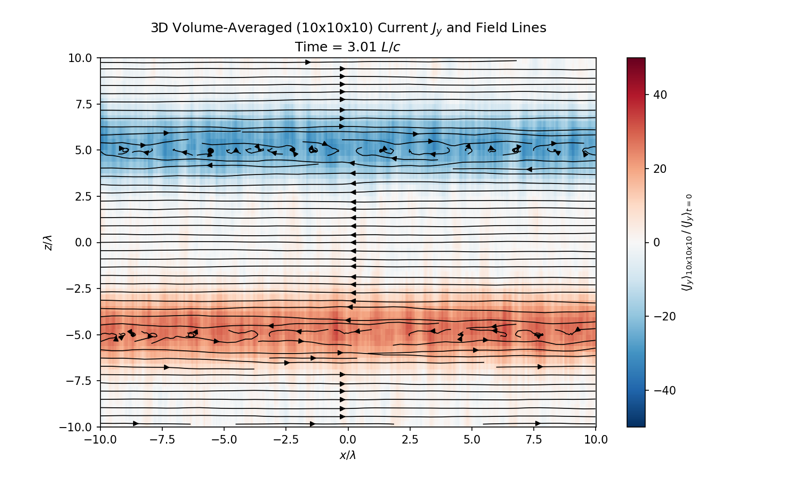}\vline
\includegraphics[width=.33\linewidth]{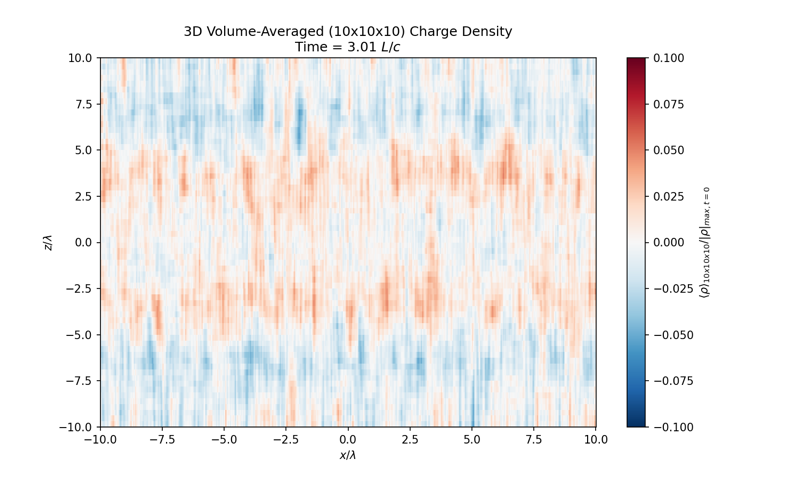}\vline
\includegraphics[width=.3\linewidth]{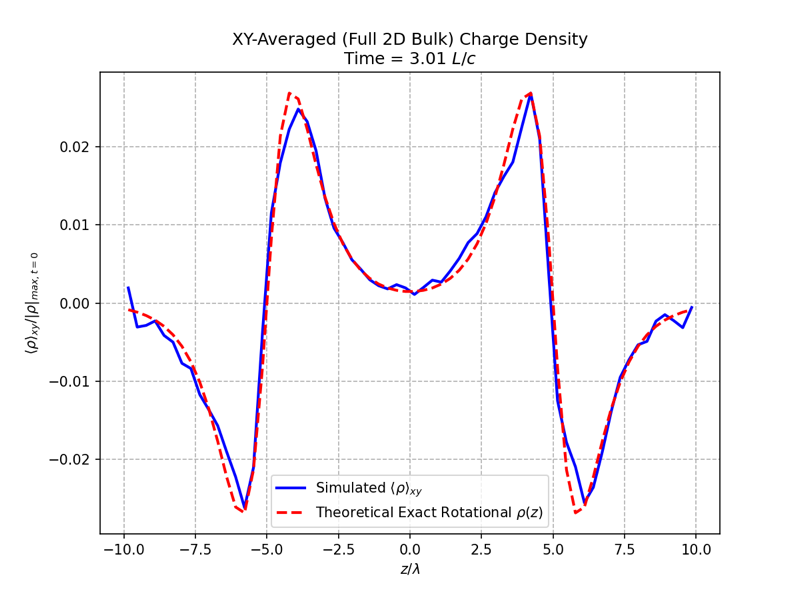}\\
\includegraphics[width=.33\linewidth]{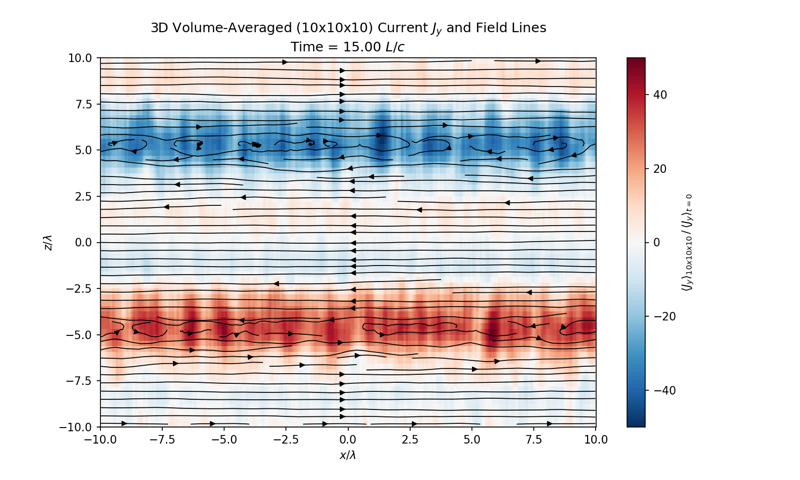}\vline
\includegraphics[width=.33\linewidth]{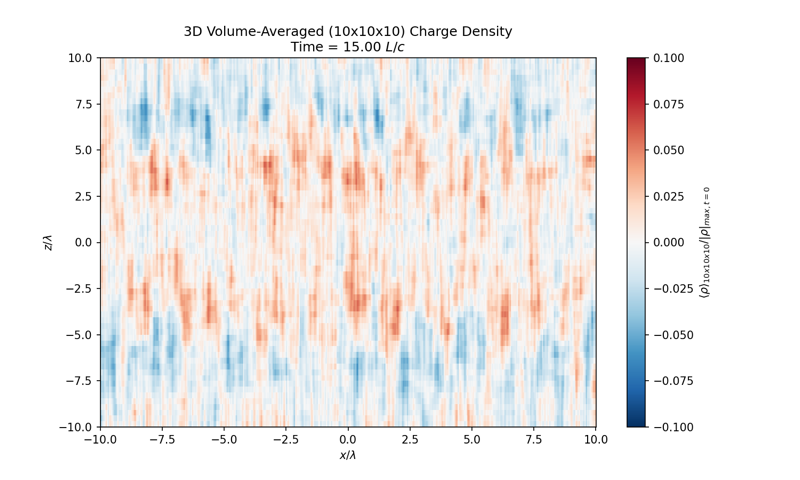}\vline
\includegraphics[width=.3\linewidth]{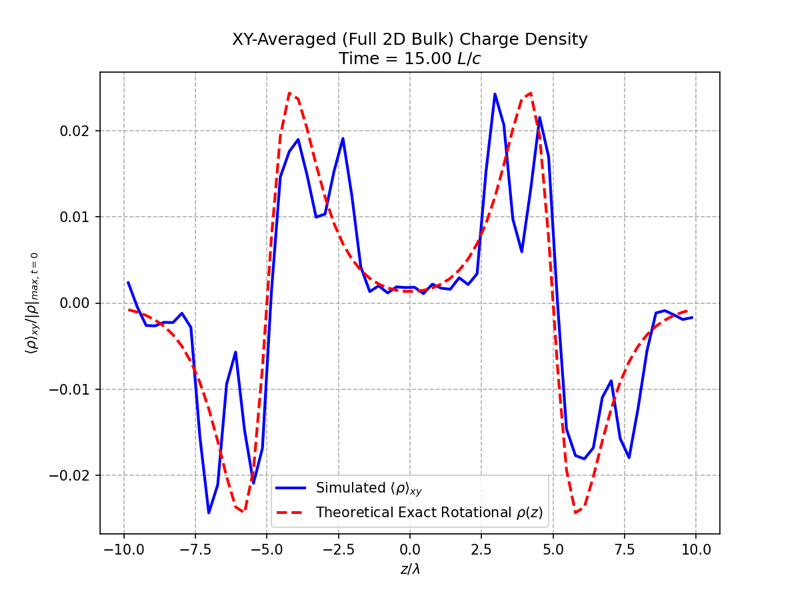}\\
\includegraphics[width=.33\linewidth]{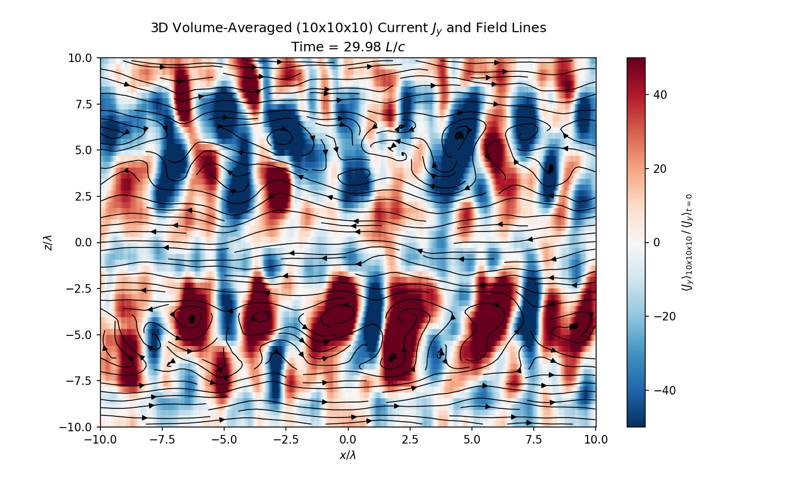}\vline
\includegraphics[width=.33\linewidth]{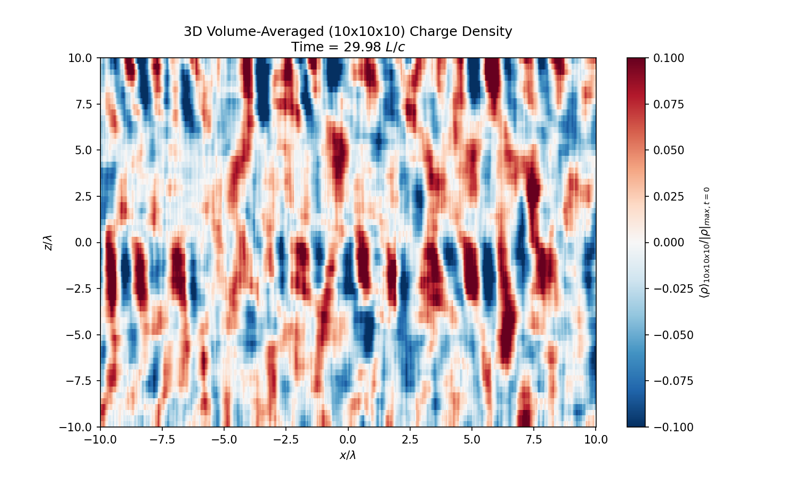}\vline
\includegraphics[width=.3\linewidth]{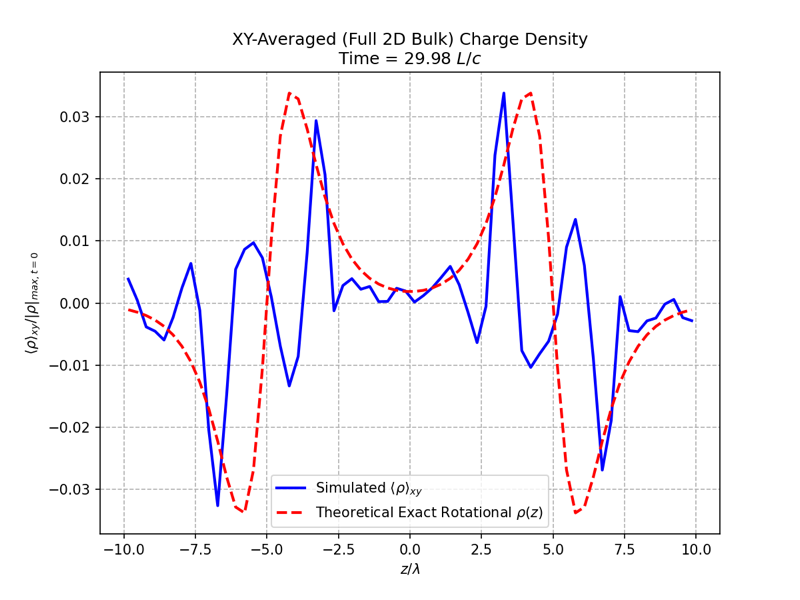}
\caption{3D simulations of charged rotational current sheet. Large charge density fluctuations appear near $L/c \sim 20-30$, consistent with 2D simulations. (Parameters Warm-1, Table \ref{table},  resolution $512\times 512\times64$, 10 particles per cell.
}
\label{3D}
\end{figure}

Finally, as the results of 2D simulations  of rotational current layers are somewhat surprising, we repeat the setup in 3D runs, Fig. \ref{3D}.  There is a general consistency of the results between 2D and 3D.

\section{Discussion}
  \label{exaplain}
  
  Summarizing the results of simulations, we observe
  \begin{itemize}
  \item  In the Harris-CHSC, fast {\it electrostatic} relaxation occurs, which overall  does affect the slower tearing rate  
  \item In rotation-CRCS, 
  	\begin{itemize}
	 \item  large charge fluctuations develop even in the initially charge-neutral set-up
	 \item overall tearing rate is greatly enhanced in charged configurations, of compared with uncharged ones
  	  \end {itemize}  
  \end {itemize}
  Below we discuss our results separating Harris-CHSC, \S \ref{CHSC}, and  rotation-CRCS, \S \ref{CRCS}
  
  \subsection{Charge Harris current sheet}
  \label{CHSC}
  
Charged waves  in   Harris-CHSC  set-up are simpler to exaplain. Initially, opposite signs of charge are in electric potential minimum or maximum. The system is in force balance, but an electrostatically unstable one. For example, in the middle of the layer the \Bf\ is vanishing, so charges can easily move across the layer. The overall situation is  of the kind "Bernstein waves in inhomogeneous \Bf.
Thus,    initial configuration consisted of a certain high background level of charge, plus a small charge wave that decayed into two traveling waves BWs.

 Since plasma has everywhere non-zero temperature BW are natural eigenmodes.  BWs are {\it not} the flows of charge: charges oscillate  (in the linear regime), but are not producing   bulk moving. It is the separation of the initial coherent configuration into many modes that gives the impression of charges moving across the \Bf\  field. In a BW charges gyrate  in phase with the electrostatic wave,  creating  periodic charge accumulation with a wave
length four times the gyro-radius. Even though the electric field is perpendicular to the magnetic
field the electrons will not express any E-cross-B drift since the wave frequency is larger than the
cyclotron frequency, and thus the average electric field will cancel during the gyro-motion. (BWs are the short-wavelength modes associated with the X mode wave branch.)

 Typical   phase speed  of  BWs  is  on the order of the thermal velocity - hence in hotter plasmas charge relaxation occurs faster. Also, even in the constant \Bf\ case BWs  exhibit strong dispersion, especially near the harmonics of the cyclotron frequency. More so in the inhomogeneous case. This explains quick phase mixing, which shows up as a reduced charge density. 
  
  BWs are reflected at the upper hybrid frequency (which decreases  away from the center of the current layer)  -  this exaplain their trapping in the Harris-CHSC.  This reflection sets up an internal standing  wave, that is slowly dissipating $\equiv$ smoothing out the charged density.  
  
  Mild plasma heating  by BWs  is mainly  due to cyclotron damping.
  
The upper-hybrid frequency acts as a resonance layer where electrostatic waves can  linearly  couple to  electromagnetic waves. Typically, this is used to heat the plasma (but converting in-falling \EM\ waves into BW, that then dissipate inside plasma \cite{2007PPCF...49....1L},  \citep[ see also capacitive coupled plasma (CCP) discharges][]{2006ApPhL..89z1502G}. In our case, a reverse process is expected: coherently generated BWs will convert to \EM\ waves at the upper  hybrid resonance.

Typically, {\it  linear}  BWs  do not  create a "permanent" increase in electric charge density on their own in the sense of a non-neutral, static charging. We associated the small remaining charge with the BW-tearing interplay: it is the tearing that ``captures'' the tails of the BW charge oscillation, leaving the final plasmoids mildly charged. 

BWs can trap and accelerate plasma particles. Relevant simulation would require much higher resolution/resources. We hope to perform  the relevant simulation in the future.

 \subsection{Charge rotational  current sheet}    
\label{CRCS}

For rotation-CRCS,  our two important observation are: (i)  even in the initially charge-neutral case, temporarily, large charge fluctuations develop; (ii) in charged  rotation-CRCS tearing develops much faster than in the uncharged case

  Initial set-up of   rotation-CRCS  allows zero temperature, hence, formally no BWs. But even in the case of zero initial temperature, numerical heating introduced some finite (small) temperature that allows BWs to propagate.

  \subsection{Astrophysical application: Michel's solution of pulsar winds}    
\label{Michelnst}

  One of the implications is that  the Michel's  solution \citep{1973ApJ...180L.133M} for pulsar winds's equatorial current sheet  is electrostatically unstable.
  Michel's  solution, an elegant mathematical construction, turns out to be the  asymptotical limit for radii much larger than the light cylinder radius 
  \citep{1999ApJ...511..351C,2023ApJ...943..105H}. Our Harris-sheet simulations are applicable to the pulsar equatorial current sheet (except they are done in a special frame moving nearly with the \EM\ velocity of the bulk flow). Our result, that charged Harris configurations are electrostatically unstable, imply that a proper charged Michel current sheet never gets assembles in pulsar winds. 
  
  On the other hand, requirement of charge equatorial curent sheet, and electric  potential extremum there comes from the global structure of the \ms. We envision then that as the equatorial current gets assembled, efficient {\it electrostatic}  dissipation  of BWs occurs  at upper hybrid resonance. (Global  PIC simulations may under-resolve the    structure of the current sheet.)

\bibliographystyle{apsrev}
 \bibliography{/Users/lyutikov/Library/CloudStorage/Dropbox/Research/BibTex,/Users/lyutikov/Library/CloudStorage/Dropbox/Research/BibTexShort.bib,//Users/lyutikov/Library/CloudStorage/Dropbox/Research/NASA_FRB.bib,/Users/lyutikov/Library/CloudStorage/Dropbox/Research/references}

\appendix

   \section{Simulation set-up} 
   \label{EPOCH}
We use user-modified PIC code EPOCH \citep{Arber:2015hc}.  Importantly, we  treat the  initial velocities and charge distribution as current in the first step. 
As a result,  in the initial equilibrium  $\curl \B = {\bf j} $. This is not a standard EPOCH set-up.

Direction perpendicular to the current layer is called $y$ in 2D and $z$ in 3D; this is the standard set-up of the EPOCH code.

 For numerical implementation, 
        there are two important physical  limitations. First, for a given \Bf\ $B_0$ and current layer thickness $L$, there is minimal density to carry the required current
        \be
        n^\ast    \approx  \frac{B_0}{2  L}  \times \frac{1}{4\pi}
        \ee
        (factor of two comes from two components equally contributing to the current).  Factor $1/(4\pi)$ is separated  explicitly to connect to the cgs unites. The corresponding requirement can be expressed in terms of 
        \be
        \beta^\ast =  \frac{B_0}{2  L n^\ast },
        \ee
         diminutional  velocity. Parameter $   \beta^\ast$ must be smaller than unity:  for a given \Bf\ and scale $L$ this set minimal density $n^\ast  $.
         
For $   \beta^\ast  \ll 1$ the development of the tearing mode becomes very slow. As a results, realistic simulations are limited to $    \beta^\ast  \leq 1$. This limitation precludes detailed scan of parameters.
   
   Importance of charger density is incorporated into parameter $\gamma_0 = 1/\sqrt{1-\beta_0^2}$, Eqns (\ref{fieldsHarris} -(\ref{fieldsrotational}). In all charged simulations we use $\gamma_0=2$.

Table \ref{table} lists the parameters used in simulations
   \begin{table}
  \begin{tabular}{cccccc}
     {\rm Run} & $\sigma$ & $\delta/L$ &  $\Theta $ & $n/n_0$ & $r_D/L$ \\
     Cold-1 & $5 \times 10^{-6}$ & 5.3 & 0& 120  & 0\\
      Cold-2 & $5 \times 10^{-7}$ & 1.7 & 0& $1.2 \times 10^3$ &  0\\
     Warm-1 & $5 \times 10^{-6}$ & 5.3 & 0.01& 120 & 0.5\\
       High-B-Cold& $5 \times 10^{-4}$ & 1.7 & 0.0& 120 & 0.5\\
     High-B& $5 \times 10^{-4}$ & 5.3 & 0.01& 12 & 0.5\\
       \end{tabular}
  \caption{Simulation parameters.}
  \label{table}
  \end{table}
In practice, we use $L=10^{-6}$ m, \Bf\ field $b_0=10-100-1000$ Tesla, plasma densities $n= 10^{24-25} $ particles per meter cubed. All 2D simulations use 30 particles per cell, resolution $2048\times 1024$.  For parameter scan, Fig. \ref{plot_b0_scan} right plot, $b_0$ is the \Bf\ in Tesla.

\section{Double-sheets set-up}
\label{needfordouble}

Using analytical set-ups for rotational and charged current sheets \S \ref{setup}, we set {\it  double}  current sheet configurations,  Fig. \ref{configurations} as 
it  turns out that  reasonable-size  simulations of both the charged Harris  and  rotational  sheets  require douse-sheet configuration. Reasons for the double-sheet configurations  are subtle, and different for Harris and rotational set-up.

In the rational configuration, where \Bf\ experiences 180 degrees rotation and \Ef\ is non-switching, there is  a potential difference between top and bottom boundaries. This leads to large-scale charge drifts that quickly interferes with reflecting or open boundary conditions. We
then set-up double rotational configuration, with no potential difference between top and bottom boundaries. 
(There is mild difference between the case of full rotation and forward-back cases.)

In case of Harris configuration, 
while the Harris sheet analytically maintains force balance everywhere across an infinite domain,  it requires finite background pressure. As a result,  truncating that continuum  with reflecting boundaries forces three distinct numerical violations that launch powerful displacement waves inwards
\begin{itemize}
\item  At the  $z$  boundary, the residual current is small but not zero. Since reflecting  boundary behaves as a perfect conductor, applying a Neumann boundary condition ($\partial B_x / \partial z = 0$) at the boundary ghost cells. Because the boundary enforces $\nabla \times \mathbf{B} = 0$, but the particles are initialized with $J_y \neq 0$, Ampere's Law is instantly violated exclusively at the wall. This violently launches a high-frequency $E_y$ electromagnetic wave directly from the boundary inward at the speed of light. 
\item 
At the reflecting boundary, macro-particles bounce back exactly at the interface. However, because the computational boundary cell only captures half exactly "half" the volumetric weighting fraction of a bulk cell during particle-to-grid shape deposition, the measured density drops artificially at the wall. This small numerical grid artifact generates an explicitly localized pressure gradient  purely at the wall, which launches strong longitudinal acoustic/fast-magneto-sonic waves
\item 
 Thermal Statistical Noise. 
We mathematically initialized perfectly smooth continuous magnetic gradients ($\partial B_x / \partial z$), but instantiated the current ($J_{PIC}$) using discrete macro-particles. The statistical positioning thermal noise ($\propto 1/\sqrt{N_{particles} }$) across the massive   background density means the localized particle current $J_{PIC}$ fluctuates wildly around the true mean. This bulk mismatch  instantly creates high-frequency $\partial E / \partial t$ waves everywhere, which aggressively slam into and reflect off the rigid boundary matrices, doubling the visible boundary chaos.
\end{itemize}

\end{document}